\documentclass[prb,twocolumn,notitlepage,superscriptaddress,longbibliography]{revtex4-1}
\usepackage{amsmath,amsfonts,amssymb,amsthm,epsfig,array}
\usepackage{dsfont}
\usepackage{slashed}
\usepackage{graphics}
\usepackage{float}
\usepackage{verbatim}%\begin{comment} ... \end{comment}
\usepackage{color}
\usepackage[dvipsnames]{xcolor}
\usepackage[hidelinks,colorlinks,linkcolor=blue,
citecolor=blue,urlcolor=blue]{hyperref}
\usepackage[titletoc,title]{appendix}
\usepackage{multirow}
\usepackage{bibentry}

\setcitestyle{number,sort,open={(},close={)}}

\newcommand{\eq}[1]{\begin{equation} #1 \end{equation}}
\newcommand{\eqarr}[1]{\begin{eqnarray} #1 \end{eqnarray}}

\let\oldAA\AA
\renewcommand{\AA}{\text{\normalfont\oldAA}}

\usepackage[mathscr]{euscript}

\begin{document}
\title{Phonon helicity and Nieh-Yan Anomaly in the Kramers-Weyl semimetals of Chiral Crystals}
\author{Chao-Xing Liu}
\email{cxl56@psu.edu}
\affiliation{Department of Physics, the Pennsylvania State University, University Park, PA 16802}
\begin{abstract}
Nieh-Yan anomaly describes the non-conservation of chiral charges induced by the coupling between Dirac fermions and torsion fields. Since the torsion field is beyond general relativity, this effect remains hypothetical and its relevance to our universe is unclear in the context of high-energy physics. In this work, we propose that the phonons can induce a torsion field for the Kramers-Weyl fermions through electron-phonon interaction in a non-magnetic chiral crystal, thus leading to the occurrence of the Nieh-Yan anomaly in this condensed matter system. As a consequence, the Nieh-Yan term can strongly influence the phonon dynamics and lead to the helicity of acoustic phonons, namely, two transverse phonon modes mix with each other to form a circular polarization with a non-zero angular momentum at a finite phonon momentum and the phonon angular momentum reverses its sign for opposite momenta due to time reversal symmetry. The phonon helicity can be probed through measuring the total phonon angular momentum driven by a temperature gradient.
%In particular, the induced phonon angular momentum will depend on $T^2$ for the temperature $T$ as a direct result of the thermal Nieh-Yan anomaly.
%Chiral crystals are materials with a well-defined handedness and non-magnetic chiral crystals can generally host the so-called "Kramers-Weyl" fermions at time-reversal-invariant momenta when taking into account spin-orbit coupling [cite]. In this work, we show that the phonons can induce a torsion field for the Kramers-Weyl fermions through electron-phonon coupling in chiral crystals. This is in sharp contrast to the conventional Weyl semimetals, in which the phonons mainly play the role of pseudo-gauge fields. As a result, we find that the Nieh-Yan anomaly due to the torsion field can be driven by phonons and can give rise to the phonon angular momentum of both acoustic (elastic wave) and optical phonons in the Kramers-Weyl semimetal phase of chiral crystals.
\end{abstract}
\maketitle

{\it Introduction -}
Weyl fermion is a two-component relativistic fermion with a definite chirality and serves as the building block for fermions in quantum field theory \cite{peskin2018introduction}. The chirality of Weyl fermions can give rise to a variety of physical effects, including chiral anomaly \cite{fujikawa2004path,bertlmann2000anomalies}, chiral magnetic effect\cite{fukushima2008chiral}, mixed axial-gravitational anomaly\cite{delbourgo1972gravitational,eguchi1976quantum}, chiral torsion effect \cite{khaidukov2018chiral,sumiyoshi2016torsional,imaki2019lattice} and Nieh-Yan anomaly\cite{nieh1982quantized,nieh1982identity,chandia1997topological,nieh2007torsional}. Weyl fermions can also emerge as low-energy quasiparticles in condensed matter systems. These systems are dubbed "Weyl semimetals"\cite{armitage2018weyl,yan2017topological,hosur2013recent,vafek2014dirac,hasan2017discovery}, which have been demonstrated in a number of materials through observing the surface Fermi arc\cite{lv2015experimental,yang2015weyl,yang2015weyl,liu2019magnetic,xu2015discovery,xu2015experimental}, a negative magnetoresistance\cite{huang2015observation,burkov2015chiral}, a negative magneto-thermoelectric resistance\cite{gooth2017experimental}, {\it et al}\cite{armitage2018weyl,yan2017topological,hosur2013recent,vafek2014dirac,hasan2017discovery}.
%The Kramers-Weyl (KW) fermion is one type of Weyl fermion that exists in chiral crystals [cite paper] that possess a well-defined handedness. It was shown that the KW fermion is a universal topological electronic property of all non-magnetic chiral crystals with spin-orbit coupling and protected by time-reversal (TR) symmetry. Due to the maximum separation of the location of Weyl nodes in the momentum space in chiral crystals, the KW fermions possess several advantages in exploring Weyl physics, as compared to other conventional Weyl semimetals [cite].

In Weyl semimetals, position-dependent and time-dependent perturbations, such as magnetic fluctuations\cite{liu2013chiral}, strain\cite{pikulin2016chiral,grushin2016inhomogeneous} and structure inhomogeneity\cite{jia2019observation,peri2019axial}, can shift the position of Weyl nodes and thus act as an emergent gauge field, dubbed the "pseudo-gauge field"\cite{ilan2020}, which provides a unified description of various physical phenomena in Weyl semimetals. The chiral zero Landau levels induced by pseudo-gauge fields have been observed experimentally in photonic and acoustic Weyl metamaterials\cite{jia2019observation,peri2019axial}. Furthermore, these perturbations can also give corrections to the Fermi velocity of Weyl fermions and thus play the role of the frame fields (also called tetrad or vierbein)\cite{weinberg1972gravitation,parker2009quantum}, allowing to mimic Weyl fermions in the curved spacetime\cite{liang2019curved,weststrom2017designer,gooth2017experimental,jia2020chiral,zubkov2015emergent,cortijo2016emergent}. An exciting theoretical proposal is to realize the Nieh-Yan anomaly\cite{nieh1982quantized,nieh1982identity,chandia1997topological,nieh2007torsional} due to the torsion field in Weyl semimetals\cite{hughes2013torsional,parrikar2014torsion,huang2020nieh,nissinen2019thermal,nissinen2020emergent,nissinen2020thermal,laurila2020torsional,huang2020hamiltonian,huang2020torsional}, which may be probed through topological magnetotorsional effect\cite{liang2020topological}, anomalous thermal Hall conductance\cite{huang2020nieh}, or the sound-wave-induced current oscillations in tilted Weyl semimetal interfaces under magnetic fields\cite{ferreiros2019mixed}. However, one challenging still remains: since the perturbation terms normally shift the Weyl node location and modify the Fermi velocity simultaneously, both the chiral anomaly and Nieh-Yan anomaly can be induced by these perturbation terms, and thus their roles as the pseudo-gauge fields and the frame fields are difficult to distinguish. Furthermore, it is still unclear how the quasiparticle excitations due to these perturbations, such as elastic wave, phonon dynamics and magnetic dynamics, are influenced by the Nieh-Yan anomaly of Weyl fermions.

\begin{figure}[!htbp]
	\centering
	\includegraphics[width=1\linewidth]{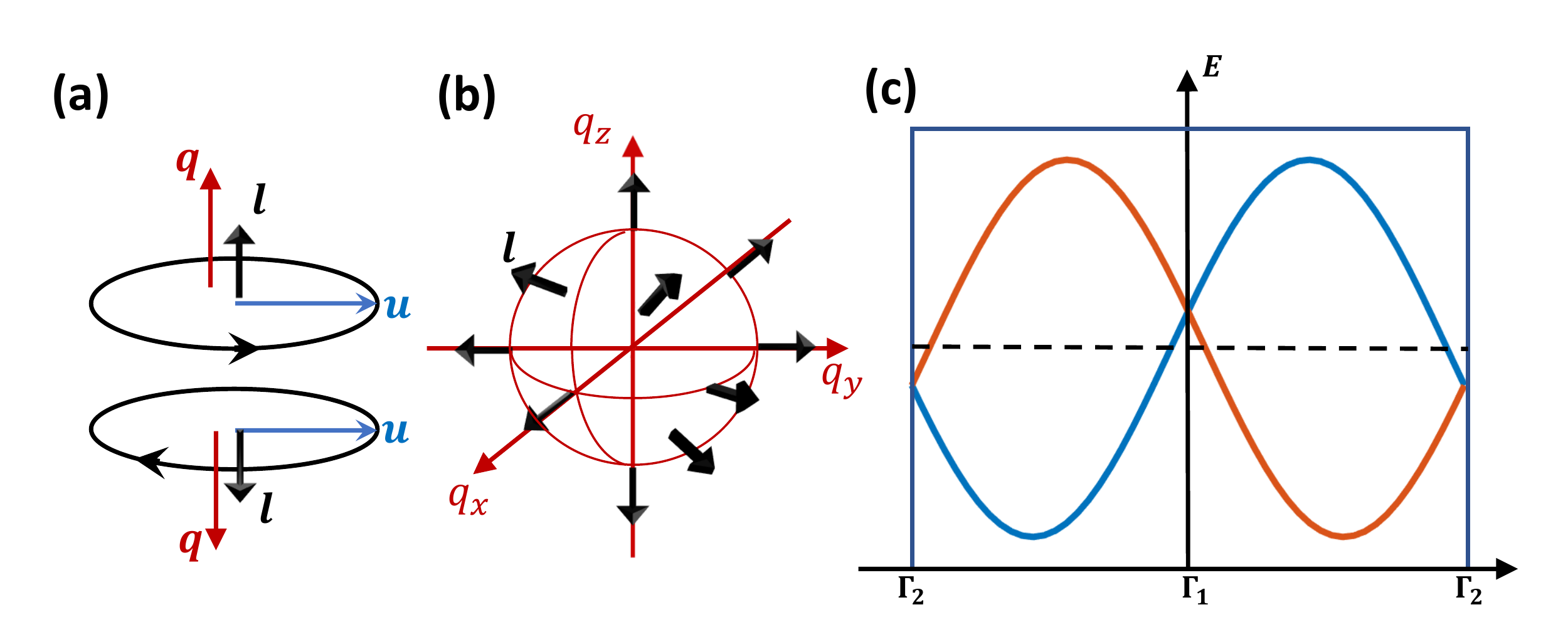}
	\caption{ (a) The PAM ${\bf l}$ is parallel with the phonon momentum ${\bf q}$ and its sign reverses for opposite ${\bf q}$ for one branch of phonon modes. This feature of transverse phonon modes is referred as "phonon helicity". Here ${\bf u}$ labels the displacement vector. (b) The hedgehog texture of the PAM ${\bf l}$ is formed in the ${\bf q}$-space. (c) Energy dispersion between two time-reversal invariant momenta ${\bf \Gamma}_1$ and ${\bf \Gamma}_2$ for the KW semimetal phase of chiral crystals. The black dashed line depicts the Fermi energy, near which the linear term dominates.  }
	\label{fig1}
\end{figure}

To address these challenges, we turn to the so-called "Kramers-Weyl" (KW) fermions in chiral crystals. Chiral crystals are the crystal structures with a well-defined handedness and non-magnetic chiral crystals can generally host the KW fermions when taking into account spin-orbit coupling (SOC) \cite{chang2018topological,sanchez2019topological,schroter2019chiral,schroter2020observation,li2019chiral}. The location of KW fermions is pinned at time-reversal (TR) invariant momenta by TR symmetry. Consequently, any TR-preserving perturbations cannot shift the KW fermions in the momentums space and thus will not induce the vector potential of pseudo-gauge field for KW fermions. In this work, we will consider the role of acoustic phonons in the KW semimetal phase of chiral crystals and demonstrate that the phonons will induce a torsion field for the KW fermions. By integrating out the KW fermions, we prove that the phonon self-energy contains an off-diagonal term that originates from the Nieh-Yan anomaly of the torsion fields. While this self-energy correction has no influence on the longitudinal phonon mode, it will mix two transverse modes and give rise to the phonon angular momentum (PAM) at a finite phonon momentum. In particular, the induced PAM reverse its sign for opposite phonon momentum, as shown in Fig. \ref{fig1}a, and form a hedgehog texture in the momentum space for one phonon branch (Fig. \ref{fig1}b), analogous to the helical spin texture in spin-orbit coupled electronic band structures. Thus, we term it as "phonon helicity"\cite{hu2021phonon}. The phonon helicity can be probed through measuring the total PAM induced by a temperature gradient \cite{hamada2018phonon}.

{\it Electron-phonon Coupling in Chiral Crystals -}
In non-magnetic chiral crystals, all the energy bands are at least doubly degenerate (spin degeneracy) at high symmetry momenta, labelled as ${\bf \Gamma}_i$ ($i=1,2...,8$), and SOC can lift this spin degeneracy for the momentum away from ${\bf \Gamma}_i$, thus giving rise to the KW fermions. If the full rotation symmetry is assumed in the low-energy sector of the KW fermions (see Appendix \ref{App_Sec:kpTheory} and \ref{App_Sec:SymmetryHam}), the low energy physics can be described by an isotropic Hamiltonian $H_0=\frac{\hbar^2 k^2}{2m_0^*}+\hbar v_f {\bf k\cdot \sigma}-\mu$ expanded around ${\bf \Gamma}_i$ up to the $k^2$ terms, where $\sigma$ labels the Pauli matrix for spin, $m_0^*$ is the effective mass, $v_f$ is the Fermi velocity and $\mu$ is the chemical potential. As discussed in Ref [\onlinecite{he2019kramers}], this isotropic Hamiltonian remains valid for the chiral point groups $T$ and $O$, while for other chiral point groups, the Fermi velocity $v_f$ becomes anisotropic and should be replaced by a tensor. We only focus on the isotropic Hamiltonian in this work. In chiral crystals, there are normally multiple Fermi surfaces, and here we only focus on the parameter regime where the linear-$k$ terms dominate over the $k^2$ terms at the Fermi surfaces around different ${\bf \Gamma}_i$, as depicted in Fig. \ref{fig1}c. We call this parameter regime as the "KW semimetal phase", in which the $k^2$ term only limit the valid regime in the momentum space for Weyl physics and thus can be replaced by a momentum cut-off $\Lambda=\frac{2m_0^*v_f}{\hbar}$ (assuming $\mu\ll \frac{\hbar^2 \Lambda^2}{2m_0^*}$).
%Thus, we only keep the linear term and chemical potential in $H_0$ with the momentum cut-off below.

The electron-phonon coupling in this system can be derived from the standard ${\bf k\cdot p}$ theory up to the second order perturbation terms with the help of Schrieffer-Wolf transformation, as discussed in Appendix \ref{App_Sec:kpTheory}. The TR symmetry $\hat{T}=i\sigma_y\mathcal{K}$, where $\mathcal{K}$ is the complex conjugate, gives a strong constraint on the form of the resulting effective Hamiltonian. In this work, we focus on the acoustic phonons, which couple to electrons through an internal strain, described by the strain tensor $u_{ij}=\frac{1}{2}(\partial_iu_j+\partial_ju_i)$ with the displacement vector $u_i$ ($i,j=x,y,z$). Since the strain tensor is even under $\hat{T}$, it cannot directly couple to either electron momentum ${\bf k}$ or spin ${\bf \sigma}$, both of which are odd under $\hat{T}$. This implies that the strain tensor {\it cannot} play the role of the vector potential of pseudo-gauge field. Based on the symmetry construction in Appendix \ref{App_Sec:SymmetryHam}, the symmetry-allowed strain-electron coupling Hamiltonian for an isotropic KW fermion is written as $H_{ep}=(C_1+g_0k_j\sigma_j) u_{ii}+g_1 u_{ij}k_i\sigma_j$ up to the order of $k_i\sigma_j$, since both $u_{ij}$ and $T_{ij}=k_i\sigma_j$ are rank-2 tensors. We have assumed the summation over the duplicated indices. $C_1, g_0, g_1$ are three independent parameters. The $C_1$ term provides a correction to the chemical potential, while both $g_0$ and $g_1$ terms give the corrections to the Fermi velocity tensor. Microscopically, both $g_0$ and $g_1$ terms are from the second order perturbation, combining the ${\bf k\cdot p}$ term and the bare electron-phonon coupling term, as shown in Appendix \ref{App_Sec:kpTheory}.

Putting $H_0$ and $H_{ep}$ together, one can derive the effective action for the KW semimetals around ${\bf \Gamma}_i$, given by $S_{eff}=\int d\tau d^3r\mathcal{L}_{eff}$ with (See Appendix \ref{App_Sec:Action})
\eqarr{\label{eq:Leff1}
\mathcal{L}_{eff}=\hat{\psi}^{\dag}_{\bf \Gamma_i}\left(\frac{\partial}{\partial\tau}-\mu-A_0({\bf r})+\frac{\chi}{2}\{e^j_a,(-i\partial_j)\}\sigma^a\right)\hat{\psi}_{\bf \Gamma_i},\nonumber\\
}
where $\hat{\psi}_{\bf \Gamma_i}$ is the field operator for the KW fermions, $\tau$ is the imaginary time, $A_0=C_1u_{ii}$, $\chi=sign(\hbar v_f)$ and $e^j_a=\delta^j_a+\Delta^j_a$ with the ${\bf \Delta}$ field given by
\eq{\Delta^j_a=g_0u_{ii}\delta_{ja}+g_1 u_{ja}\label{eq:Delta}}
where $i,j,a=x,y,z$. Here the imaginary-time formalism is used for the study of finite temperature effect. The Fermi velocity $\hbar v_f$ is absorbed by re-scaling the spatial coordinate, while we still keep track of the chirality $\chi$ (the sign of Fermi velocity). The $e^j_a$ describes the frame field\cite{weinberg1972gravitation,parker2009quantum} and it only involves non-trivial spatial part due to TR symmetry. The effective Lagrangian (\ref{eq:Leff1}) describes the Weyl fermions in a space with non-trivial frame field but zero spin connection, which is known as the "Weitzenb$\ddot{o}$ck spacetime" studied in "teleparallel gravity theory"\cite{hayashi1979new}. Zero spin connection means vanishing Riemann curvature while non-trivial frame field implies non-zero torsion field, given by $T^a_{ij}=\partial_i \bar{e}^a_j-\partial_j \bar{e}^a_i$, where $\bar{e}^a_i$ is the coframe field define by $\bar{e}^a_i e^j_a=\delta^j_i$ (See Appendix \ref{App_Sec:NYanomaly}). Thus, strain or phonon can induce a torsion field for the KW fermions in chiral crystals.

{\it Stress-stress correlation function and Nieh-Yan anomaly -}
Our next step is to integrate out Weyl fermions to obtain an effective action for the ${\bf \Delta}$ field that is directly related to the strain field. To do that, we separate the effective Lagrangian into $\mathcal{L}_{eff}=\mathcal{L}_0+\mathcal{L}_1$, where $\mathcal{L}_0=\hat{\psi}^{\dag}_{\bf \Gamma_i}\left(\frac{\partial}{\partial\tau}-\mu+\chi((-i{\bf \nabla})\cdot {\bf \sigma})\right)\hat{\psi}_{\bf \Gamma_i}$ and $\mathcal{L}_1=\hat{\psi}^{\dag}_{\bf \Gamma_i}\left(-A_0+\frac{\chi}{2}\{\Delta^j_a,(-i\partial_j)\}\sigma^a\right)\hat{\psi}_{\bf \Gamma_i}$. Here both the $A_0$ and ${\bf \Delta}$ fields in $\mathcal{L}_1$ are related to the strain field induced by acoustic phonons and thus treated as perturbations. By integrating out the KW fermions in $\mathcal{L}_{eff}$ (see Appendix \ref{App_Sec:Action}), we find the effective action $W[A_0,\Delta]$ for the $A_0$ and ${\bf \Delta}$ fields (Eq. \ref{App_eq:Weff1} in Appendix \ref{App_Sec:Action}), which provide the corrections to the phonon effective action. Here we focus on the term $W_{NY}$ for the ${\bf \Delta}$ field
\eq{\label{eq:WNY}
W_{NY}[{\bf \Delta}]=\frac{1}{2}\sum_{\tilde{q}}\Delta^i_a(\tilde{q})\Delta^j_b
(-\tilde{q})\Phi^{ab}_{ij}(\tilde{q})
}
where $\tilde{q}=({\bf q},i\nu_m)$ and $\sum_{\tilde{q}}=\frac{1}{\beta V}\sum_{{\bf q},i\nu_m}$. The stress-stress correlation function $\Phi^{ab}_{ij}$ is defined as
\eq{
\Phi^{ab}_{ij}(\tilde{q})=\sum_{\tilde{k}}Tr_{\sigma}\left(\mathcal{T}^{a}_i(\tilde{k},\tilde{k}-\tilde{q})
\mathcal{G}_0(\tilde{k}-\tilde{q})\mathcal{T}^{b}_j(\tilde{k}-\tilde{q},\tilde{k})\mathcal{G}_0(\tilde{k})\right),\label{eq:Phi0}\\
}
where $\tilde{k}=({\bf k},i\omega_n)$, $\mathcal{G}_0=\left(i\omega_n+\mu-\chi({\bf k\cdot\sigma})\right)^{-1}$ and the stress tensor $\mathcal{T}^{a}_i({\bf k, k'})=\frac{\chi}{2}(k_i+k'_i)\sigma^a$ with $i,a=x,y,z$.
%the single-particle Green's function is

Our next task is to show that $\Phi^{ab}_{ij}$ includes a contribution from the Nieh-Yan anomaly. To see that, we consider $i\nu_m=0$ and treat both the momentum ${\bf q}$ and the chemical potential $\mu$ as the perturbations. In this limit, we can first expand $\Phi^{ab}_{ij}({\bf q})\approx\Phi^{ab}_{ij}(0)+(\partial_{q_l}\Phi^{ab}_{ij})_{\bf q=0}q_l$ up to the linear order in ${\bf q}$. Since ${\bf \Delta}$ is directly proportional to the strain tensor, the zero order term $\Phi^{ab}_{ij}(0)$ just provides the corrections to the elastic moduli (See Appendix \ref{App_Sec:phonon}). Thus, below we focus on the linear-${\bf q}$ term with its coefficient denoted as $\Phi^{ab}_{ij,l}=(\partial_{q_l}\Phi^{ab}_{ij})_{\bf q=0}$, which can be further expanded as $\Phi^{ab}_{ij,l}(\mu)\approx\Phi^{ab}_{ij,l}(\mu=0)+(\partial_{\mu}\Phi^{ab}_{ij,l})_{\mu=0}\mu$ up to the linear order in $\mu$. In Appendix \ref{App_Sec:Analytical}, direct calculations show that $\Phi^{ab}_{ij,l}(\mu=0)=0$ and $(\partial_{\mu}\Phi^{ab}_{ij,l})_{\mu=0}=i\epsilon^{alb}\delta_{ij}\chi \mathcal{F}_0$ with
$\mathcal{F}_0=\frac{1}{3\pi^2\beta}\sum_{i\omega_n}\int_0^{\Lambda} k^4dk \frac{1}{D^4}\left(1-\frac{4(i\omega_n)^2}{D^2}\right)$ and $D^2=(i\omega_n)^2-k^2$. With these approximations, the effective action is
\eq{\label{eq:WNY}
W_{NY}[{\bf \Delta}]=-\epsilon^{alb}\delta_{ij}\mu_5\mathcal{F}_0\int d^3r d\tau \Delta^i_a\left(\partial_l\Delta^j_b\right)
}
after the Fourier transform into the real space, where $\mu_5=\frac{\chi\mu}{2}$ has the physical meaning of chiral chemical potential.

Now we will show that $W_{NY}$ is a manifestation of the Nieh-Yan anomaly. To see that, let's recall the Nieh-Yan anomaly equation $\partial_\mu \langle j^{5\xi}\rangle=\frac{\mathcal{F}}{4}\eta_{ab}\epsilon^{\xi\nu\lambda\rho}T^a_{\xi\nu}T^b_{\lambda\rho}$ with the axial 4-current $\langle j^{5\xi}\rangle$ for one Dirac fermion, the torsion field $T^a_{\xi\nu}=\partial_{\xi}\bar{e}^a_{\nu}-\partial_{\nu}\bar{e}^b_{\xi}$, and certain parameter $\mathcal{F}$ (See Appendix \ref{App_Sec:NYanomaly} for a short review). This anomaly equation can be obtained from $\langle j^{5\xi}\rangle=\delta S_{NY}/\delta A_{5\xi}$ with the effective action $S_{NY}=\mathcal{F} \int d^4x \eta_{ab}\epsilon^{\xi\nu\lambda\rho}A_{5\xi} \bar{e}^a_{\nu} \partial_{\lambda} \bar{e}^b_{\rho}$\cite{liang2020topological}, where $A_{5\xi}$ is the axial gauge potential. Now let's consider two KW fermions with opposite chiralities at two TR-invariant momenta, which together form one massless Dirac fermion. For this case, the chiral chemical potential $\mu_5$ in Eq. (\ref{eq:WNY}) should be given by $\mu_5=\frac{1}{2}(\mu_+-\mu_-)$ with $\pm$ for the chiralities of two KW fermions. Since $\mu_5$ is identical to $A_{50}$ and the ${\bf \Delta}$ field gives the spatial part of $\bar{e}^a_{\nu}$, one finds that $W_{NY}$ takes the exactly same form as the terms with $\xi=0$ in $S_{NY}$.

Next let's discuss the coefficient $\mathcal{F}_0$ in $W_{NY}$, which is found to take form of $\mathcal{F}_0=F_0+F_1(k_BT)^2$ for the finite temperature (See Appendix \ref{App_Sec:Analytical}). A simple dimension counting suggests that $\mathcal{F}_0$ has the dimension of $\frac{1}{L^2}$ where $L$ is the length dimension. Thus, $F_0$ is {\it not} universal and generally depends on the momentum cut-off $F_0\propto \Lambda^2$. The value of $F_0$ depends on the normalization scheme and our calculation with the momentum cut-off $\Lambda$ gives $F_0=0$ for the lowest order perturbation term. On the other hand, for the temperature-dependent term, $(k_BT)^2$ exactly carries the dimension of $\frac{1}{L^2}$. Here we follow the convention in high energy physics with $e=\hbar=v_f=1$, so that energy and momentum have the same dimension. Consequently, the parameter $F_1$ is dimensionless. It was argued that this parameter is universal and reflects the central charge of Dirac fermions\cite{huang2020nieh}. Our calculations gives $F_1=-\frac{1}{12}$, consistent with the results in literature obtained from different approaches\cite{huang2020nieh,nissinen2019thermal,nissinen2020thermal}, and this $T^2$-dependent term is called "thermal Nieh-Yan anomaly".

\begin{figure}[!htbp]
	\centering
	\includegraphics[width=1\linewidth]{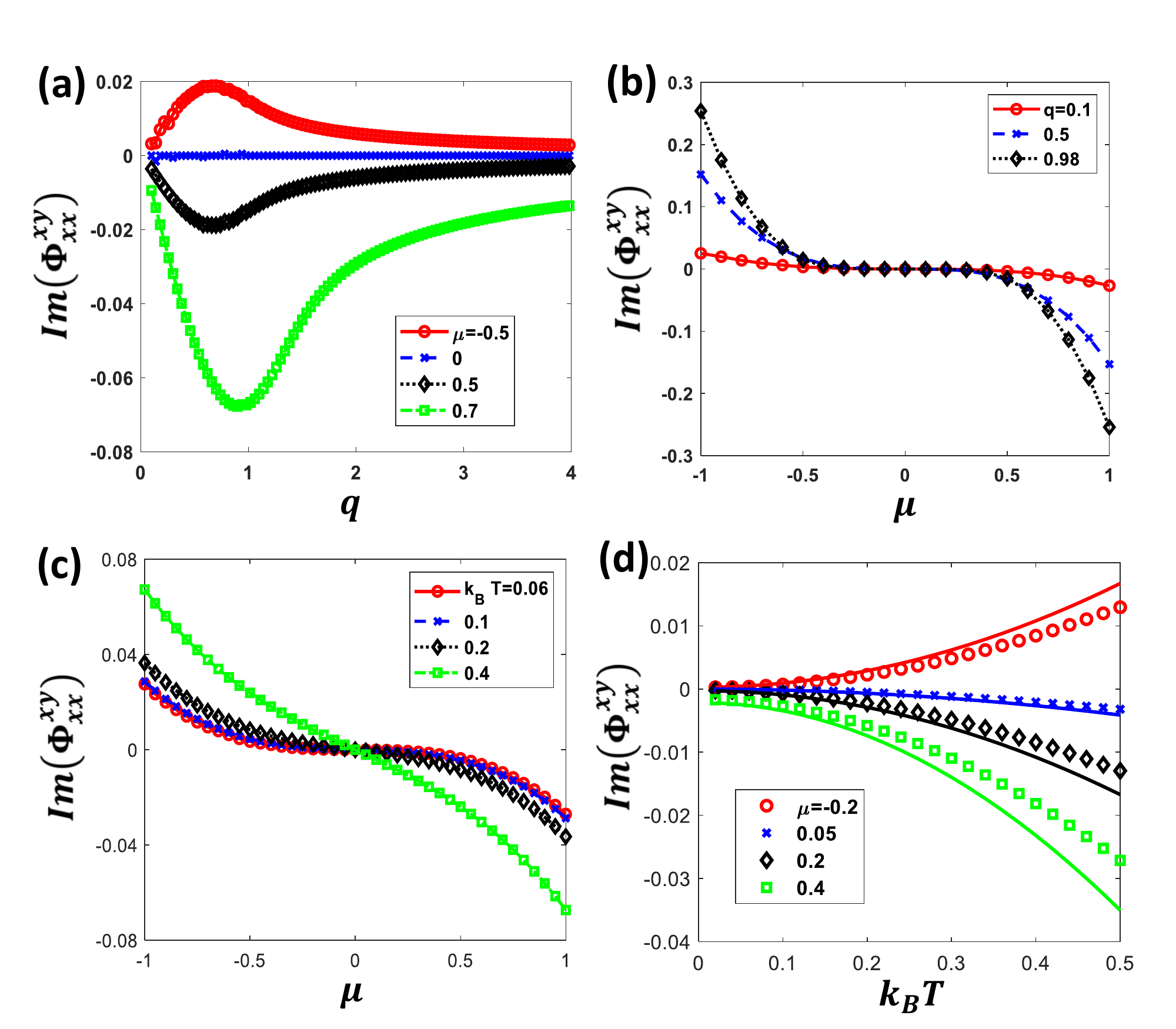}
	\caption{ (a) $Im(\Phi^{xy}_{xx})$ as a function of $q$ for $\omega=0$, $k_B T=0.05$ and $\mu=-0.5, 0, 0.5, 0.7$; (b) $Im(\Phi^{xy}_{xx})$ as a function of $\mu$ for $\omega=0$, $k_B T=0.05$ and $q=0.1, 0.5, 0.98$; (c) $Im(\Phi^{xy}_{xx})$ as a function of $\mu$ for $k_B T=0.06, 0.1, 0.2, 0.4$, $\omega=0$ and $q=0.1$; (d) $Im(\Phi^{xy}_{xx})$ as a function of $k_B T$ for $\omega=0$, $q=0.1$ and $\mu=-0.2, 0.05, 0.2, 0.4$. Here the lines are from the analytical expression (\ref{eq:Phixyxx}). The length is in unit of the lattice constant $a_0$, so that the momentum is in unit of $1/a_0$. The energy is in unit of $\hbar v_f/a_0$ and $\Phi^{xy}_{xx}$ is in unit of $\hbar v_f/a_0^4$.      }
	\label{fig2}
\end{figure}

The above discussion is for a single Dirac fermion with $\mu\rightarrow 0$. In realistic systems, multiple KW fermions can exist at different TR-invariant momenta ${\bf \Gamma}_i$ near the Fermi energy. Since our formalism here treats each KW fermion separately, we can simply introduce a single chiral chemical potential $\mu_5=\frac{1}{2}\sum_{{\bf \Gamma}_i}\chi_{{\bf \Gamma}_i}\mu_{{\bf \Gamma}_i}$ for different ${\bf \Gamma}_i$ into Eq. (\ref{eq:WNY}). Thus, below we just focus on a single KW fermion. For each KW fermion, the chemical potential $\mu$ with respect to the Weyl point can be far away from zero. This requires us to numerically evaluate the full expression (\ref{eq:Phi0}) for $\Phi^{ab}_{ij}({\bf q},i\omega_n)$. Here we consider the momentum ${\bf q}$ along the z-direction (${\bf q}=q\hat{e}_z$) and numerically compute the component $\Phi^{xy}_{xx}$ (See Appendix \ref{App_Sec:Numerical} for the numerical methods), which is shown in Fig. \ref{fig2}. $\Phi^{xy}_{xx}$ is found to be purely imaginary and $Im(\Phi^{xy}_{xx})$ as a function of $q$ is shown in Fig. \ref{fig2}a for different $\mu$. $Im(\Phi^{xy}_{xx})$ is linearly proportional to $q$ for $q\ll 1$. Furthermore, $Im(\Phi^{xy}_{xx})$ vanishes at $\mu=0$ and its sign reverses for opposite $\mu$. Fig. \ref{fig2}b shows the dependence of $Im(\Phi^{xy}_{xx})$ on $\mu$ for different $q$. $Im(\Phi^{xy}_{xx})$ depends on $\mu$ asymmetrically, and for $\mu$ close to zero, one can see that $Im(\Phi^{xy}_{xx})$ is flat and almost zero. This suggests that the linear $\mu$ term in $Im(\Phi^{xy}_{xx})$ should almost vanish, consistent with our analytical result $F_0=0$. Fig. \ref{fig2}c shows $Im(\Phi^{xy}_{xx})$ as a function of $\mu$ for different temperatures $k_B T$. With increasing temperature, the linear-$\mu$ term in $Im(\Phi^{xy}_{xx})$ gradually appears. Finally, we plot $Im(\Phi^{xy}_{xx})$ as a function of $k_B T$ for different $\mu$ in Fig. \ref{fig2}d, which shows a $T^2$-dependence, as required by thermal Nieh-Yan anomaly. At small $\mu$ and low $k_B T$, this temperature dependence can be well captured by the expression
\eq{\label{eq:Phixyxx}
Im(\Phi^{xy}_{xx})\approx -q\left(\frac{\mu^3}{12\pi^2}+\frac{\mu (k_B T)^2}{12}\right)
}
from the perturbation expansion (See Appendix \ref{App_Sec:Numerical}), as shown by the blue curve in Fig.\ref{fig2}d for $\mu=0.05$. When increasing $\mu$, a derivation from Eq. (\ref{eq:Phixyxx}) is found for the red, black and green curves in Fig.\ref{fig2}d for $\mu=-0.2, 0.2, 0.4$, respectively, implying that the terms with higher orders of $\mu$ come into play.

%Since $\mu$ only appears in $\mathcal{G}_0$, we may first expand $\mathcal{G}_0(\mu)=\mathcal{G}_0(\mu=0)+\mu(\partial_{\mu}\mathcal{G}_0)_{\mu=0}+...$ with $\partial_{\mu}\mathcal{G}_0=-(\mathcal{G}_0)^2$.

{\it Helicity of acoustic phonons -}
Since the ${\bf \Delta}$ field in the effective action $W_{NY}$ is determined by the strain field ${\bf u}$ in Eq. (\ref{eq:Delta}), one may naturally expect this term will influence the acoustic phonon dynamics (elastic wave). As discussed in Appendix \ref{App_Sec:phonon}, this term will provide a correction $\int d^3r dt\xi_{ijklm}(\partial_i u_{jk})u_{lm}$ to the effective action of acoustic phonons. In a uniform isotropic system with TR symmetry, there is only a single independent parameter, labelled by $\xi_0$, for the rank-5 tensor $\xi_{ijklm}$. Thus, the effective phonon action is taken the form
\eq{
S_{ph,NY}=\xi_0\int dt d^3r \epsilon_{ikl}(\partial_i u_{jk})u_{jl},
}
where $\xi_0=\mu_5 \mathcal{F}_0 g_1^2$. Here we have changed from imaginary time to real time. This term gives rise to the correction to the equation of motion for acoustic phonons as
\eq{\label{eq:EOMphonon}
\frac{d^2}{dt^2}{\bf u}=c_t^2\nabla^2{\bf u}+(c_l^2-c_t^2)\nabla(\nabla\cdot{\bf u})+\frac{\xi_0}{2}\nabla\times(\nabla^2{\bf u}),
}
where ${\bf u}$ is the displacement field, and $c_{t(l)}$ is the velocity of transverse (longitudinal) modes. We can decompose ${\bf u}$ into the longitudinal and transverse components, ${\bf u}={\bf u}_l+{\bf u}_t$, and this new term only contributes to the transverse part and has no influence on the longitudinal component. With the wave solution ansatz ${\bf u}_t={\bf u}_0 e^{i({\bf q\cdot r}-\omega t)}$, the equation of motion of the transverse modes is changed to $\omega^2{\bf u}_0=c_t^2 q^2{\bf u}_0+\frac{\xi q^2}{2} i{\bf q}\times{\bf u}_0$, which possess the eigen-frequencies $\omega^t_s=\sqrt{c_t^2q^2+s\frac{|\xi_0|}{2}q^3}$ for the $s$-mode ($s=\pm$). More importantly, these two transverse modes carry angular momentum ${\bf l}_{s}({\bf q})$, defined as $l_{s,i}({\bf q})=\hbar {\bf u}^{\dag}_{0,s}({\bf q})M_i{\bf u}_{0,s}({\bf q})$ \cite{hamada2018phonon,zhang2014angular,hamada2020phonon,zhu2018observation,zhang2015chiral}, where ${\bf u}_{0,s}({\bf q})$ is the polarization vector for the mode $s=\pm$ and the ${\bf M}$ matrix is given by $(M_i)_{jk}=(-i)\epsilon_{ijk}$ with $i,j,k=x,y,z$. Direct calculation gives ${\bf l}_{s}({\bf q})=s \frac{\hbar {\bf q}}{q}$, which satisfies the relation ${\bf l}_{s}({\bf q})=-{\bf l}_{s}(-{\bf q})$ from the TR symmetry. Thus, the PAM ${\bf l}_{s}$ forms a hedgehog texture in the momentum space, as shown in Fig. \ref{fig1}b, resembling the helical spin texture of spin-orbit coupled electronic systems. Thus, we term this PAM texture ${\bf l}_{s}({\bf q})$ as "phonon helicity"\cite{hu2021phonon}.

\begin{figure}[!htbp]
	\centering
	\includegraphics[width=1\linewidth]{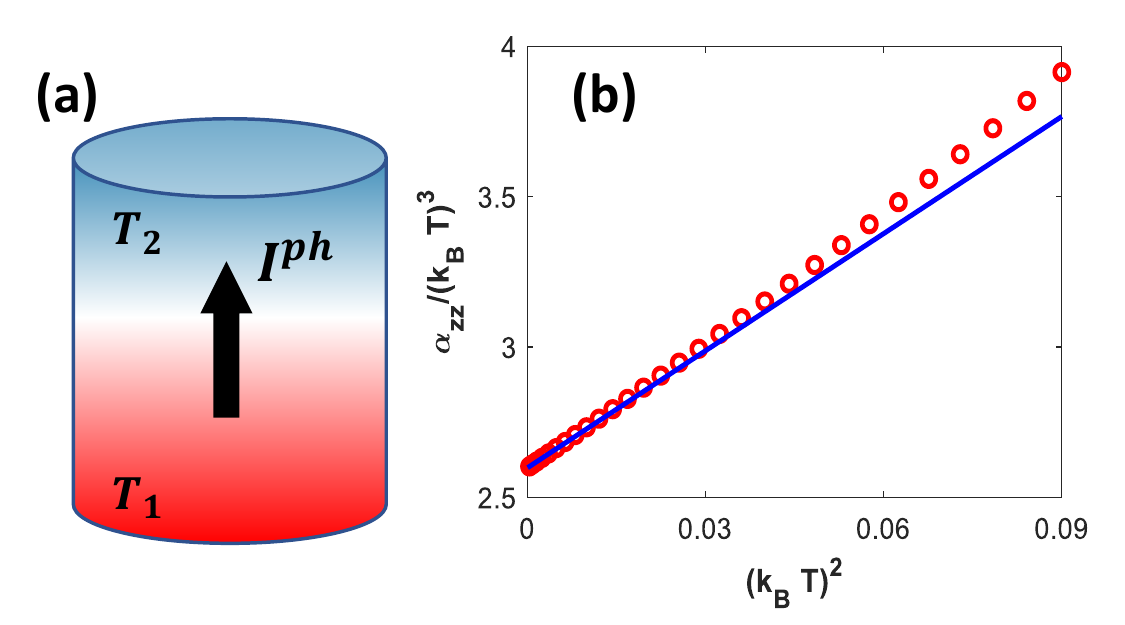}
	\caption{ (a) When there is a temperature gradient in the sample, $T_1>T_2$, a total PAM ${\bf I}^{ph}$ is generated. (b) $\alpha_{zz}/(k_B T)^3$ as a function of $(k_B T)^2$. Here the red circles are for the full numerical calculations of Eq. (\ref{App_eq:alphazz1}) in Appendix \ref{App_Sec:phonon} and the blue lines are from the analytical expression (\ref{eq:alphazz}). }
	\label{fig3}
\end{figure}

The total PAM is defined as ${\bf I}^{ph}=\sum_{s,{\bf q}} {\bf l}_s({\bf q}) (f(\omega^t_s)+\frac{1}{2})$, where $f(\omega^t_s)$ describes the distribution function for phonons. At the equilibrium state, the total PAM vanishes due to the TR symmetry. The phonon helicity ${\bf l}_{s}({\bf q})$ can give rise to a total phonon angular momentum ${\bf I}^{ph}$ by driving a thermal current with a temperature gradient\cite{hamada2018phonon}, as shown in Fig. \ref{fig3}a. This is in analog to the Edelstein effect of electrons due to spin-orbit coupling \cite{edelstein1990spin}. In the linear response regime, the response tensor $\alpha_{ij}$, defined as $I^{ph}_i=\alpha_{ij}\frac{\partial T}{\partial x_j}$, can be evaluated from the formula $\alpha_{ij}=-\frac{\tilde{\tau}}{V}\sum_{{\bf q},s}l_{s,i}v^{ph}_{s,j}\frac{\partial f_0(\omega^t_{s})}{\partial T}$ \cite{hamada2018phonon}, where $\tilde{\tau}$ is the phonon relaxation time, ${\bf v}^{ph}_{s}=\partial \omega^t_{s}/\partial {\bf q}$ is the group velocity of $s$-phonon mode and $f_0(\omega^t_{s})$ is the Bose-Einstein distribution function. Since our model is isotropic, the non-zero component $\alpha_{xx}=\alpha_{yy}=\alpha_{zz}$ is calculated as
\eq{\label{eq:alphazz}
\alpha_{zz}%\approx \frac{2\pi^2 \tau\hbar |\xi_0|k_B}{45c_t^5}\left(k_B T\right)^3\nonumber\\
%&&
\approx\frac{2\tilde{\tau}\hbar \pi^2 k_B g_1^2}{45 c_t^5}\mu_5 \left(F_0 \left(k_B T\right)^3+F_1\left(k_B T\right)^5\right).
}
Thus, by plotting $\alpha_{zz}/T^3$ as a function of $T^2$, our theory predicts a linear line dependence, as shown in Fig.\ref{fig3}b, and the intercept and the slope of the curve determine the contribution of the normal and thermal Nieh-Yan anomaly. Furthermore, we notice that the chiral chemical potential $\mu_5$ will determine the sign of $\alpha_{zz}$, which provides a testable signature in experiments.

{\it Discussion and Conclusion -}
To summarize, we have demonstrated that acoustic phonons can induce a torsion field for the KW fermions and consequently, the Nieh-Yan anomaly of the KW fermions can lead to the phonon helicity, which can be probed through measuring phonon angular momentum driven by a temperature gradient. Phonon dynamics in Weyl/Dirac semimetals was previously studied in the context of chiral anomaly \cite{rinkel2017signatures,rinkel2019influence,heidari2019hall,chernodub2019chiral,yuan2020discovery,sengupta2020phonon,sukhachov2021anomalous} and Kohn anomaly \cite{nguyen2020topological,yue2019soft}. The concept of PAM has recently been explored in the studies of the conversion between electron spin and mechanical rotation\cite{zhang2014angular}, selection rules for phonon-assisted electronic inter-valley scattering\cite{zhang2015chiral,mentink2019quantum}, phonon-induced spin relaxation\cite{garanin2015angular,bauke2014electron,nakane2018angular}, phonon-magnon interaction\cite{holanda2018detecting,thingstad2019chiral}, thermal Hall effect\cite{park2020phonon}, phonon roto-electric effect\cite{hamada2020phonon} and phonon orbital magnetic moments\cite{juraschek2019orbital}, and experimental evidence of PAM has been found\cite{holanda2018detecting,zhu2018observation}. The PAM due to temperature gradient can be measured through the rigid-body rotation or the phonon orbital magnetic moments, and the theoretical estimate for several realistic materials suggests these phenomena are detectable in experiments\cite{hamada2018phonon}. Temperature gradient can also drive electron spin polarization in spin-orbit-coupled systems\cite{xiao2016thermoelectric,dyrdal2013thermally,dyrdal2018thermally}, which shows a different temperature dependence from Eq. (\ref{eq:alphazz}) for PAM. In Appendix \ref{App_Sec:Interaction}, we also demonstrate that Coulomb interaction will not affect the Nieh-Yan term for the relevant phonon modes. A large number of chiral crystal materials for the KW fermions have been proposed in Ref \onlinecite{chang2018topological}, and to identify the best candidate materials requires more sophisticated first principles calculations of both electronic band structure and phonon spectrum in these compounds, which will be left for the future study.

{\it Acknowledgments--.}
We thank Z. Bi, I. Garate, L.H. Hu, Z.M. Huang, H.W. Jia, J. Jain, J.B. Yu, J.H. Zhou and S. Zhang for helpful discussions.
C.-X.~Liu acknowledges the support of the Office of Naval Research (Grant No. N00014-18-1-2793) and Kaufman New Initiative research Grant No. KA2018-98553 of the Pittsburgh Foundation.

%%---------------------------------------------------------------
%\begin{figure}[t]
%	\centering
%	\includegraphics[width=	\linewidth]{fig3.png}
%	\caption{
%		(a) and (b) show the spectrum and circular polarization of optical phonons with $m_0=25$ meV.
%        The phonon dispersions are labeled by black dashed lines, while the color represents the polarization.
%        (c) shows Im($\Pi_{AB}$) (black line) and the phonon circular polarization (red and blue lines) as a function of $m_0$ with $q_y=m_0/2v_0$ and $\omega=10$ meV.
%        The inset is the schematics for the elliptical vibration of phonon modes.
%		Parameters: $v_0=100$ meV$\cdot$nm, $\mu=0$, $\omega_{A_1}=20$ meV, $\omega_{B_1}=30$ meV, $g_0^o=g_1^o=80$ meV$\cdot$nm$^{-1}$. And $q_0=m_0/v_0$.
%	}
%	\label{fig3}
%\end{figure}
%%----------------------------------------------------------------
%

%%%%%%%%%%%%%%%%%%%%%%%%%%%%%%%%%%%%%%%%%%%%%%%%%%%%
%\bibliographystyle{prsty}
%\bibliography{ref.bib}% Produces the bibliography via BibTeX.
\bibliographystyle{apsrev4-1}
%\bibliography{ref}% Produces the bibliography via BibTeX.
%%%%%%%%%%%%%%%%%%%%%%%%%%%%%%%%%%%%%%%%%%%%%%%%%%%%
%merlin.mbs apsrev4-1.bst 2010-07-25 4.21a (PWD, AO, DPC) hacked
%Control: key (0)
%Control: author (72) initials jnrlst
%Control: editor formatted (1) identically to author
%Control: production of article title (-1) disabled
%Control: page (0) single
%Control: year (1) truncated
%Control: production of eprint (0) enabled
%

\clearpage
\begin{appendix}
\begin{widetext}
\section{$k\cdot p$ Hamiltonian and electron-phonon interaction of the Kramers-Weyl fermions}
\label{App_Sec:kpTheory}
In this section, we will describe the $k\cdot p$ theory for the KW fermions in chiral crystals. To be general, we start from the Schr$\ddot{o}$dinger equation
\eq{
\label{App_eq:Schrodinger_1}
\left(\frac{\hat{\bf p}^2}{2m_0}+\hat{V}({\bf r})\right)\Psi({\bf r})=E\Psi({\bf r})
}
where $m_0$ is the electron mass, $\hat{\bf p}=-i\hbar{\bf \nabla}$ and the periodic potential $\hat{V}({\bf r+R_n})=\hat{V}({\bf r})$ including both the lattice potential and spin-orbit coupling. The Bloch theorem requires $\Psi({\bf r})=e^{i{\bf K\cdot r}}\mathfrak{u}({\bf r})$ where $\mathfrak{u}({\bf r+R_n})=\mathfrak{u}({\bf r})$ and ${\bf K}$ is the crystal momentum.

Let's consider the perturbation expansion around a time-reversal (TR) invariant momentum ${\bf \Gamma}_i$ ($i=1,2,...$ labels different TR invariant momenta) and assume that the Schr$\ddot{o}$dinger equation (\ref{App_eq:Schrodinger_1}) has been solved at ${\bf \Gamma}_i$ with the eigen-energy $E_{{\bf \Gamma}_i,n}$ and eigen-wave function $\psi_{{\bf \Gamma}_i,n}({\bf r})=e^{i{\bf \Gamma_i\cdot r}}\mathfrak{u}_{{\bf \Gamma}_i,n}({\bf r})$. We are interested in the momentum ${\bf K=\Gamma_i+k}$ where ${\bf k}$ is considered as a perturbation, and expand the wave function $\mathfrak{u}_{\bf K}({\bf r})=\sum_n C_n \mathfrak{u}_{\bf \Gamma_i}({\bf r})$. Thus, $\psi_{\bf K}({\bf r})=e^{i{\bf K\cdot r}}\mathfrak{u}_{\bf K}({\bf r})=e^{i{\bf (\Gamma_i+k)\cdot r}}\sum_n C_n \mathfrak{u}_{\bf \Gamma_i}({\bf r})=e^{i{\bf k\cdot r}}\sum_n C_n \psi_{\bf \Gamma_i,n}({\bf r})$. The Schr$\ddot{o}$dinger equation at ${\bf K=\Gamma_i+k}$ can be expanded as
\eq{
\label{App_eq:Schrodinger_2}
\left(E_{{\bf \Gamma_i},m}+\frac{\hbar^2k^2}{2m_0}\right)\delta_{nm}C_m+\sum_n{\bf \mathcal{P}}_{mn}\cdot {\bf k}C_n=EC_m
}
and ${\bf \mathcal{P}}_{mn}=\frac{\hbar}{m_0}\langle \psi_{{\bf \Gamma_i},m}|{\bf \hat{p}}|\psi_{{\bf \Gamma_i},n}\rangle$, and the corresponding %effective ${\bf k\cdot p}$
Hamiltonian can be written as
\eq{\label{App_eq:kpHam1}
H=H_0+H_{\bf k\cdot p}
}
where
\eq{\label{App_eq:kpHam10}
(H_0)_{mn}=\left(E_{{\bf \Gamma_i},m}+\frac{\hbar^2k^2}{2m_0}\right)\delta_{nm}
}
and
\eq{\label{App_eq:kpHam1kp}
(H_{\bf k\cdot p})_{mn}={\bf \mathcal{P}}_{mn}\cdot {\bf k}.
}

Since we focus on TR symmetry, we denote the quantum number $m=(\alpha,s)$, where $s$ labels different spin states, and $\alpha$ labels other band indices. With the TR operator $\hat{T}$, we define the bands of a Kramers' pair as $\hat{T}\psi_{{\bf \Gamma}_i,\alpha,s}=s\psi_{{\bf \Gamma}_i,\alpha,\bar{s}}$ and $E_{{\bf \Gamma}_i,\alpha,s}=E_{{\bf \Gamma}_i,\alpha,\bar{s}}$, where $\bar{s}=-s$. We may assume around ${\bf \Gamma}_i$, only one Kramers' pair of bands, denoted as $\psi_{{\bf \Gamma}_i,0,s}$ ($\alpha=0$), contribute to the low energy physics and thus the low energy effective Hamiltonian can be written as
\eq{\label{App_eq:kpHam2}
(H_{eff})_{ss'}=\left(E_{{\bf \Gamma}_i,0}+\frac{\hbar^2k^2}{2m^*_0}\right)\delta_{ss'}+{\bf \mathcal{P}}_{0,ss'}\cdot {\bf k},
}
with $s,s'=\pm$. Here we only keep the correction to the electron mass ($m_0\rightarrow m^*_0$) for the terms of the second order in ${\bf k}$. In general, the effective mass $m^*_0$ should be a tensor, but we only consider the quadratic term to provide a cut-off for the valid regime of the KW fermion physics and thus simply treat $m^*_0$ as a scalar. We can decompose the parameter matrix ${\bf \mathcal{P}}_{0}$ as $\mathcal{P}^j_{0,ss'}=\sum_{a}\mathcal{P}^j_{0,a}(\sigma^{a})_{ss'}$ where $\sigma^{a=x,y,z}$ are three Pauli matrices. Here TR symmetry requires that the linear ${\bf k}$ term cannot couples to identity matrix. With $Tr(\sigma^{a}\sigma_{b})=2\delta^a_b$, we have $\mathcal{P}^j_{0,a}=\frac{1}{2}Tr(\mathcal{P}^j_0\sigma_a)$. Correspondingly, the effective Hamiltonian can be written as
\eq{\label{App_eq:kpHam3}
H_{eff}=E_{{\bf \Gamma}_i,0}+\frac{\hbar^2k^2}{2m^*_0}+\sum_{a,j}\mathcal{P}^j_{0,a}\sigma^a k_j.
}

The parameters $\mathcal{P}^j_{0,a}$ will be constrained by the crystal symmetry of the system. Since the chiral crystals only involve rotation symmetry, we consider the full rotation symmetry here. As discussed in \ref{App_Sec:SymmetryHam}, one can show that up to a unitary transformation, the effective Hamiltonian can be written as
\eq{\label{App_eq:kpHam4}
H_{eff,0}({\bf k})=E_{{\bf \Gamma}_i,0}+\frac{\hbar^2k^2}{2m^*_0}+\hbar v_f {\bf k\cdot \sigma},
}
where the Fermi velocity is given by $\hbar v_f=\frac{\hbar}{2m_0}Tr(\mathcal{P}^x_0\sigma_x)=\frac{\hbar}{2m_0}Tr(\mathcal{P}^y_0\sigma_y)=\frac{\hbar}{2m_0}Tr(\mathcal{P}^z_0\sigma_z)$ (with full rotation symmetry) and its sign determines the chirality of the KW fermions.  It should be mentioned that the above Hamiltonian is also valid for the crystals with the $O$ and $T$ chiral point groups\cite{he2019kramers}. In the second quantization language, we can write down $H_{eff,0}=\frac{1}{V}\sum_{\bf k}\hat{c}^{\dag}_{0,{\bf k}}H_{eff,0}({\bf k})\hat{c}_{0,{\bf k}}$. In the real space, the Hamiltonian is written as
\eq{\label{App_eq:kpHam6}
H_{eff,0}=\int d^3r \hat{\psi}^{\dag}_{\Gamma_i}({\bf r})\left(E_{{\bf \Gamma}_i,0}-\frac{\hbar^2\nabla^2}{2m^*_0}+\hbar v_f (-i{\bf \nabla})\cdot {\bf \sigma}\right)\hat{\psi}_{\Gamma_i}({\bf r}).
}
The Fourier transform is given by
\eqarr{
%\begin{eqnarray}
\label{App_eq:Fourier}
&\hat{\psi}({\bf r})=\frac{1}{V}\sum_{\bf k}e^{i{\bf k\cdot r}}\hat{c}_{\bf k} \\
&\hat{c}_{\bf k}=\int d^3r e^{-i{\bf k\cdot r}}\hat{\psi}({\bf r}),
%\end{eqnarray}
}
with $V$ the volume of the system, where we have suppressed the lower indices.

\subsection{Deformation potential for electron-acoustic phonon interaction}
The electron-phonon coupling for acoustic phonons should vanish in the long-wave length limit with the phonon momentum ${\bf q}\rightarrow 0$, while there is no such constraint for optical phonons. Thus, our starting deformation Hamiltonians for acoustic phonons and optical phonons are slightly different. For the acoustic phonons, we will start from the continuous model with the electron field described by $\hat{\psi}_{\Gamma_i}({\bf r})$ while the acoustic phonon modes can be described by the displacement field ${\bf u}({\bf r})$. In the case of more than one atoms in one unit cell, the acoustic phonon modes correspond to the polarization vector with the same amount of shift in the same direction for all the atoms in one unit cell and thus we can still use one vector field. The displacement field ${\bf u}$ can give rise to the polarization ${\bf P}$ which couples to electron density through its divergence $\nabla\cdot{\bf P}$. Let's assume this coupling is local as the lowest order perturbation. Thus, our starting point will be the volume deformation Hamiltonian
\eq{\label{App_eq:epHam1acoustic}
H_{ep}=g_a\int d^3r \hat{\psi}_{\Gamma_i}^{\dag}({\bf r})({\bf r})\left(\nabla\cdot{\bf u}\right)\hat{\psi}_{\Gamma_i}({\bf r}),
}
with the coupling constant $g_a$. Here the spin index $s$ does not appear because volume deformation potential does not involve spin. $\nabla\cdot{\bf u}=\sum_i u_{ii}$ is the trace part of the strain tensor (volume dilation), which is defined as $\bar{u}$ below. In the momentum space, we have
\eq{\label{App_eq:epHam1acoustic_k0}
H_{ep}=\frac{g_a}{V^2}\sum_{\bf k,q} \hat{c}^{\dag}_{0,{\bf k}} \bar{u}_{\bf q}\hat{c}_{0,{\bf k-q}},
}
where $\bar{u}_{\bf q}=\int d^3r \bar{u}({\bf u}) e^{-i{\bf q\cdot r}}$. Here we only consider the lowest order bands, and thus only the volume dilation enters into the effective Hamiltonian. If we consider all the other bands, other components of the strain tensor can also appear to couple different bands and the most general form of the electron-acoustic phonon coupling is given by
\eq{\label{App_eq:epHam1acoustic_k1}
H_{ep}=\sum_{\alpha\beta,ij} \frac{g_{a,\alpha\beta}^{ij}}{V^2}\sum_{\bf k,q} \hat{c}^{\dag}_{\alpha,{\bf k}} u_{ij}({\bf q})\hat{c}_{\beta, {\bf k-q}},
}
where the form of the coupling parameter $g^{ij}_{a,\alpha\beta}$ can be constrained by the crystal symmetry, which will be discussed in Sec. \ref{App_Sec:SymmetryHam}.

\subsection{Deformation potential for electron-optical phonon interaction}
The optical phonons involve the relative motions of atoms within one unit cell and thus we cannot directly start from the continuous model.
Let's next consider the deformation potential
\eq{\label{App_eq:epHam1}
H_{ep}=\int d^3r \hat{\Psi}^{\dag}({\bf r})\left(\sum_{\bf n,\tau}{\bf u_{n\tau}}\cdot\frac{\partial U({\bf r-R_{n\tau}})}{\partial {\bf R_{n\tau}}}\right)\hat{\Psi}({\bf r}),
}
where ${\bf R_{n\tau}=R_{n}+\mathfrak{r}_{\tau}}$ with the lattice vector ${\bf R_{n}}$ and the position ${\bf \mathfrak{r}_{\tau}}$ for the atom $\tau$ in one unit cell. Here the electron field operator can be expanded as $\hat{\Psi}({\bf r})=\frac{1}{V}\sum_{\alpha s {\bf K}}e^{i{\bf K\cdot r}}\mathfrak{u}_{\alpha s{\bf K}}({\bf r})\hat{c}_{\alpha s{\bf K}}$, where $s$ labels spin and $\alpha$ labels other band index, while ${\bf u_{n\tau}}$ is the displacement field, which can be expanded as ${\bf u_{n\tau}}=\frac{1}{\sqrt{NM_{\tau}}}\sum_{\lambda {\bf q}}Q_{\lambda,{\bf q}}{\bf \mathfrak{e}}^{\lambda}_{\tau}e^{i{\bf q\cdot R_n}}$. Here $M_\tau$ is the atom mass, $Q_{\lambda,{\bf q}}$ labels the normal mode of the $\lambda$-phonon and the polarization vector ${\bf \mathfrak{e}}^{\lambda}_{\tau}$ satisfies the equation of motion
\eq{
\omega_{\lambda}^2\mathfrak{e}^{\lambda}_{\tau,i}=\sum_{\xi j}D_{\bf k}(\tau i;\xi j)\mathfrak{e}^{\lambda}_{\xi,j},
}
where $i,j=x,y,z$ and $D_{\bf k}(\tau i;\xi j)$ is the dynamical matrix for phonons in the momentum space. Since the displacement ${\bf u_{n\tau}}$ is real, $Q_{\lambda,{\bf q}}=Q^\dag_{\lambda,-{\bf q}}$. By substituting the expansion of $\hat{\Psi}({\bf r})$ and ${\bf u_{n\tau}}$ into Eq. (\ref{App_eq:epHam1}) and after some straightforward simplifications, we obtain
\eq{\label{App_eq:epHam2}
H_{ep}=\frac{1}{V^2}\sum_{{\bf K,q},\alpha s,\beta t,\lambda} g^{\lambda}_{\alpha s,\beta t}({\bf K, q})Q_{\lambda}({\bf q})\hat{c}^{\dag}_{\alpha s, {\bf K}}\hat{c}_{\beta t,{\bf K-q}},
}
where
\eq{\label{App_eq:ep_q1}
g^{\lambda}_{\alpha s,\beta t}({\bf K, q})=\sum_{\tau}\sqrt{\frac{N}{M_\tau}}{\bf \mathfrak{e}}^{\lambda}_{\tau}\cdot\langle\mathfrak{u}_{{\bf K},\alpha s}|e^{-i{\bf q\cdot r}}\frac{\partial U({\bf r-\mathfrak{r}_\tau})}{\partial \mathfrak{r}_\tau}|\mathfrak{u}_{{\bf K-q},\beta t}\rangle.
}
We can again expand $g^{\lambda}_{\alpha s,\beta t}({\bf K, q})$ in terms of the 2 by 2 matrices as $g^{\lambda}_{\alpha s,\beta t}({\bf K, q})=\sum_{\mu}g^{\lambda}_{\alpha\beta,\mu}({\bf K, q})(\sigma^\mu)_{st}$ where $\mu=0,x,y,z$. The electron-phonon coupling Hamiltonian is written as
\eq{\label{App_eq:epHam3}
H_{ep}=\frac{1}{V^2}\sum_{{\bf K,q},\alpha\beta, \mu,\lambda} g^{\lambda}_{\alpha\beta,\mu}({\bf K, q})Q_{\lambda}({\bf q})\hat{c}^{\dag}_{\alpha, {\bf K}}\sigma^\mu\hat{c}_{\beta,{\bf K-q}}.
}
The hermitian property of the Hamiltonian $H_{ep}^{\dag}=H_{ep}$ requires $(g^{\lambda}_{\alpha\beta,\mu}({\bf K, q}))^*=g^{\lambda}_{\beta\alpha,\mu}({\bf K, -q})$.
%\eq{\label{App_eq:epHam3}
%H_{ep}^{\dag}=\sum_{{\bf K',q},\alpha\beta, \mu,\lambda} (g^{\lambda}_{\alpha\beta,\mu}({\bf K, q}))^*Q_{\lambda}({\bf -q})\hat{c}^{\dag}_{\beta,{\bf K'}}\sigma^\mu \hat{c}_{\alpha, {\bf K'+q}}.
%}

The time reversal symmetry provides a constraint on the form of $g^{\lambda}_{\alpha s,\beta t}({\bf K, q})$. With $\hat{T}Q_{\lambda}({\bf q})\hat{T}^{-1}=Q_{\lambda}({\bf -q})$ and $\hat{T}\hat{c}^{\dag}_{\alpha}({\bf K})\hat{T}^{-1}=\hat{c}^{\dag}_{\alpha}({\bf -K})(i\sigma_y\mathcal{K})$, we have
\eqarr{
\hat{T}H_{ep}\hat{T}^{-1}&=&\sum_{{\bf K,q},\alpha\beta, \mu,\lambda} (g^{\lambda}_{\alpha\beta,\mu}({\bf K, q}))^*Q_{\lambda}({\bf -q})\hat{c}^{\dag}_{\alpha, {\bf -K}}(i\sigma_y\mathcal{K})\sigma^\mu(i\sigma_y\mathcal{K})\hat{c}_{\beta,{\bf q-K}}\\
&=&\sum_{{\bf K,q},\alpha\beta, \mu,\lambda} (g^{\lambda}_{\alpha\beta,\mu}({\bf -K, -q}))^*Q_{\lambda}({\bf q})\hat{c}^{\dag}_{\alpha, {\bf K}}(i\sigma_y\mathcal{K})\sigma^\mu(-i\sigma_y\mathcal{K})\hat{c}_{\beta,{\bf K-q}}.
}
Since $(i\sigma_y\mathcal{K})\sigma^0(-i\sigma_y\mathcal{K})=\sigma^0$ and $(i\sigma_y\mathcal{K})\sigma^{a}(-i\sigma_y\mathcal{K})=-\sigma^{a}$ ($a=x,y,z$), the time reversal symmetry $\hat{T}H_{ep}\hat{T}^{-1}=H_{ep}$ gives rise to $(g^{\lambda}_{\alpha\beta,0}({\bf -K, -q}))^*=g^{\lambda}_{\alpha\beta,0}({\bf -K, -q})$ and $(g^{\lambda}_{\alpha\beta,a}({\bf -K, -q}))^*=-g^{\lambda}_{\alpha\beta,a}({\bf K, q})$.

%Since $\langle \hat{T}\phi|\hat{T}\psi\rangle =\langle \psi|\phi\rangle$, we have
%\eq{
%\langle\mathfrak{u}_{{\bf K},\alpha s}|e^{-i{\bf q\cdot r}}\frac{\partial U({\bf r-\mathfrak{r}_\tau})}{\partial \mathfrak{r}_\tau}|\mathfrak{u}_{{\bf K-q},\beta t}\rangle=\langle\mathfrak{u}_{{\bf K},\alpha s}|e^{-i{\bf q\cdot r}}\frac{\partial U({\bf r-\mathfrak{r}_\tau})}{\partial \mathfrak{r}_\tau}|\mathfrak{u}_{{\bf K-q},\beta t}\rangle
%}
In the spirit of the ${\bf k\cdot p}$ theory, we consider the electron-phonon coupling around time-reversal invariant momenta ${\bf \Gamma}_i=-{\bf \Gamma}_i$ and thus take the approximation ${\bf K=\Gamma_i}$ for the parameter $g^{\lambda}_{\alpha\beta,\mu}({\bf K, q})$, so that $(g^{\lambda}_{\alpha\beta,0}({\bf \Gamma_i, -q}))^*=g^{\lambda}_{\alpha\beta,0}({\bf \Gamma_i, -q})$ and $(g^{\lambda}_{\alpha\beta,a}({\bf \Gamma_i, -q}))^*=-g^{\lambda}_{\alpha\beta,a}({\bf \Gamma_i, -q})$. Below we will drop the index ${\bf \Gamma}_i$ in the coupling constant $g$. Furthermore, we can assume the phonon momentum ${\bf q}$ is a small number and expand $g$ as $g^{\lambda}_{\alpha\beta,\mu}({\bf q})=g^{\lambda}_{\alpha\beta,\mu}({\bf q=0})+(\partial_{\bf q} g^{\lambda}_{\alpha\beta,\mu})_{\bf q=0}\cdot {\bf q}+...$, and for optical phonons, we may only focus on the lowest order term $g^{\lambda}_{\alpha\beta,\mu}=g^{\lambda}_{\alpha\beta,\mu}({\bf q=0})$.
%, but for acoustic phonons, since the ${\bf q=0}$ term must vanish since it corresponds to the shift of the whole lattice by a constant value of distance and should not influence the energy of the system. Thus, we should consider the linear ${\bf q}$ dependent terms.
%For optical phonons,
Thus, we should have $(g^{\lambda}_{\alpha\beta,0})^*=g^{\lambda}_{\alpha\beta,0}$ and $(g^{\lambda}_{\alpha\beta,a})^*=-g^{\lambda}_{\alpha\beta,a}$ ($a=x,y,z$) and thus $g^{\lambda}_{\alpha\beta,0}$ is real while $g^{\lambda}_{\alpha\beta,a}$ is pure imaginary. The hermitian condition requires $(g^{\lambda}_{\alpha\beta,\mu})^*=g^{\lambda}_{\beta\alpha,\mu}$. Therefore, if we choose $\alpha=\beta$, only $g^{\lambda}_{\alpha\alpha,0}$ does not vanish while the pure imaginary $g^{\lambda}_{\alpha\beta,a}$ only couples different bands $\alpha\neq \beta$.

%For acoustic phonons, let's denote ${\bf g'}^{\lambda}_{\alpha\beta,\mu}=(\partial_{\bf q} g^{\lambda}_{\alpha\beta,\mu})_{\bf q=0}$ and thus $({\bf g'}^{\lambda}_{\alpha\beta,0})^*=-{\bf g'}^{\lambda}_{\alpha\beta,0}$ and $({\bf g'}^{\lambda}_{\alpha\beta,a})^*={\bf g'}^{\lambda}_{\alpha\beta,a}$ ($a=x,y,z$), which means ${\bf g'}^{\lambda}_{\alpha\beta,0}$ is imaginary and ${\bf g'}^{\lambda}_{\alpha\beta,a}$ is real. The hermitian condition requires $({\bf g'}^{\lambda}_{\alpha\beta,\mu})^*=-{\bf g'}^{\lambda}_{\beta\alpha,\mu}$. Thus, if $\alpha=\beta$, only the imaginary ${\bf g'}^{\lambda}_{\alpha\alpha,0}$ does not vanish, and the real ${\bf g'}^{\lambda}_{\alpha\beta,a}$ couples different bands $\alpha\neq \beta$.

Now let's look at the effective electron-phonon coupling for the band $\psi_{{\bf \Gamma}_i,0,s}$. Since now we choose $\alpha=\beta=0$, it is clear that only $g^{\lambda}_{\alpha\alpha,0}$ is non-vanishing for the lowest order terms. Therefore, at the lowest order, the electron-phonon coupling takes the form
%it is clear that only $g^{\lambda}_{\alpha\alpha,0}$ and $g'^{\lambda}_{\alpha\alpha,0}$ are non-vanishing for the lowest order terms for optical and acoustic phonons, respectively. Therefore, at the lowest order, the electron-phonon coupling takes the form
\eq{\label{App_eq:epHam0optical}
H_{ep,{\bf \Gamma}_i}=\frac{1}{V^2}\sum_{{\bf k,q},\lambda} g^{\lambda}_{o} Q_{\lambda}({\bf q})\hat{c}^{\dag}_{0, {\bf k}}\hat{c}_{0,{\bf k-q}}
}
for optical phonons, where $g^{\lambda}_{o}$ is a real number.
% and
%\eq{\label{App_eq:epHam0acoustic}
%H_{ep,{\bf \Gamma}_i}=\sum_{{\bf k,q},\lambda} i {\bf g}^{\lambda}_{a}\cdot{\bf q} Q_{\lambda}({\bf q})\hat{c}^{\dag}_{0, {\bf k}}\hat{c}_{0,{\bf k-q}}
%}
%for acoustic phonons. Now both $g^{\lambda}_{o}$ and ${\bf g}^{\lambda}_{a}$ take real values.
In the real space, we have
\eq{\label{App_eq:epHam1optical}
H_{ep,{\bf \Gamma}_i}=\sum_{\lambda}g^{\lambda}_o\int d^3r  Q_{\lambda}({\bf r})\hat{\psi}^{\dag}_{\Gamma_i}({\bf r}) \hat{\psi}_{\Gamma_i}({\bf r})
}
for optical phonons,
% and
%\eq{\label{App_eq:epHam1acoustic}
%H_{ep,{\bf \Gamma}_i}=\sum_{\lambda}\int d^3r  ({\bf g}^{\lambda}_a\cdot\partial_{\bf r}Q_{\lambda}({\bf r}))\hat{\psi}^{\dag}_{\Gamma_i}({\bf r}) \hat{\psi}_{\Gamma_i}({\bf r})
%}
%for acoustic phonons,
where we have defined $Q_{\lambda}({\bf q})=\int d^3r Q_{\lambda}({\bf r})e^{-i{\bf q\cdot r}}$. We notice that for acoustic phonon, we just need to replace the normal mode $Q_{\lambda}$ by the strain tensor $u_{ij}$, and all the formalism remains valid.

%$\partial_{\bf r}Q_{\lambda}({\bf r})$ is essentially the strain of the system and thus we may also rewrite ${\bf g}^{\lambda}_a\cdot\partial_{\bf r}Q_{\lambda}({\bf r})$ with the strain tensor, as discussed in Sec. {\bf XXX}.

\subsection{Second order perturbation terms from Schrieffer-Wolf transformation}
We next consider the second order perturbation contribution to the electron-phonon interaction through the Schrieffer-Wolf transformation. We consider the full ${\bf k\cdot p}$ Hamiltonian including the electron-phonon interaction around ${\bf \Gamma}_i$ with the form $H_{\bf \Gamma_i}=H_0+H_1$ with the perturbation Hamiltonian $H_1=H_{\bf k\cdot p}+H_{ep}$, where $H_{\bf k\cdot p}$ is given by Eq. (\ref{App_eq:kpHam1kp}) with $m=(\alpha,s), n=(\beta, t)$ and $H_{ep}$ is given by the Eq. (\ref{App_eq:epHam1acoustic_k1}) for acoustic phonons and the Eq. (\ref{App_eq:epHam2}) for optical phonons, in which ${\bf K}$ should be replaced by ${\bf k}$ since all the momenta are expanded around ${\bf \Gamma_i}$.

To get the second order perturbation terms, we consider the Schrieffer-Wolf transformation
\eq{
H_{eff}=e^{-S}H_{\bf \Gamma_i}e^{S}\approx H_0+H_1-[S,H_0]-[S,H_1]+\frac{1}{2}[S,[S,H_0]]+...,
}
and we require the first order term to vanish, $H_1-[S,H_0]=0$. The corresponding second order term is given by
\eq{
H_{eff}\approx H_0-\frac{1}{2}[S,H_1].
}

Since $H_1=H_{\bf k\cdot p}+H_{ep}$, we can also decompose $S=S_1+S_2$ with $[S_1,H_0]=H_{\bf k\cdot p}$ and $[S_2,H_0]=H_{ep}$. One can show that
\eq{
S_1=\sum_{{\bf k},\alpha\neq \beta, st}\frac{1}{\Delta_{\beta\alpha}}\hat{c}^{\dag}_{\alpha s,{\bf k}} ({\bf \mathcal{P}}_{\alpha s,\beta t}\cdot {\bf k}) \hat{c}_{\beta t,{\bf k}}
}
and
\eq{
S_2=\frac{1}{V}\sum_{{\bf k,q},\alpha\neq \beta, st}\frac{1}{\Delta_{\beta\alpha}}g^{\lambda}_{\alpha s,\beta t}Q_{\lambda}({\bf q})\hat{c}^{\dag}_{\alpha s, {\bf k}}\hat{c}_{\beta t,{\bf k-q}},
}
where $\Delta_{\beta\alpha}=E_{\beta}-E_{\alpha}$. Here we choose optical phonons as the example and keep the lowest order terms, so we can drop the index ${\bf K,q}$ for the coupling constant $g^{\lambda}_{\alpha s,\beta t}$. For acoustic phonons, we just need to replace $Q_{\lambda}({\bf q})$ by $\bar{u}_{ij}({\bf q})$. With the decomposition of $S$, the effective Hamiltonian includes
\eq{
H_{eff}\approx H_0-\frac{1}{2}([S_1,H_{\bf k\cdot p}]+[S_2,H_{\bf k\cdot p}]+[S_1,H_{ep}]+[S_2,H_{ep}]).
}
The first perturbation term is given by
\eq{
-\frac{1}{2}[S_1,H_{\bf k\cdot p}]=\frac{1}{2V}\sum_{{\bf k},\mu\nu,\alpha\beta\gamma}\left(\frac{1}{\Delta_{\alpha\gamma}}+\frac{1}{\Delta_{\beta\gamma}}\right)({\bf \mathcal{P}}^{\mu}_{\alpha\gamma}\cdot {\bf k})({\bf \mathcal{P}}^{\nu}_{\gamma\beta}\cdot {\bf k})\hat{c}^{\dag}_{\alpha,{\bf k}}\sigma_{\mu}\sigma_{\nu}\hat{c}_{\beta,{\bf k}},
}
which provides a correction to the effective mass. The second and the third term together give
\eq{
-\frac{1}{2}([S_2,H_{\bf k\cdot p}]+[S_1,H_{ep}])=\frac{1}{2V^2}\sum_{{\bf k,q},\mu\nu,\alpha\beta\gamma,\lambda}\left(\frac{1}{\Delta_{\alpha\gamma}}+\frac{1}{\Delta_{\beta\gamma}}\right)\left(g^{\lambda, \mu}_{\alpha\gamma}({\bf \mathcal{P}}^{\nu}_{\gamma\beta}\cdot ({\bf k-q}))+({\bf \mathcal{P}}^{\mu}_{\alpha\gamma}\cdot {\bf k})g^{\lambda,\nu}_{\gamma\beta}\right)Q_{\lambda}({\bf q})\hat{c}^{\dag}_{\alpha,{\bf k}}\sigma_{\mu}\sigma_{\nu}\hat{c}_{\beta,{\bf k-q}},
}
where $g^{\lambda}_{\alpha s,\beta t}=\sum_{\mu}g^{\lambda,\mu}_{\alpha\beta}(\sigma_{\mu})_{st}$. The last term $-\frac{1}{2}[S_2,H_{ep}]$ is of order $g^2$ and we neglect this term.

Now we apply the above second order perturbation formalism to the Kramers-Weyl fermions to project into the lowest-energy subspace with $\alpha=\beta=0$, and obtain the effective Hamiltonian
\eq{\label{App_eq:Ham_final_1}
H_{eff,{\bf \Gamma}_i}=H_{eff,0}+H_{eff,ep}
}
where
\eq{
H_{eff,0}=\frac{1}{V}\sum_{\bf k}\hat{c}^{\dag}_{0,{\bf k}}\left(E_{{\bf \Gamma}_i,0}+\frac{\hbar^2k^2}{2m^*_0}+\hbar v_f({\bf k\cdot\sigma})\right)\hat{c}_{0,{\bf k}}
}
and
\eqarr{
&&H_{eff,ep}=H_{eff,ep}^{(0)}+H_{eff,ep}^{(2)}=\frac{1}{V^2}\sum_{{\bf k,q},\lambda} g^{\lambda}_{o} Q_{\lambda}({\bf q})\hat{c}^{\dag}_{0, {\bf k}}\hat{c}_{0,{\bf k-q}}\\
&&+\frac{1}{V^2}\sum_{{\bf k,q},\mu\nu,\gamma,\lambda}\frac{1}\Delta_{0\gamma}\left(g^{\lambda, \mu}_{0\gamma}({\bf \mathcal{P}}^{\nu}_{\gamma 0}\cdot ({\bf k-q}))+({\bf \mathcal{P}}^{\mu}_{0\gamma}\cdot {\bf k})g^{\lambda,\nu}_{\gamma 0}\right)Q_{\lambda}({\bf q})\hat{c}^{\dag}_{0,{\bf k}}\sigma_{\mu}\sigma_{\nu}\hat{c}_{0,{\bf k-q}}.
}
Here we have made the approximation of isotropic effective mass
\eq{
\frac{\hbar^2}{2m^*_0}=\frac{\hbar^2}{2m_0}+\sum_{\mu,\gamma\neq 0}\frac{1}{\Delta_{0\gamma}}\left\langle({\bf \mathcal{P}}^{\mu}_{0\gamma}\cdot {\bf k})({\bf \mathcal{P}}^{\mu}_{\gamma 0}\cdot {\bf k})\right\rangle_{\hat{k}},
}
where $\langle ...\rangle_{\hat{k}}$ means the average over all different directions of the momentum ${\bf k}$.

Since $(\sigma_\mu)^2=1, \sigma_0\sigma_a=\sigma_a\sigma_0=\sigma_a, \sigma_a\sigma_b=\delta_{ab}+i\epsilon_{abc}\sigma_c$ with $a,b,c=x,y,z$, we may define
\eq{
\mathcal{C}^{\lambda,j}_{0}=\sum_{\mu,\gamma\neq 0}\frac{1}{\Delta_{0\gamma}}g^{\lambda\mu}_{0\gamma}\mathcal{P}^{\mu,j}_{\gamma 0}
}
and
\eq{
\mathcal{C}^{\lambda,j}_{a}=\sum_{\gamma\neq 0}\frac{1}{\Delta_{0\gamma}}\left(g^{\lambda 0}_{0\gamma}\mathcal{P}^{a,j}_{\gamma 0}+g^{\lambda a}_{0\gamma}\mathcal{P}^{0,j}_{\gamma 0}+i\sum_{bc}\epsilon_{abc}g^{\lambda b}_{0\gamma}\mathcal{P}^{c,j}_{\gamma 0}\right),
}
and $H_{eff,ep}^{(2)}$ can be re-written as
\eq{
H_{eff,ep}^{(2)}=\frac{1}{V^2}\sum_{{\bf k,q},\mu,\lambda}\left(\mathcal{C}^{\lambda,j}_{\mu}Q_{\lambda}({\bf q})\hat{c}^{\dag}_{0,{\bf k}}\sigma^{\mu}(k_j-q_j)\hat{c}_{0,{\bf k-q}}+h.c.\right).
}
Transforming back to the real space, we find that $H_{eff,0}$ is given by Eq. (\ref{App_eq:kpHam6}), $H_{eff,ep}^{(0)}$ is given by Eq. (\ref{App_eq:epHam1optical}) and
\eq{\label{App_eq:epHameff2}
H_{eff,ep}^{(2)}=\sum_{\mu,\lambda,j}\int d^3r Q_{\lambda}({\bf r})\left(\mathcal{C}^{\lambda,j}_{\mu}\hat{\psi}^{\dag}_{\bf \Gamma_i}({\bf r})\sigma^{\mu}(-i\partial_j\hat{\psi}_{\bf \Gamma_i}({\bf r}))+(\mathcal{C}^{\lambda,j}_{\mu})^*(i\partial_j \hat{\psi}^{\dag}_{\bf \Gamma_i}({\bf r}))\sigma^\mu\hat{\psi}_{\bf \Gamma_i}({\bf r})\right).
}

Now let's impose the TR symmetry on the above expression and we find $\mathcal{C}^{\lambda,j}_{0}=-(\mathcal{C}^{\lambda,j}_{0})^*$ so that $\mathcal{C}^{\lambda,j}_{0}$ should be pure imaginary while $\mathcal{C}^{\lambda,j}_{a}=(\mathcal{C}^{\lambda,j}_{a})^*$ ($a=x,y,z$) so that $\mathcal{C}^{\lambda,j}_{a}$ should be real. This means that for the $\mathcal{C}^{\lambda,j}_{0}$ term, the Hamiltonian will only depend on $\partial_jQ_{\lambda}({\bf r})$, while the $\mathcal{C}^{\lambda,j}_{a}$ terms ($a=x,y,z$) depend on $Q_\lambda({\bf r})$. In the long wavelength limit, the $\mathcal{C}^{\lambda,j}_{0}$ term will be much smaller than the $\mathcal{C}^{\lambda,j}_{a}$ terms. Thus, we only consider the $\mathcal{C}^{\lambda,j}_{a}$ terms below.

The Hamiltonians (\ref{App_eq:kpHam6}), (\ref{App_eq:epHam1optical}) and (\ref{App_eq:epHameff2}) together form the starting point for our study on the electron-phonon coupling in the Kramers-Weyl semimetals, and we can rewrite them in a more compact form
\eqarr{\label{App_eq:Heff1}
&&H_{eff}=\int d^3r \left[\hat{\psi}^{\dag}_{\bf \Gamma_i}({\bf r})\left(-\frac{\hbar^2}{2m_0^*}\nabla^2-\mu-A_0({\bf r})\right)\hat{\psi}_{\bf \Gamma_i}({\bf r})\right.\nonumber\\
&&\left.+\frac{\hbar v_f}{2}\sum_{a,j}\left(\hat{\psi}^{\dag}_{\bf \Gamma_i}({\bf r})e^{ j}_a\sigma^{a}(-i\partial_j\hat{\psi}_{\bf \Gamma_i}({\bf r}))+(i\partial_j \hat{\psi}^{\dag}_{\bf \Gamma_i}({\bf r}))e^{j}_a\sigma^a\hat{\psi}_{\bf \Gamma_i}({\bf r})\right)\right],
}
where the chemical potential $\mu$ has included the energy $E_{{\bf \Gamma}_i,0}$, $A_0=g^{\lambda}_o Q_{\lambda}({\bf r})$ is from the zero order term of electron-phonon coupling and behaves as a chemical potential fluctuation (or scalar potential), and the frame field $e^{j}_a=\delta^j_a+\Delta^j_a({\bf r})$ with $\Delta^j_a({\bf r})=\frac{2}{\hbar v_f}\sum_{\lambda}\mathcal{C}^{\lambda,j}_{a} Q_\lambda({\bf r})$ comes from the second order perturbation involving electron-phonon coupling. In deriving the above Hamiltonian, we only used the TR symmetry and the spatial rotation symmetry will give additional constraints on the parameters $\mathcal{C}^{\lambda,j}_{a}$ and reduce the number of independent parameters, as discussed in the Sec. \ref{App_Sec:SymmetryHam}. From the above form of the Hamiltonian, it is clear that the phonon provides a background frame field of the curved space for the KW fermions.

The above derivation is directly applied to optical phonons and for acoustic phonons (elastic wave), we just need to change the normal mode $Q_{\lambda}({\bf r})$ to the strain tensor $u_{ij}=\frac{1}{2}(\partial_iu_j+\partial_ju_i)$, as discussed above. In the next section, we will construct the Hamiltonian coupled to the strain tensor by analyzing the symmetry of the Hamiltonian.

\section{Symmetry construction of the Hamiltonian}
\label{App_Sec:SymmetryHam}
In this section, we will consider the construction of the effective Hamiltonian for the KW fermions from the symmetry principle\cite{winkler2003spin,dresselhaus2007group}. Here we will focus on the acoustic phonons, which couple to electrons through the strain tensor $u_{ij}$. Similar discussion can also be applied to the optical phonons, which is described by the normal modes $Q_\lambda$.

The chiral crystals only involve rotation symmetries and to simplify the problem, let's first consider the isotropic systems with the full rotation symmetry ($SO(3)$) and TR symmetry. In this case, we can classify all the physical operators in terms of their angular momentum. The Hamiltonian should be invariant under the full rotation and thus should carry the angular momentum 0. Both the momentum ${\bf k}$ and the spin $\sigma$ carry the angular momentum 1, and thus allow to construct two terms $k^2$ and ${\bf k\cdot \sigma}$. As discussed above, the acoustic phonons couple to electrons through the strain tensor $u_{ij}$, which is a rank-2 tensor and symmetric with respect to $i$ and $j$ ($u_{ij}=u_{ji}$). For the momentum ${\bf k}$ and spin $\sigma$, one can define another rank-2 tensor $T_{ij}=k_i\sigma_j$, which is actually the stress tensor of Weyl fermions. Now we need to construct the invariant terms based on two rank-2 tensors $u_{ij}$ and $T_{ij}$ and there are two ways: (1) $\sum_{ij}u_{ii} T_{jj}$ and $\sum_{ij}u_{ij}T_{ij}$. Thus, we can write down the Hamiltonian as
\eq{\label{App_eq:Hamsym1}
H_{eff}=C_0+C_1\bar{u}+C_3k^2+\left(C_2+g_0\bar{u}\right)({\bf k\cdot \sigma})+g_1\sum_{ij}u_{ij}T_{ij},
}
up to the order of $k^2$ and $u_{ij}k_l$, where $\bar{u}=\sum_i u_{ii}$ is the trace of strain tensor. For TR symmetry $\hat{T}$, ${\bf k}$ and ${\bf \sigma}$ are TR-odd while $u_{ij}$ and $T_{ij}$ is TR-even. Therefore, all the terms above are allowed for a TR-invariant Hamiltonian. We may further consider the strain as a field with spatial and temporal variations and thus require to change $u_{ij}T_{ij}=u_{ij}k_i \sigma_j$ to $\frac{1}{2}\{u_{ij},(-i\partial_i)\}\sigma_j$ and $\bar{u}({\bf k\cdot \sigma})$ to $\frac{1}{2}\{\bar{u},({\bf -i\nabla\cdot \sigma})\}$. The corresponding Hamiltonian can be written as \eq{
H_{eff}=\int d^3 r \hat{\psi}^{\dag}({\bf r})\left(C_0+ C_1\bar{u}-C_3\nabla^2 +C_2 (-i{\bf \nabla}\cdot\sigma)+\frac{g_0}{2} \{\bar{u},(-i{\bf \nabla\cdot \sigma})\}+\frac{g_1}{2}\{u_{ij},(-i\partial_i)\}\sigma_j \right)\hat{\psi}({\bf r})
}

As compared with Eq. (\ref{App_eq:Heff1}), one finds $\mu=-C_0$, $A_0=C_1\bar{u}$, $C_3=\frac{\hbar^2}{2m_0^*}$, $C_2=\hbar v_f$, and $\Delta^j_a=\frac{1}{\hbar v_f}(g_0\bar{u}\delta_{ja}+g_1 u_{ja})$. Since we only focus on the spatial component, the upper and lower indices do not have specific meaning as in general relativity.

Here we notice that there are two independent parameters that characterize the coupling between strain tensor $u_{ij}$ and the stress tensor $T_{ij}$. This conclusion can also be obtained by classifying the rank-2 tensors according to their angular momentum. The strain tensor can be decomposed into two parts, $u_{J=0;M=0}=\sum_{i}u_{ii}=\bar{u}$ with the angular momentum 0 and $u_{J=2;M=\pm 2,\pm 1, 0}$ with the angular momentum 2. The explicit form of $u_{J=2;M}$ is given by
\eqarr{\label{App_eq:straintensor1}
&&u_{2,2}=\frac{1}{2}\left(u_{xx}-u_{yy}+i(u_{xy}+u_{yx}\right)\\
&&u_{2,1}=\left(-\frac{1}{2}\right)\left(u_{xz}+u_{zx}+i(u_{yz}+u_{zy})\right)\\
&&u_{2,0}=\sqrt{\frac{1}{6}}\left(2u_{zz}-u_{xx}-u_{yy}\right)\\
&&u_{2,-1}=\frac{1}{2}\left(u_{xz}+u_{zx}-i(u_{yz}+u_{zy})\right)\\
&&u_{2,-2}=\frac{1}{2}\left(u_{xx}-u_{yy}-i(u_{xy}+u_{yx})\right)\label{App_eq:straintensor2}
}
Similarly, the stress tensor component with the angular momentum 2 can also be written as
\eqarr{\label{App_eq:stresstensor1}
&&T_{2,2}=\frac{1}{2}\left(T_{xx}-T_{yy}+i(T_{xy}+T_{yx}\right)\\
&&T_{2,1}=\left(-\frac{1}{2}\right)\left(T_{xz}+T_{zx}+i(T_{yz}+T_{zy})\right)\\
&&T_{2,0}=\sqrt{\frac{1}{6}}\left(2T_{zz}-T_{xx}-T_{yy}\right)\\
&&T_{2,-1}=\frac{1}{2}\left(T_{xz}+T_{zx}-i(T_{yz}+T_{zy})\right)\\
&&T_{2,-2}=\frac{1}{2}\left(T_{xx}-T_{yy}-i(T_{xy}+T_{yx})\right)\label{App_eq:stresstensor2}
}
The term $\sum_{m} u_{2,m}T_{2,-m}$ is invariant under the rotation. Collecting all the invariant terms, we obtain the form of the Hamiltonian
\eq{\label{App_eq:Hamsym2}
H_{eff}=C_1u_{0,0}+C_3k^2+\left(C_2+\tilde{g}_{0}u_{0,0}\right)({\bf k\cdot \sigma})+\tilde{g}_1\sum_{m=\pm 2, \pm 1, 0}u_{2,m}T_{2,-m}.
}
One can easily check that this Hamiltonian is the same as Eq. (\ref{App_eq:Hamsym1}).

%This effective Hamiltonian takes the same form as Eq. (\ref{App_eq:Heff1}) with the replacement of $Q$ field by the strain field $u_{ij}$, and it also gives the number of independent parameters in the Hamiltonian (\ref{App_eq:Heff1}) when taking into account the full rotation symmetry.

When reducing the symmetry group from $SO(3)$ to chiral point group symmetry, the number of independent parameters will increase. Here we take the $O$ group as an example and other chiral symmetry groups can be worked out in a similar manner. According to the character table of $O$ group\cite{dresselhaus2007group}, there are five irreducible representations (IRRs), labelled by $A_1, A_2, E, T_1$ and $T_2$. The Hamiltonian should be invariant and thus belongs to the $A_1$ IRR. Both ${\bf k}$ and ${\bf \sigma}$ belong to the $T_1$ and since $T_1\otimes T_1=A_1+E+T_1+T_2$, we have one term ${\bf k\cdot \sigma}$ can be constructed to be invariant. For the strain tensor, $u_{0,0}=\sum_i u_{ii}$ belongs to the $A_1$ IRR, $(u_{xx}-u_{yy},2u_{zz}-u_{xx}-u_{yy})$ belongs to the $E$ IRR, and $(u_{xy},u_{yz},u_{zx})$ belongs to the $T_2$ IRR. This allows us to construct the following invariant terms: (1) $u_{0,0}{\bf k\cdot\sigma}$ and $u_{0,0}k^2$; (2) $(2u_{zz}-u_{xx}-u_yy)k_z\sigma_z+(2u_{xx}-u_{yy}-u_{zz})k_x\sigma_x+(2u_{yy}-u_{xx}-u_{zz})k_y\sigma_y$; (3) $u_{xy}(k_x\sigma_y+k_y\sigma_x)+u_{yz}(k_y\sigma_z+k_z\sigma_y)+u_{zx}(k_z\sigma_x+k_x\sigma_z)$. Therefore, the first four terms in the Hamiltonian (\ref{App_eq:Hamsym1}) remain the same while the last term is changed to
\eqarr{ &&g_{1}\left((2u_{zz}-u_{xx}-u_yy)k_z\sigma_z+(2u_{xx}-u_{yy}-u_{zz})k_x\sigma_x+(2u_{yy}-u_{xx}-u_{zz})k_y\sigma_y\right)+\\
&&g_{2}\left(u_{xy}(k_x\sigma_y+k_y\sigma_x)+u_{yz}(k_y\sigma_z+k_z\sigma_y)+u_{zx}(k_z\sigma_x+k_x\sigma_z)\right),
}
which possesses two independent parameters $g_1$ and $g_2$.

\section{Effective action and Correlation function}
\label{App_Sec:Action}
In this section, we will show the derivation of the formalism for the effective action for the phonon dynamics and the corresponding correlation functions.

From the Hamiltonian (\ref{App_eq:Heff1}), the effective action can be written as
\eqarr{\label{App_eq:Seff2}
&&S_{eff}=\int_0^{\beta} d\tau \int d^3r \left[\hat{\psi}^{\dag}_{\bf \Gamma_i}\frac{\partial}{\partial\tau}\hat{\psi}_{\bf \Gamma_i}+\mathcal{H}_{eff}\right]\nonumber\\
&&=\int_0^{\beta} d\tau \int d^3r\left[\hat{\psi}^{\dag}_{\bf \Gamma_i}\left(\frac{\partial}{\partial\tau}-\mu-A_0({\bf r})-\frac{\hbar^2}{2m_0^*}\nabla^2\right)\hat{\psi}_{\bf \Gamma_i}+\frac{\hbar v_f}{2}\sum_{a,j}\left(\hat{\psi}^{\dag}_{\bf \Gamma_i}e^{ j}_a\sigma^{a}(-i\partial_j\hat{\psi}_{\bf \Gamma_i})+(i\partial_j \hat{\psi}^{\dag}_{\bf \Gamma_i})e^{j}_a\sigma^a\hat{\psi}_{\bf \Gamma_i}\right)\right],
}
where we have used the imaginary time $\tau=i t$ and $e^{j}_a=\delta^j_a+\Delta^j_a$.
%The corresponding partition function is given by $Z=\int \mathcal{D}(\hat{\psi}^{\dag}\hat{\psi}) exp(-S_{eff})$.

To implement the perturbation calculations, we may separate the full action into two parts $S_{eff}=S_0+S_1$, where \eq{
S_0=\int_0^{\beta} d\tau \int d^3r \hat{\psi}^{\dag}_{\bf \Gamma_i}\left(\frac{\partial}{\partial\tau}-\mu+\hbar v_f(-i{\bf \nabla\cdot\sigma})\right)\hat{\psi}_{\bf \Gamma_i}
}
and
\eq{\label{App_eq:actionS1}
S_1=\int_0^{\beta} d\tau \int d^3r\left[-\hat{\psi}^{\dag}_{\bf \Gamma_i}A_0({\bf r})\hat{\psi}_{\bf \Gamma_i}+\frac{\hbar v_f}{2}\sum_{a,j}\left(\hat{\psi}^{\dag}_{\bf \Gamma_i}\Delta^{ j}_a\sigma^{a}(-i\partial_j\hat{\psi}_{\bf \Gamma_i})+(i\partial_j \hat{\psi}^{\dag}_{\bf \Gamma_i})\Delta^{j}_a\sigma^a\hat{\psi}_{\bf \Gamma_i}\right)\right]
}
Now let's define the Fourier transform as
\eqarr{
\hat{\psi}_{\bf \Gamma_i}({\bf r},\tau)=\frac{1}{\beta V}\sum_{i\omega_n,{\bf k}} e^{i{\bf k\cdot r}-i\omega_n\tau} \hat{\psi}_{\bf \Gamma_i}({\bf k},i\omega_n),\\
A_0({\bf r},\tau)=\frac{1}{\beta V}\sum_{i\nu_m,{\bf q}} e^{i{\bf q\cdot r}-i\nu_m\tau} A_0({\bf q},i\nu_m),\\
\Delta^j_a({\bf r},\tau)=\frac{1}{\beta V}\sum_{i\nu_m,{\bf q}} e^{i{\bf q\cdot r}-i\nu_m\tau} \Delta^j_a({\bf q},i\nu_m). }
Then, we have
\eq{
S_0=\frac{1}{\beta V}\sum_{i\omega_n,{\bf k}}\hat{c}^{\dag}({\bf k},i\omega_n)\left(-i\omega_n-\mu+\hbar v_f({\bf k\cdot\sigma})\right)\hat{c}({\bf k},i\omega_n)
}
and
\eqarr{
&&S_1=\frac{1}{(\beta V)^2}\sum_{i\omega_n,i\nu_m, {\bf k,q}}\left(-\hat{\psi}^{\dag}_{\bf \Gamma_i}({\bf k},i\omega_n)A_0({\bf q},i\nu_m)\hat{\psi}_{\bf \Gamma_i}({\bf k-q},i\omega_n-i\nu_m)\right.\nonumber\\
&&\left.+\frac{\hbar v_f}{2}\sum_{a,j}\hat{\psi}^{\dag}_{\bf \Gamma_i}({\bf k},i\omega_n)\Delta^{ j}_a({\bf q},i\nu_m)\sigma^{a}(2k_j-q_j)\hat{\psi}_{\bf \Gamma_i}({\bf k-q},i\omega_n-i\nu_m)\right).
}
Here $i\omega_n=i\frac{(2n+1)\pi}{\beta}$ and $i\nu_m=i\frac{2m \pi}{\beta}$. We have dropped the quadratic term $\frac{\hbar^2k^2}{2m_0^*}$ in $S_0$ and instead, we limit the momentum summation within the range set by the cut-off $\Lambda=\frac{2m_0^*v_f}{\hbar}$, where the quadratic term $\frac{\hbar^2\Lambda^2}{2m_0^*}$ is at the same order as the linear term $\hbar v_f \Lambda$.

Below, we introduce the notation $\tilde{k}=(i\omega_n,{\bf k})$ for short so that the summation over both the frequency and the momentum can be simplified as $\sum_{\tilde{k}}=\frac{1}{\beta V}\sum_{i\omega_n,{\bf k}}$. One should keep in mind the value of $i\omega_n$ is for boson or for fermion operators. The full action can be written as
\eq{
S=S_0+S_1=\sum_{\tilde{k}}\hat{\psi}^{\dag}_{\tilde{k}}(-\mathcal{G}_0^{-1})
\hat{\psi}_{\tilde{k}}+\sum_{\tilde{k},\tilde{k}'}\hat{\psi}^{\dag}_{\tilde{k}}
\mathfrak{X}(\tilde{k},\tilde{k'})\hat{\psi}_{\tilde{k}'},
}
where
\eq{\label{App_eq:G0_1}
\mathcal{G}_0=\left(i\omega_n+\mu-\hbar v_f({\bf k\cdot\sigma})\right)^{-1}
}
and
\eq{
\mathfrak{X}(\tilde{k},\tilde{k'})=-A_0(\tilde{q}=\tilde{k}-\tilde{k}')+\sum_{a,i}\mathcal{T}^{a}_i(\tilde{k},\tilde{k}')\Delta^i_a(\tilde{q}=\tilde{k}-\tilde{k}')
}
with the stress tensor operator
$\mathcal{T}^{a}_i(\tilde{k},\tilde{k}')=\frac{\hbar v_f}{2}(k_i+k'_i)\sigma^a$ for Weyl fermions.

Since the full action is quadratic in fermion operators, one can directly integrate out the Weyl fermions to get the effective action for the $A_0$ and $\Delta$ fields. Let's consider the partition function
\eq{
Z=\int \mathcal{D}(\hat{\psi}^{\dag}\hat{\psi}) exp(-S)=Det(-\mathcal{G}_0^{-1}+\mathfrak{X})
}
and the zero order partition function
\eq{
Z_0=\int \mathcal{D}(\hat{\psi}^{\dag}\hat{\psi}) exp(-S_0)=Det(-\mathcal{G}_0^{-1}),
}
where $Det$ is the determinant. The effective action $W[A_0,\Delta]$ is defined as $Z=Z_0 e^{-W}$ and thus
\eqarr{
&&W[A_0,\Delta]=-ln(Det(-\mathcal{G}_0^{-1}+\mathfrak{X}))+ln(Det(-\mathcal{G}_0^{-1}))
=-Tr(ln(-\mathcal{G}_0^{-1}+\mathfrak{X})-ln(-\mathcal{G}_0^{-1}))\nonumber\\
&&=-Tr(ln(1-\mathfrak{X}\mathcal{G}_0))=\sum_n\frac{1}{n}Tr\left((\mathfrak{X}\mathcal{G}_0)^n\right)
}
which gives us the perturbation expansion. The first order term is given by $Tr(\mathfrak{X}\mathcal{G}_0)=\sum_{\tilde{k}}\mathfrak{X}(\tilde{k},\tilde{k})\mathcal{G}_0(\tilde{k})$. This involve only the terms $A_0(\tilde{q}=0)$ and $\Delta^i_a(\tilde{q}=0)$ in $\mathfrak{X}$. Since we only consider the fluctuation for $\mathfrak{X}$ and thus can choose $A_0(\tilde{q}=0)=0$ and $\Delta^i_a(\tilde{q}=0)=0$ when defining $S_1$. Thus, the first order term will be zero.

Next we consider the second order term, given by
\eqarr{\label{App_eq:Weff1}
&&W[A_0,\Delta]=\frac{1}{2}Tr(\mathfrak{X}\mathcal{G}_0\mathfrak{X}\mathcal{G}_0)=\frac{1}{2}\sum_{\tilde{k}_1,\tilde{k}_2}Tr_{\sigma}\left(
\mathfrak{X}(\tilde{k}_1,\tilde{k}_2)\mathcal{G}_0(\tilde{k}_2)\mathfrak{X}(\tilde{k}_2,\tilde{k}_1)
\mathcal{G}_0(\tilde{k}_1)\right)\nonumber\\
&&=\frac{1}{2}\sum_{\tilde{q}}\left(A_0(\tilde{q})A_0(-\tilde{q})\Pi_0(\tilde{q})+\sum_{ij,ab}\Delta^i_a(\tilde{q})\Delta^j_b
(-\tilde{q})\Phi^{ab}_{ij}(\tilde{q})-\sum_{i,a}A_0(\tilde{q})\Delta^i_a(-\tilde{q})\Theta^a_i(\tilde{q})
-\sum_{i,a}\Delta^i_a(\tilde{q})A_0(-\tilde{q})\Theta^a_i(-\tilde{q})\right),\nonumber\\
}
where three types of correlation functions are defined as
\eqarr{
&&\Pi_0(\tilde{q})=\sum_{\tilde{k}}Tr_{\sigma}\left(\mathcal{G}_0(\tilde{k}-\tilde{q})\mathcal{G}_0(\tilde{k})\right),\label{App_eq:Pi0}\\
&&\Phi^{ab}_{ij}(\tilde{q})=\sum_{\tilde{k}}Tr_{\sigma}\left(\mathcal{T}^{a}_i(\tilde{k},\tilde{k}-\tilde{q})
\mathcal{G}_0(\tilde{k}-\tilde{q})\mathcal{T}^{b}_j(\tilde{k}-\tilde{q},\tilde{k})\mathcal{G}_0(\tilde{k})\right),\label{App_eq:Phi0}\\
&&\Theta^a_i(\tilde{q})=\sum_{\tilde{k}}Tr_{\sigma}\left(\mathcal{G}_0(\tilde{k}-\tilde{q})
\mathcal{T}^{a}_i(\tilde{k}-\tilde{q},\tilde{k})\mathcal{G}_0(\tilde{k})\right)\label{App_eq:Theta0}.
}
Here $\Pi_0$ is the density-density correlation function, $\Phi^{ab}_{ij}$ is the stress-stress correlation function, while $\Theta^a_i$ is the stress-density correlation function.

\section{A brief review of Nieh-Yan anomaly}
\label{App_Sec:NYanomaly}
In this section, we will first review the Weyl/Dirac fermions in the curved space and the Nieh-Yan anomaly. Then we will discuss the connection of the stress-stress correlation function $\Phi^{ab}_{ij}$ and Nieh-Yan anomaly in our formalism. In a curved space, the action for Dirac fermions can be written as \cite{weinberg1972gravitation,parker2009quantum}
\eq{
S_D=\int d^4x \bar{\psi}\left(\frac{i}{2}\gamma^a\{e_a^\mu,\nabla_\mu\}-m\right)\psi,
}
where $\gamma^a$s are the standard $\gamma$-matrices with $\{\gamma^a,\gamma^b\}=2\eta^{ab}$ and $\eta^{ab}$ is the metric in the Minkowski space, and $e_a^\mu$ is the frame field. Here the Einstein summation rule has been assumed. The frame fields satisfy $\eta^{ab}e^\mu_a e^\nu_b=g^{\mu\nu}$ with the metric $g^{\mu\nu}$ in the curved space, so that $\{\gamma^\mu,\gamma^\nu\}=2g^{\mu\nu}$ where $\gamma^\mu=\gamma^a e^\mu_a$. $\nabla_\mu=\partial_\mu+\Gamma_\mu$ is the covariant derivative with the spin connection $\Gamma_\mu=\frac{i}{2}\eta_{ac}\Gamma^c_{b\mu}J^{ab}$ and the generator $J^{ab}=-\frac{i}{4}[\gamma^a,\gamma^b]$ of the Lorentz group.

It is convenient to introduce the so-called spin connection 1-form ${\bf \Gamma}^c_b=\Gamma^c_{b\mu}d{\bf x}^\mu$ and the coframe field 1-form ${\bf e}^a=\bar{e}^a_{\mu}d{\bf x}^\mu$, where the coframe field is the inverse of the frame field, $e^\nu_a \bar{e}^a_{\mu}=\delta^{\nu}_{\mu}$. The Cartan's structure equations of the Riemann-Cartan spacetime \cite{eguchi1980gravitation} define the curvature 2-form ${\bf R}^a_b=d{\bf \Gamma}^a_b+{\bf \Gamma}^a_c\wedge {\bf \Gamma}^c_b$ and the torsion 2-form ${\bf T}^a=d{\bf e}^a+{\bf \Gamma}^a_c\wedge {\bf e}^c$. In the standard general relativity, the torsion is assumed to be zero (zero torsion constraint), leading to the relation between the coframe field and the spin connection ($d{\bf e}^a+{\bf \Gamma}^a_c\wedge {\bf e}^c=0$). However, in the Einstein-Cartan theory\cite{hehl1976general,eguchi1980gravitation}, the torsion can be non-zero and consequently, the frame field (or coframe field) and the spin connection should be treated as two independent fields.

By comparing the effective action (\ref{App_eq:Heff1}) for our system with the action of Dirac field in the curved space, it is clear that the electron-phonon coupling can create a non-trivial coframe field ${\bf e}^a$ but the spin connection ${\bf \Gamma}^c_b$ is still zero. Consequently, the Weyl fermions in our system will feel nonzero torsion, given by ${\bf T}^a=d{\bf e}^a$, but zero curvature ${\bf R}^a_b=0$. This is in sharp contrast to the standard general relativity with non-zero curvature but zero torsion, and is known as the "Weitzenb${\ddot o}$ck spacetime", which was studied in the "teleparallel gravity theory"\cite{hayashi1979new}.

When the Dirac fermion is coupled to an electromagnetic field, quantum correction can give rise to the anomalous non-conservation of the chiral charge, known as the "chiral anomaly", given by
\eq{
\partial_\mu \langle j^{5\mu}\rangle=\frac{1}{16\pi^2}\epsilon^{\mu\nu\lambda\rho}F_{\mu\nu}F_{\lambda\rho}.
}
where $F_{\mu\nu}=\partial_\mu A_{\nu}-\partial_\nu A_{\mu}$ is the strength of the gauge field. The chiral anomaly effect plays an essential role in predicting and understanding a number of physical phenomena. %, including negative magnetoresistance, chiral zero Landau levels, chiral magnetic field effect, planar Hall effect and optical activity, in Weyl semimetals.
In our system, the phonon will induce the torsion field for the KW fermions, instead of the pseudo-gauge field. It turns out that the torsion field can also contribute to the non-conservation of the chiral charge, which is known as the Nieh-Yan anomaly\cite{nieh1982quantized,nieh1982identity,chandia1997topological,nieh2007torsional} and given by
\eq{
\partial_\mu \langle j^{5\mu}\rangle=\frac{\mathcal{F}}{4}\eta_{ab}\epsilon^{\mu\nu\lambda\rho}T^a_{\mu\nu}T^b_{\lambda\rho}.
}
Since the spin connection vanishes, the torsion field ${\bf T}^a=d{\bf e}^a$ gives the field strength of the coframe field ${\bf e}^a$ and its components are given by $T^a_{\mu\nu}=\partial_{\mu} \bar{e}^a_{\nu}-\partial_{\nu} \bar{e}^a_{\mu}$ and the Nieh-Yan anomaly equation is also written as
\eq{\label{App_eq:NYanomaly1}
\partial_\mu \langle j^{5\mu}\rangle=\mathcal{F}\eta_{ab}\epsilon^{\mu\nu\lambda\rho}\partial_\mu \bar{e}^a_{\nu} \partial_{\lambda} \bar{e}^b_{\rho}.
}
Unlike the chiral anomaly with the universal dimensionless coefficient $\frac{1}{16\pi^2}$ (Here we follow the convention of high energy physics and set $e=\hbar=1$), the coefficient $\mathcal{F}$ in the Nieh-Yan anomaly has the dimension $[1/L]^2$, where $[L]$ labels the dimension of length. This can be easily obtained from the dimension counting as follows. In the Dirac action, the fermion field has the dimension $[1/L]^{3/2}$ and the chiral current $\langle j^{5\mu}\rangle=\langle \bar{\psi}\gamma^5\psi \rangle$ has the dimension $[1/L]^3$. Thus, the left hand side of the anomaly equation ($\partial_\mu \langle j^{5\mu}\rangle$) has the dimension $[1/L]^4$. On the other hand, the coframe field $\bar{e}^a_{\nu}$ is dimensionless and thus $\partial_\mu \bar{e}^a_{\nu} \partial_{\lambda} \bar{e}^b_{\rho}$ has the dimension $[1/L]^2$. This immediately means that the coefficient $\mathcal{F}$ has the dimension $[1/L]^2$. It turns out that $\mathcal{F}$ depends on the cut-off $\Lambda$ in the momentum space and is given by $\mathcal{F}=\frac{\Lambda^2}{4\pi^2}$\cite{huang2020nieh,chandia1997topological} (the factor $4\pi^2$ is not important since one can always re-scale the cut-off $\Lambda$). In contrast, it was recently proposed\cite{huang2020nieh,nissinen2020thermal,nissinen2019thermal} that at a finite temperature, the coefficient takes the form $\mathcal{F}=F_0+F_1(k_BT)^2$ with $F_0=\frac{\Lambda^2}{4\pi^2}$ and $F_1=-\frac{1}{12}$. The temperature dependent term is proportional to $(k_BT)^2$, which absorbs the dimension $[1/L]^2$, and thus, the coefficient $F_1$ becomes dimensionless and universal in the sense that it is only proportional to the central charge of (1+1)-dimensional Dirac fermion\cite{huang2020nieh}. Thus, the anomaly induced by this temperature dependent term is dubbed "thermal Nieh-Yan anomaly".

In order to compare with our results, it is convenient to derive the effective action that corresponds to the Nieh-Yan anomaly. Since the right hand side of Eq. (\ref{App_eq:NYanomaly1}) is actually a total derivative, the chiral current can be given by $\langle j^{5\mu}\rangle=\mathcal{F}\eta_{ab}\epsilon^{\mu\nu\lambda\rho}\bar{e}^a_{\nu} \partial_{\lambda} \bar{e}^b_{\rho}$. With $\langle j^{5\mu}\rangle=\frac{\delta S_{eff}}{\delta A_{5\mu}}$, we have
\eq{
S_{NY}=\mathcal{F} \int d^4x \eta_{ab}\epsilon^{\mu\nu\lambda\rho}A_{5\mu} \bar{e}^a_{\nu} \partial_{\lambda} \bar{e}^b_{\rho},
}
where $A_{5\mu}$ is the chiral gauge potential. Since the phonon can only induce chiral chemical potential in our case, we only keep the $A_{50}$ term in the above action and thus obtain
\eq{\label{App_eq:SNY1}
S_{NY}=\mathcal{F} \int d^4x \eta_{ab}\epsilon^{\nu\lambda\rho}A_{50} \bar{e}^a_{\nu} \partial_{\lambda} \bar{e}^b_{\rho},
}
where $\nu,\lambda,\rho=x,y,z$.

In our effective action (\ref{eq:Leff1}) of the main text, the frame field is given by $e^{j}_a=\delta^j_a+\Delta^j_a$ and the corresponding coframe field is $\bar{e}^{a}_j=\delta^a_j-\Delta^j_a$ with $a,j=x,y,z$. Since we only concern the spatial coordinate here, the upper and lower indices do not have much meaning. Furthermore, $\Delta^j_a$ is proportional to the strain tensor $u_{ja}$ which is symmetric with respect to $j$ and $a$. We next focus on the stress-stress correlation function term (\ref{App_eq:Phi0}) in the effective action (\ref{App_eq:Weff1}). We may expand $\Phi^{ab}_{ij}(\tilde{q})$ as a function of $\tilde{q}$, $\Phi^{ab}_{ij}(\tilde{q})=\Phi^{ab}_{ij}(0)+(\partial_{\bf q}\Phi^{ab}_{ij})_{\tilde{q}=0}\cdot {\bf q}+(\partial_{\omega_n}\Phi^{ab}_{ij})_{\tilde{q}=0}\omega_n +...$. Here we only focus on the term that is linearly proportional to ${\bf q}$ and denote $\Phi^{ab}_{ij,l}=(\partial_{q_l}\Phi^{ab}_{ij})_{\tilde{q}=0}$. Then, the corresponding term in the action is given by
\eqarr{\label{App_eq:WeffNY2}
&&W_{NY}[\Delta]=\frac{1}{2}\sum_{\tilde{q},ijl,ab}\Phi^{ab}_{ij,l}\left(\Delta^i_a(\tilde{q})\Delta^j_b
(-\tilde{q})q_l\right),\nonumber\\
&&=\frac{1}{2}\sum_{ijl,ab}\int d^3r d\tau \Phi^{ab}_{ij,l}\Delta^i_a({\bf r},\tau)\left( i\partial_l\Delta^j_b({\bf r},\tau)\right).
}
As demonstrated in the next Sec. \ref{App_Sec:Analytical}, the coefficient $\Phi^{ab}_{ij,l}$ is purely imaginary and proportional to $\delta_{ij}\epsilon^{abl}$, where $\epsilon$ is the Levi-civita symbol. Moreover, this term is linearly proportional to the chemical potential $\mu$ for a small $\mu$. Therefore, by replacing $\bar{e}^a_{\nu}$ with $\Delta^i_a$, $W_{NY}$ takes the exact form as $S_{NY}$ for the Nieh-Yan anomaly, and we call this term as the Nieh-Yan term below. Actually, the symmetry analysis of the possible terms for the strain tensors, given in the Sec. \ref{App_Sec:phonon}, suggests that the Nieh-Yan term is the only term that is allowed by symmetry at this order of ${\bf q}$ in a uniform isotropic system.
%Furthermore, we evaluate the temperature dependence of this term and find the $T^2$ dependence, as expected from the thermal Nieh-Yan anomaly.

\section{Analytical evaluation of the Nieh-Yan term}
\label{App_Sec:Analytical}
In this section, we will evaluate the coefficient $\Phi^{ab}_{ij,l}$ analytically. With Eq. (\ref{App_eq:Phi0}), we have
\eqarr{\label{App_eq:pPhipq1}
&&\Phi^{ab}_{ij,l}=\left(\frac{\partial\Phi^{ab}_{ij}}{\partial q_l}\right)_{\tilde{q}=0}=\sum_{\tilde{k}}Tr_{\sigma}\left((\partial_{q_l}\mathcal{T}^{a}_i(\tilde{k},\tilde{k}-\tilde{q}))_{\tilde{q}=0}
\mathcal{G}_0(\tilde{k})\mathcal{T}^{b}_j(\tilde{k},\tilde{k})\mathcal{G}_0(\tilde{k})+\mathcal{T}^{a}_i(\tilde{k},\tilde{k})
(\partial_{q_l}\mathcal{G}_0(\tilde{k}-\tilde{q}))_{\tilde{q}=0}\mathcal{T}^{b}_j(\tilde{k},\tilde{k})\mathcal{G}_0(\tilde{k})\right.\nonumber\\
&&\left.+\mathcal{T}^{a}_i(\tilde{k},\tilde{k})
\mathcal{G}_0(\tilde{k})(\partial_{q_l}\mathcal{T}^{b}_j(\tilde{k}-\tilde{q},\tilde{k}))_{\tilde{q}=0}\mathcal{G}_0(\tilde{k})\right)\nonumber\\
&&=\sum_{\tilde{k}}Tr_{\sigma}\left(\left(-\frac{\hbar v_f}{2}\right)(\delta_{il}\sigma^a\mathcal{G}_0(\tilde{k})\mathcal{T}^{b}_j(\tilde{k},\tilde{k})\mathcal{G}_0(\tilde{k})+\mathcal{T}^{a}_i(\tilde{k},\tilde{k})
\mathcal{G}_0(\tilde{k})\delta_{jl}\sigma^b \mathcal{G}_0(\tilde{k}))+\mathcal{T}^{a}_i(\tilde{k},\tilde{k})
(\partial_{k_l}\mathcal{G}_0(\tilde{k}))\mathcal{T}^{b}_j(\tilde{k},\tilde{k})\mathcal{G}_0(\tilde{k})\right)\nonumber\\
}
The Green's function $\mathcal{G}_0$ is given in Eq. (\ref{App_eq:G0_1}) and here we focus on the situation with a small chemical potential $\mu$, so we expand $\mathcal{G}_0$ up to the linear order in $\mu$ as
$\mathcal{G}_0(\mu)=\mathcal{G}_0(\mu=0)+(\partial_{\mu}\mathcal{G}_0)_{\mu=0}\mu$ with
\eq{
\mathcal{G}_0(\mu=0)=\left(i\omega_n-\hbar v_f({\bf k\cdot\sigma})\right)^{-1}=\frac{i\omega_n+\hbar v_f({\bf k\cdot\sigma})}{(i\omega_n)^2-(\hbar v_f k)^2}
}
and
\eq{
(\partial_{\mu}\mathcal{G}_0)_{\mu=0}=-\left(i\omega_n-\hbar v_f({\bf k\cdot\sigma})\right)^{-2}=-(\mathcal{G}_0(\mu=0))^2.
}

Below we absorb $\hbar v_f$ into the definition of the momentum ${\bf k}$, but keep track on the chirality of the Weyl fermions. Thus, we define the chirality as $\chi= sign (\hbar v_f)$, and obtain
\eq{
\mathcal{G}_0(\mu=0)=\frac{i\omega_n+\chi({\bf k\cdot\sigma})}{D^2}
}
where $D^2=(i\omega_n)^2-k^2$, and
\eqarr{
&&(\partial_{\mu}\mathcal{G}_0)_{\mu=0}=-(\mathcal{G}_0(\mu=0))^2=-\frac{(i\omega_n+\chi({\bf k\cdot\sigma}))^2}{D^4}\nonumber\\
&&=-\frac{(i\omega_n)^2+k^2+2i\omega_n \chi{\bf k\cdot\sigma}}{D^4}.
}
Thus, the Green's function is given by
\eq{
\mathcal{G}_0({\bf k},i\omega_n)=\frac{i\omega_n+\chi({\bf k\cdot\sigma})}{D^2}-\frac{((i\omega_n)^2+k^2+2i\omega_n \chi{\bf k\cdot\sigma})\mu}{D^4}.
}

Let's first look at the first term in Eq. (\ref{App_eq:pPhipq1}) and the direct calculation gives
\eqarr{
&&\sum_{\tilde{k}}Tr_{\sigma}(\delta_{il}\sigma^a\mathcal{G}_0(\tilde{k})\mathcal{T}^{b}_j(\tilde{k},\tilde{k})\mathcal{G}_0(\tilde{k}))\nonumber\\
&&=\sum_{\tilde{k}}\frac{\delta_{il}k_j}{D^4}\left(i\omega_n+\mu-\frac{2(i\omega_n)^2\mu}{D^2}\right)\left(1-\frac{2i\omega_n\mu}{D^2}\right)
(\sigma^a\sigma^b\chi({\bf k\cdot \sigma})+\sigma^a\chi({\bf k\cdot\sigma})\sigma^b)\nonumber\\
&&=\sum_{\tilde{k}}\frac{\chi\delta_{il}k_j^2}{D^4}\left(i\omega_n+\mu-\frac{2(i\omega_n)^2\mu}{D^2}\right)\left(1-\frac{2i\omega_n\mu}{D^2}\right)
Tr_{\sigma}(\sigma^a\sigma^b\sigma^j+\sigma^a\sigma^j\sigma^b)\nonumber\\
&&=\sum_{\tilde{k}}\frac{\chi\delta_{il}k_j^2}{D^4}\left(i\omega_n+\mu-\frac{2(i\omega_n)^2\mu}{D^2}\right)\left(1-\frac{2i\omega_n\mu}{D^2}\right)
Tr_{\sigma}(\sigma^a 2\delta^{bj})=0
}
In the second step of the above derivation, we collect all the terms that have even order in the momentum ${\bf k}$ since these are the nonzero term under the integral of the momentum angle. Similar calculation also shows the second term in Eq. (\ref{App_eq:pPhipq1}) is also zero. Thus, only the last term is left non-zero. To evaluate this term, we first need to calculate $\partial_{k_l}\mathcal{G}_0(\tilde{k})$, which is given by
\eqarr{
&&\partial_{k_l}\mathcal{G}_0(\tilde{k})=\frac{\chi\sigma^l}{D^2}+(i\omega_n+\chi({\bf k\cdot \sigma}))\frac{\partial (1/D^2)}{\partial k_l}-\mu\frac{2k^l+2i\omega_n \chi\sigma^l}{D^4}-\mu((i\omega_n)^2+k^2+2i\omega_n \chi{\bf k\cdot\sigma})\frac{\partial (1/D^4)}{\partial k_l}\nonumber\\
&&=\frac{\chi\sigma^l}{D^2}+(i\omega_n+\chi({\bf k\cdot \sigma}))\frac{2k^l}{D^4}-\mu\frac{2k^l+2i\omega_n \chi\sigma^l}{D^4}-\mu((i\omega_n)^2+k^2+2i\omega_n \chi{\bf k\cdot\sigma})\frac{4k^l}{D^6}\nonumber\\
&&=\frac{\chi\sigma^l}{D^2}+(i\omega_n+\chi({\bf k\cdot \sigma}))\frac{2k^l}{D^4}-\frac{\mu}{D^6}\left((2k^l+2i\omega_n \chi\sigma^l)((i\omega_n)^2-k^2)+4k^l((i\omega_n)^2+k^2+2i\omega_n \chi{\bf k\cdot\sigma})\right)\nonumber\\
&&=\frac{\chi\sigma^l}{D^2}+(i\omega_n+\chi({\bf k\cdot \sigma}))\frac{2k^l}{D^4}-\frac{2\mu i\omega_n \chi\sigma^l}{D^4}-\frac{\mu}{D^6}\left(6k^l(i\omega_n)^2+2k^2k^l+8k^l(i\omega_n) \chi{\bf k\cdot\sigma}\right).
}
By substituting these terms into the last term in Eq. (\ref{App_eq:pPhipq1}), we find
\eqarr{
&&\Phi^{ab}_{ij,l}=\sum_{\tilde{k}}Tr_{\sigma}\left(\mathcal{T}^{a}_i(\tilde{k},\tilde{k})
(\partial_{k_l}\mathcal{G}_0(\tilde{k}))\mathcal{T}^{b}_j(\tilde{k},\tilde{k})\mathcal{G}_0(\tilde{k})\right)\nonumber\\
&&=\sum_{\tilde{k}}Tr_{\sigma}\left(\sigma^a k_i
\left(\frac{\chi\sigma^l}{D^2}+(i\omega_n+\chi({\bf k\cdot \sigma}))\frac{2k^l}{D^4}-\frac{2\mu i\omega_n \chi\sigma^l}{D^4}-\frac{\mu}{D^6}\left(6k^l(i\omega_n)^2+2k^2k^l+8k^l(i\omega_n) \chi{\bf k\cdot\sigma}\right)\right)\right.\nonumber\\
&&\left.\sigma^b k_j\left(\frac{i\omega_n+\chi({\bf k\cdot\sigma})}{D^2}-\frac{\mu}{D^4}((i\omega_n)^2+k^2+2i\omega_n \chi{\bf k\cdot\sigma})\right)\right)\nonumber\\
&&=\sum_{\tilde{k}}Tr_{\sigma}\left(\sigma^a k_i
\left(\frac{\chi\sigma^l}{D^2}+\chi({\bf k\cdot \sigma})\frac{2k^l}{D^4}-\frac{2\mu i\omega_n \chi\sigma^l}{D^4}-\frac{\mu}{D^6}8k^l(i\omega_n) \chi{\bf k\cdot\sigma}\right)\sigma^b k_j\left(\frac{i\omega_n}{D^2}-\frac{((i\omega_n)^2+k^2)\mu}{D^4}\right)\right.\nonumber\\
&&\left.+\sigma^a k_i
\left(i\omega_n\frac{2k^l}{D^4}-\frac{\mu}{D^6}\left(6k^l(i\omega_n)^2+2k^2k^l\right)\right)\sigma^b k_j\left(\frac{\chi({\bf k\cdot\sigma})}{D^2}-\frac{(2i\omega_n \chi{\bf k\cdot\sigma})\mu}{D^4}\right)\right).
}
In the above, since only the terms with even number of the momentum ${\bf k}$ can be non-zero, we thus obtain two terms, and let's calculate them separately. For the first term, we have
\eqarr{
&&\sum_{\tilde{k}}Tr_{\sigma}\left(\sigma^a k_i
\left(\frac{\chi\sigma^l}{D^2}\left(1-\frac{2i\omega_n\mu}{D^2}\right)+\chi({\bf k\cdot \sigma})\frac{2k^l}{D^4}\left(1-\frac{4i\omega_n\mu}{D^2} \right)\right)\sigma^b k_j\left(\frac{i\omega_n}{D^2}-\frac{((i\omega_n)^2+k^2)\mu}{D^4}\right)\right)\nonumber\\
&&=\sum_{\tilde{k}}\chi k_i k_j\left(\frac{i\omega_n}{D^2}-\frac{((i\omega_n)^2+k^2)\mu}{D^4}\right)
\left(\frac{1}{D^2}\left(1-\frac{2i\omega_n\mu}{D^2}\right)Tr_{\sigma}\left(\sigma^a\sigma^l\sigma^b\right)+\frac{2k^l}{D^4}\left(1-\frac{4i\omega_n\mu}{D^2} \right)Tr_{\sigma}\left(\sigma^a({\bf k\cdot \sigma})\sigma^b\right)\right)\nonumber\\
&&=\sum_{\tilde{k}}\chi k_i k_j\left(\frac{i\omega_n}{D^2}-\frac{((i\omega_n)^2+k^2)\mu}{D^4}\right)
\left(\frac{1}{D^2}\left(1-\frac{2i\omega_n\mu}{D^2}\right)2i\epsilon^{alb}+\frac{2k^l k_m}{D^4}\left(1-\frac{4i\omega_n\mu}{D^2} \right)2i\epsilon^{amb}\right)\nonumber\\
&&=-\sum_{\tilde{k}}\frac{2\chi\mu k_i k_j}{D^6}\left(2(i\omega_n)^2\left(i\epsilon^{alb}+4i\epsilon^{amb}\frac{k^l k_m}{D^2}\right)+((i\omega_n)^2+k^2)\left(i\epsilon^{alb}+2i\epsilon^{amb}\frac{k^l k_m}{D^2}\right)\right).
}
In the last step above, we have dropped all the terms with odd number of $i\omega_n$ since the frequency summation will make them vanishing.
For the second term, we have
\eqarr{
&&\sum_{\tilde{k}}Tr_{\sigma}\left(\sigma^a k_i
\left(i\omega_n\frac{2k^l}{D^4}-\frac{\mu}{D^6}\left(6k^l(i\omega_n)^2+2k^2k^l\right)\right)\sigma^b k_j\left(\frac{\chi({\bf k\cdot\sigma})}{D^2}-\frac{(2i\omega_n \chi{\bf k\cdot\sigma})\mu}{D^4}\right)\right)\nonumber\\
&&=\sum_{\tilde{k}}\chi k_i k_j\left(i\omega_n\frac{2k^l}{D^4}-\frac{\mu}{D^6}\left(6k^l(i\omega_n)^2+2k^2k^l\right)\right)\left(\frac{1}{D^2}-\frac{2i\omega_n\mu}{D^4}\right)Tr_{\sigma}\left(\sigma^a
\sigma^b ({\bf k\cdot\sigma})\right)\nonumber\\
&&=\sum_{\tilde{k}}\frac{2\chi k_i k_jk^l}{D^6}\left(i\omega_n-\frac{\mu}{D^2}\left(3(i\omega_n)^2+k^2\right)\right)\left(1-\frac{2i\omega_n\mu}{D^2}\right)Tr_{\sigma}\left(\sigma^a
\sigma^b ({\bf k\cdot\sigma})\right)\nonumber\\
&&=-\sum_{\tilde{k}}\frac{2\chi\mu k_i k_j k^l}{D^8} \left(2(i\omega_n)^2+\left(3(i\omega_n)^2+k^2\right)\right)Tr_{\sigma}(\sigma^a
\sigma^b ({\bf k\cdot\sigma})) \nonumber\\
&&=-\sum_{\tilde{k}}\frac{4\chi\mu k_i k_j k^lk_m}{D^8} \left(5(i\omega_n)^2+k^2\right)i\epsilon^{abm}.
}
Again here we only pick up the terms with even order of $i\omega_n$. Finally, we can add these two terms together and obtain
\eqarr{
&&\left(\frac{\partial\Phi^{ab}_{ij}}{\partial q_l}\right)_{\tilde{q}=0}=-\sum_{\tilde{k}}\frac{2\chi\mu k_i k_j}{D^6}\left((3(i\omega_n)^2+k^2)i\epsilon^{alb}+(5(i\omega_n)^2+k^2)2i\epsilon^{amb}\frac{k^l k_m}{D^2}\right)\nonumber\\
&&-\sum_{\tilde{k}}\frac{4\chi\mu k_i k_j k^lk_m}{D^8} \left(5(i\omega_n)^2+k^2\right)i\epsilon^{abm}\nonumber\\
&&=-\sum_{\tilde{k}}\frac{2\chi\mu k_i k_j}{D^6}\left((3(i\omega_n)^2+k^2)i\epsilon^{alb}+(10(i\omega_n)^2+2k^2)i\epsilon^{amb}\frac{k^l k_m}{D^2}+\left(10(i\omega_n)^2+2k^2\right)i\epsilon^{abm}\frac{k^l k_m}{D^2}\right)\nonumber\\
&&=-\sum_{\tilde{k}}\frac{2\chi\mu k_i k_j}{D^6}(3(i\omega_n)^2+k^2)i\epsilon^{alb}=2i\epsilon^{alb}\chi\mu \sum_{\tilde{k}}\frac{k_i k_j}{D^4}\left(1-\frac{4(i\omega_n)^2}{D^2}\right).
}
In the above expression, the integral over the momentum ${\bf k}$ can only be non-zero when $i=j$. Furthermore, the integral over the polar and azimuthal angles of the momentum gives rise to
\eq{
\int d(cos\theta) d\varphi k_x^2=\int d(cos\theta) d\varphi k_y^2= \int d(cos\theta) d\varphi k_z^2=\frac{4\pi}{3}k^2
}
with ${\bf k}=(k,\theta,\varphi)$ in the spherical coordinate. Thus, let's define the function
\eq{\label{App_eq:F01}
\mathcal{F}_0=\frac{2}{\beta V}\sum_{{\bf k},i\omega_n}\frac{k_x^2 }{D^4}\left(1-\frac{4(i\omega_n)^2}{D^2}\right)
=\frac{1}{3\pi^2\beta}\sum_{i\omega_n}\int_0^{\Lambda} k^4dk \frac{1}{D^4}\left(1-\frac{4(i\omega_n)^2}{D^2}\right),
}
where a cut-off $\Lambda$ in the momentum space has been assumed, and
\eq{\label{App_eq:parPhql1}
\Phi^{ab}_{ij,l}=\left(\frac{\partial\Phi^{ab}_{ij}}{\partial q_l}\right)_{\tilde{q}=0}=i\epsilon^{alb}\delta_{ij}\chi\mu \mathcal{F}_0,
}
from which one can see that $\Phi^{ab}_{ij,l}$ is indeed pure imaginary ($\mathcal{F}_0$ is real, as shown below), and proportional to $\epsilon^{alb}\delta_{ij}$. Thus, we demonstrate the effective action (\ref{App_eq:Weff2}) together with Eq. (\ref{App_eq:parPhql1}) indeed takes the form of Nieh-Yan term.

Next we will evaluate the coefficient $\mathcal{F}_0$ in Eq. (\ref{App_eq:F01}) analytically. Let's separate $\mathcal{F}_0$ into $\mathcal{F}_{01}$ and $\mathcal{F}_{02}$ with
\eq{
\mathcal{F}_{01}=\frac{1}{3\pi^2\beta}\sum_{i\omega_n}\int_0^{\Lambda} dk \frac{k^4}{((i\omega_n)^2-k^2)^2}
}
and
\eq{
\mathcal{F}_{02}=\frac{4}{3\pi^2\beta}\sum_{i\omega_n}\int_0^{\Lambda} dk \frac{k^4(i\omega_n)^2}{((i\omega_n)^2-k^2)^3},
}
so that
\eq{\mathcal{F}_0=\mathcal{F}_{01}-\mathcal{F}_{02}.
}
Here we have substitute the expression for $D^2$.

We perform the frequency summation in $\mathcal{F}_{01}$ and consider the contour integral
\eq{\label{App_eq:contourint1}
I=\oint_{R} \frac{dz}{2\pi i} f(z) n_F(z)
}
with
\eq{
f(z)=\frac{1}{(z^2-k^2)^2}
}
in the complex-z plane with the radius $R$ of $z$. In the limit $R\rightarrow \infty$, since the integrand $f(z) n_F(z)$ decays to zero fast enough, this integral should be zero. On the other hand, this integral can re-expressed as the integral around the poles of the integrand in the complex-z plane, and thus is determined by the residuals of these poles according to the residual theorem. We next need to figure out all the poles and the corresponding residuals for the integrand $f(z) n_F(z)$.

For the Fermi function $n_F(z)=\frac{1}{e^{\beta z}+1}$, the poles are $z_n=i\omega_n=i\frac{(2n+1)\pi}{\beta}$ and the corresponding residuals for $f(z) n_F(z)$ are $Res[f(z) n_F(z),z=z_n]=-\frac{1}{\beta}f(i\omega_n)$.

There are another two poles $z_{\pm}=\pm k$ from the function $f(z)$. Since they are not simple poles, we need a bit more work to extract the residuals. For $z=z_+=k$, let's denote $f(z) n_F(z)=\frac{g(z)}{(z-k)^2}$ with $g(z)=\frac{n_F(z)}{(z+k)^2}$. We need to expand $g(z)$ around $z_+=k$ and pick up the first order term $\left(\frac{\partial g(z)}{\partial z}\right)_{z=k}(z-k)$ since the coefficient in this term will give the residual of $f(z) n_F(z)$. Direct calculations give rise to
\eqarr{
&&\frac{\partial g(z)}{\partial z}=-\frac{2}{(z+k)^3}n_F(z)+\frac{1}{(z+k)^2}\partial_z n_F(z)\nonumber\\
&&\partial_z n_F(z)=-\frac{\beta e^{\beta z}}{(e^{\beta z}+1)^2}=-\frac{\beta}{4(\cosh(\beta z/2))^2},
}
and thus the corresponding residual is
\eqarr{
&&Res[f(z) n_F(z),z=z_+]=\left(\frac{\partial g(z)}{\partial z}\right)_{z=k}=-\frac{2}{(2k)^3}n_F(k)-\frac{1}{(2k)^2}\frac{\beta}{4(\cosh(\beta k/2))^2}\nonumber\\
&&=-\frac{1}{4k^3}n_F(k)-\frac{1}{16k^2}\frac{\beta}{(\cosh(\beta k/2))^2}.
}

For $z=z_-=-k$, we denote $f(z) n_F(z)=\frac{g(z)}{(z+k)^2}$ with $g(z)=\frac{n_F(z)}{(z-k)^2}$. With
\eqarr{
&&\frac{\partial g(z)}{\partial z}=-\frac{2}{(z-k)^3}n_F(z)+\frac{1}{(z-k)^2}\partial_z n_F(z),
}
we obtain
\eqarr{
&&Res[f(z) n_F(z),z=z_-]=\left(\frac{\partial g(z)}{\partial z}\right)_{z=-k}=-\frac{2}{(-2k)^3}n_F(-k)-\frac{1}{(-2k)^2}\frac{\beta}{4(\cosh(\beta k/2))^2}\nonumber\\
&&=\frac{1}{4k^3}n_F(-k)-\frac{1}{16k^2}\frac{\beta}{(\cosh(\beta k/2))^2}.
}
Putting all the results together, we have
\eqarr{
&&I=-\frac{1}{\beta}\sum_{i\omega_n}f(i\omega_n)+Res[f(z) n_F(z),z=z_+]+Res[f(z) n_F(z),z=z_-]=0\nonumber\\
&&\rightarrow  \frac{1}{\beta}\sum_{i\omega_n}f(i\omega_n)=Res[f(z) n_F(z),z=z_+]+Res[f(z) n_F(z),z=z_-]\nonumber\\
&&=\frac{1}{4k^3}(n_F(-k)-n_F(k))-\frac{1}{8k^2}\frac{\beta}{(\cosh(\beta k/2))^2}\nonumber\\
&&=\frac{1}{4k^3}\tanh(\beta k/2)-\frac{1}{8k^2}\frac{\beta}{(\cosh(\beta k/2))^2},
}
and
\eqarr{
&&\mathcal{F}_{01}=\frac{1}{3\pi^2\beta}\sum_{i\omega_n}\int_0^{\Lambda} dk \frac{k^4}{((i\omega_n)^2-k^2)^2}=\frac{1}{3\pi^2}\int_0^{\Lambda} dk k^4 \left(\frac{1}{4k^3}\tanh(\beta k/2)-\frac{1}{8k^2}\frac{\beta}{(\cosh(\beta k/2))^2}\right)\nonumber\\
&&=\frac{1}{3\pi^2}\int_0^{\Lambda} dk \left(\frac{k}{4}\tanh(\beta k/2)-\frac{k^2}{8}\frac{\beta}{(\cosh(\beta k/2))^2}\right).
}
We may consider the coefficient $\mathcal{F}_{01}$ at the zero temperature and its correction from the finite temperature, $\mathcal{F}_{01}=\mathcal{F}_{01}(T=0)+\delta\mathcal{F}_{01}(T)$. At zero temperature ($\beta\rightarrow \infty$), the first term in the above integral diverges as $\Lambda^2$ since $\tanh(\beta \Lambda/2)\rightarrow 1$ while the second term converges (actually vanishes). Thus, we have
\eq{
\mathcal{F}_{01}(T=0)=\frac{1}{3\pi^2}\int_0^{\Lambda} dk\frac{k}{4}=\frac{\Lambda^2}{24\pi^2}
}
and the finite temperature correction is given by
\eqarr{
&&\delta\mathcal{F}_{01}(T)=\frac{1}{3\pi^2}\int_0^{\Lambda} dk \left(\frac{k}{4}(\tanh(\beta k/2)-1)-\frac{k^2}{8}\frac{\beta}{(\cosh(\beta k/2))^2}\right)\nonumber\\
&&=\frac{1}{3\pi^2}\int_0^{\infty} d(2x/\beta) \left(\frac{2x}{4\beta}(\tanh(x)-1)-\frac{(2x)^2}{8\beta^2}\frac{\beta}{(\cosh(x))^2}\right)\nonumber\\
&&=\frac{1}{3\pi^2 \beta^2}\int_0^{\infty} dx \left(x(\tanh(x)-1)-\frac{x^2}{(\cosh(x))^2}\right)\nonumber\\
&&=-\frac{1}{24\beta^2}=-\frac{(k_BT)^2}{24}
}
where we have used $x=\beta k/2$ and $\Lambda\rightarrow \infty$.

For $\mathcal{F}_{02}$, we should choose
\eq{
f(z)=\frac{z^2}{(z^2-k^2)^3}
}
in Eq. (\ref{App_eq:contourint1}). The poles and residuals for $n_F(z)$ remain the same. For $f(z)$, the poles are still at $z_{\pm}=\pm k$, but the residuals are changed. For $z=z_+=k$, we denote $f(z) n_F(z)=\frac{g(z)}{(z-k)^3}$ with $g(z)=\frac{z^2}{(z+k)^3}n_F(z)$. Since this is a pole of third order, we need to evaluate $\left(\frac{1}{2}\frac{\partial^2 g(z)}{\partial z^2}\right)_{z=k}$. Direct calculations give rise to
\eqarr{
&&\frac{\partial g(z)}{\partial z}=\left(\frac{2z}{(z+k)^3}-\frac{3z^2}{(z+k)^4}\right)n_F(z)+\frac{z^2}{(z+k)^3}\partial_z n_F(z),\nonumber\\
&&\frac{\partial^2 g(z)}{\partial z^2}=\left(\frac{2}{(z+k)^3}-\frac{12z}{(z+k)^4}+\frac{12z^2}{(z+k)^5}\right)n_F(z)+2\left(\frac{2z}{(z+k)^3}-\frac{3z^2}{(z+k)^4}\right)\partial_z n_F(z)+\frac{z^2}{(z+k)^3}\partial_z^2 n_F(z),\nonumber\\
&&\partial_z n_F(z)=-\frac{\beta e^{\beta z}}{(e^{\beta z}+1)^2}=-\frac{\beta}{4(\cosh(\beta z/2))^2},\nonumber\\
&&\partial_z^2 n_F(z)=-\frac{\beta^2 e^{\beta z}}{(e^{\beta z}+1)^2}+\frac{2\beta^2 e^{2\beta z}}{(e^{\beta z}+1)^3}\nonumber\\
&&=\beta^2 \frac{e^{2\beta z}-e^{\beta z}}{(e^{\beta z}+1)^3}=\beta^2 \frac{e^{\beta z}}{(e^{\beta z}+1)^2}\frac{e^{\beta z}-1}{e^{\beta z}+1}
=\frac{\beta^2 }{4(\cosh(\beta z/2))^2}\tanh(\beta z/2)
}
and the residual is
\eqarr{
&&Res[f(z) n_F(z),z=z_+]=\frac{1}{2}\left(\frac{\partial^2 g(z)}{\partial z^2}\right)_{z=k}\nonumber\\
&&=\left(\frac{1}{8k^3}-\frac{3}{8k^3}+\frac{3}{16k^3}\right)n_F(k)+\left(\frac{2}{8k^2}-\frac{3}{16k^2}\right)\partial_z n_F(k)+\frac{1}{16k}\partial_z^2 n_F(k),\nonumber\\
&&=-\frac{1}{16k^3}n_F(k)-\frac{1}{16k^2}\frac{\beta}{4(\cosh(\beta k/2))^2}+\frac{1}{16k}\frac{\beta^2 }{4(\cosh(\beta k/2))^2}\tanh(\beta k/2).
}

Similarly for For $z=z_-=-k$, we denote $f(z) n_F(z)=\frac{g(z)}{(z+k)^3}$ with $g(z)=\frac{z^2}{(z-k)^3}n_F(z)$. Direct calculations give
\eqarr{
&&\frac{\partial g(z)}{\partial z}=\left(\frac{2z}{(z-k)^3}-\frac{3z^2}{(z-k)^4}\right)n_F(z)+\frac{z^2}{(z-k)^3}\partial_z n_F(z),\nonumber\\
&&\frac{\partial^2 g(z)}{\partial z^2}=\left(\frac{2}{(z-k)^3}-\frac{12z}{(z-k)^4}+\frac{12z^2}{(z-k)^5}\right)n_F(z)+2\left(\frac{2z}{(z-k)^3}-\frac{3z^2}{(z-k)^4}\right)\partial_z n_F(z)+\frac{z^2}{(z-k)^3}\partial_z^2 n_F(z),\nonumber\\
}
and the residual is
\eqarr{
&&Res[f(z) n_F(z),z=z_-]=\frac{1}{2}\left(\frac{\partial^2 g(z)}{\partial z^2}\right)_{z=-k}\nonumber\\
&&=\left(-\frac{1}{8k^3}+\frac{3}{8k^3}-\frac{3}{16k^3}\right)n_F(-k)+\left(\frac{2}{8k^2}-\frac{3}{16k^2}\right)\partial_z n_F(-k)-\frac{1}{16k}\partial_z^2 n_F(-k),\nonumber\\
&&=\frac{1}{16k^3}n_F(-k)-\frac{1}{16k^2}\frac{\beta}{4(\cosh(\beta k/2))^2}+\frac{1}{16k}\frac{\beta^2 }{4(\cosh(\beta k/2))^2}\tanh(\beta k/2).
}
Putting all the results together, we have
\eqarr{
&&\frac{1}{\beta}\sum_{i\omega_n}f(i\omega_n)=Res[f(z) n_F(z),z=z_+]+Res[f(z) n_F(z),z=z_-]\nonumber\\
&&=\frac{1}{16k^3}(n_F(-k)-n_F(k))-\frac{1}{8k^2}\frac{\beta}{4(\cosh(\beta k/2))^2}+\frac{1}{8k}\frac{\beta^2 }{4(\cosh(\beta k/2))^2}\tanh(\beta k/2)\nonumber\\
&&=\frac{1}{16k^3}\tanh(\beta k/2)-\frac{1}{8k^2}\frac{\beta}{4(\cosh(\beta k/2))^2}+\frac{1}{8k}\frac{\beta^2 }{4(\cosh(\beta k/2))^2}\tanh(\beta k/2),
}
and
\eqarr{
&&\mathcal{F}_{02}=\frac{4}{3\pi^2}\int_0^{\Lambda} dk k^4 \frac{1}{\beta}\sum_{i\omega_n} \frac{(i\omega_n)^2}{((i\omega_n)^2-k^2)^3}\nonumber\\
&&=\frac{4}{3\pi^2}\int_0^{\Lambda} dk k^4 \left(\frac{1}{16k^3}\tanh(\beta k/2)-\frac{1}{8k^2}\frac{\beta}{4(\cosh(\beta k/2))^2}+\frac{1}{8k}\frac{\beta^2 }{4(\cosh(\beta k/2))^2}\tanh(\beta k/2)\right)\nonumber\\
&&=\frac{1}{24\pi^2}\int_0^{\Lambda} dk \left(2k\tanh(\beta k/2)-\frac{\beta k^2}{(\cosh(\beta k/2))^2}+\frac{\beta^2k^3}{(\cosh(\beta k/2))^2}\tanh(\beta k/2)\right).
}
We can decompose $\mathcal{F}_{02}$ into $\mathcal{F}_{02}=\mathcal{F}_{02}(T=0)+\delta\mathcal{F}_{02}(T)$ and find
\eqarr{
&&\mathcal{F}_{02}(T=0)=\frac{1}{12\pi^2}\int_0^{\Lambda} dk k=\frac{\Lambda^2}{24\pi^2}
}
and
\eqarr{
&&\delta\mathcal{F}_{02}(T)=\frac{1}{24\pi^2}\int_0^{\Lambda} dk \left(2k(\tanh(\beta k/2)-1)-\frac{\beta k^2}{(\cosh(\beta k/2))^2}+\frac{\beta^2k^3}{(\cosh(\beta k/2))^2}\tanh(\beta k/2)\right)\nonumber\\
&&=\frac{1}{24\pi^2\beta^2}\int_0^{\infty} d(2x) \left(4x(\tanh(x)-1)-\frac{(2x)^2}{(\cosh(x))^2}+\frac{(2x)^3}{(\cosh(x))^2}\tanh(x)\right)\nonumber\\
&&=\frac{1}{3\pi^2\beta^2}\int_0^{\infty} dx \left(x(\tanh(x)-1)-\frac{x^2}{(\cosh(x))^2}+\frac{2x^3}{(\cosh(x))^2}\tanh(x)\right)\nonumber\\
&&=\frac{1}{24\beta^2}=\frac{(k_B T)^2}{24}.
}
where we have used $x=\beta k/2$ and $\Lambda\rightarrow \infty$.

Now we can see that from $\mathcal{F}_0=\mathcal{F}_{01}-\mathcal{F}_{02}$, we have
\eq{\mathcal{F}_0(T=0)=\mathcal{F}_{01}(T=0)-\mathcal{F}_{02}(T=0)=0,
}
and
\eq{\label{App_eq:F0T}\delta\mathcal{F}_0(T)=\delta\mathcal{F}_{01}(T)-\delta\mathcal{F}_{02}(T)=-\frac{(k_B T)^2}{24}-\frac{(k_B T)^2}{24}=-\frac{(k_B T)^2}{12},
}
It is interesting to notice that the cut-off dependent term cancels between $\mathcal{F}_{01}$ and $\mathcal{F}_{02}$, while the temperature dependent term remains. Thus, $\mathcal{F}_0=F_0+F_1(k_B T)^2$ with $F_0=0$ and $F_1=-\frac{1}{12}$. It should be emphasized that the cut-off dependent term $F_0$ eventually depend on the re-normalization scheme and here we just take a simple hard cut-off and calculate the lowest order term without any re-normalization procedure. In any case, this term is not universal. In contrast, our result $F_1=-\frac{1}{12}$ recovers the value derived in literature\cite{huang2020nieh,nissinen2020thermal,nissinen2019thermal}, implying the universal property of the thermal Nieh-Yan term.

\section{Numerical Methods for Evaluating Stress-stress Correlation Function}
\label{App_Sec:Numerical}
In this section, we will describe our numerical methods to evaluate the stress-stress correlation functions. Here we only focus on the stress-stress correlation function $\Phi^{ab}_{ij}({\bf q},i\nu_m)$ in Eq. (\ref{App_eq:Phi0}). One can re-write the Green's function in Eq. (\ref{App_eq:G0_1}) as
\eq{\label{App_eq:G0_2}
\mathcal{G}_0=\sum_{s=\pm}\frac{P_s({\bf k})}{i\omega_n-\xi_{\bf k}}
}
where $\xi_{s,{\bf k}}=s\hbar v_f k-\mu$ and the projection operator $P_s({\bf k})=\frac{1}{2}\left(1+s\frac{\bf k\cdot \sigma}{k}\right)$. Then after the Matsubara frequency summation, Eq. (\ref{App_eq:Phi0}) can be simplified as
\eq{\label{App_eq:Phi0_1}
\Phi^{ab}_{ij}({\bf q},i\nu_m)=\frac{1}{V}\sum_{{\bf k},ss'}\frac{n_F(\xi_{s',{\bf k-q}})-n_F(\xi_{s,{\bf k}})}{i\nu_m+\xi_{s',{\bf k-q}}-\xi_{s,{\bf k}}}G^{ab}_{ij}(s {\bf k},s' {\bf k-q})
}
with
\eq{
G^{ab}_{ij}(s {\bf k},s' {\bf k-q})=Tr_{\sigma}\left(\mathcal{T}^{a}_i({\bf k},{\bf k-q})
P_{s'}({\bf k-q})\mathcal{T}^{b}_j({\bf k-q},{\bf k})P_s({\bf k})\right).
}
Direct evaluation of $G^{ab}_{ij}$ gives
\eqarr{
&&G^{ab}_{ij}(s {\bf k},s' {\bf k-q})=\frac{(\hbar v_f)^2}{8}(2k_i-q_i)(2k_j-q_j)\left(\delta^{ab}+i\epsilon^{abl}s\frac{k_l}{k}+i\epsilon^{alb}s'\frac{k_l-q_l}{|{\bf k-q}|}\right.\nonumber\\
&&\left.+\frac{ss'}{k|{\bf k-q}|}\left(k_b(k_a-q_a)+k_a(k_b-q_b)-\delta^{ab}{\bf k\cdot(k-q)}\right)\right).
}
Up to now, the whole formalism is general and can be calculated numerically in principles. As our purpose here is to extract the term that is related to the Nieh-Yan anomaly, we want to look for the terms $\Phi^{ab}_{ij}({\bf q},i\nu_m)\propto \delta_{ij}$. Furthermore, let's choose the momentum along the z-direction, ${\bf q}=(0,0,q_z)$, and then we should expect that the relevant component should be $\Phi^{xy}_{ii}(q_z,i\nu_m)$. Thus, our numerical simulations focus on the component $\Phi^{xy}_{xx}(q_z,i\nu_m)$ below ($\Phi^{xy}_{xx}(q_z,i\nu_m)=\Phi^{xy}_{yy}(q_z,i\nu_m)$). For this component, we have
\eqarr{
G^{xy}_{xx}(s {\bf k},s' {\bf k-q})=\frac{(\hbar v_f)^2}{2}k_x^2\left(is\frac{k_z}{k}-is'\frac{k_z-q_z}{|{\bf k-q}|}+\frac{ss'}{k|{\bf k-q}|}2k_xk_y\right).
}
Now let's consider the momentum integral in Eq. (\ref{App_eq:Phi0_1}), and choose the spherical coordinate for the momentum, ${\bf k}=(k\sin\theta_k \cos\varphi_k,k\sin\theta_k \sin\varphi_k, k\cos\theta_k)$. Furthermore, we denote ${\bf k}'={\bf k-q}=(k'\sin\theta'_k \cos\varphi'_k,k'\sin\theta'_k \sin\varphi'_k, k'\cos\theta'_k)=(k\sin\theta_k \cos\varphi_k,k\sin\theta_k \sin\varphi_k, k\cos\theta_k-q_z)$. Thus, we have $\varphi'_k=\varphi_k$, $k'\sin\theta'_k=k\sin\theta_k$ and $k'\cos\theta'_k=k\cos\theta_k-q_z$. The latter two equalities give rise to $k'=\sqrt{k^2+q_z^2-2 kq_z\cos\theta_k}$ and $\cos\theta'_k=\frac{k\cos\theta_k-q_z}{k'}$.

By choosing ${\bf q}$ along the z direction, $\xi_{s,{\bf k-q}}$ only involves $k$ and $\theta_k$ and does not depend on $\varphi_k$. Thus, we can define
\eqarr{
&&\bar{G}^{xy}_{xx}(ss';k,\cos\theta_k,q_z)=\int \frac{d\varphi_k}{2\pi} G^{xy}_{xx}(s{\bf k},s'{\bf k-q})\nonumber\\
&&= \frac{(\hbar v_f)^2}{4\pi}\int_0^{2\pi}d\varphi_k k^2 \sin^2\theta_k \cos^2\varphi_k \left(is \cos\theta_k-is'\cos\theta'_k+\frac{ss'}{k k'}2k^2\sin^2\theta_k \cos\varphi_k\sin\varphi_k \right)\nonumber\\
&&=\frac{(\hbar v_f)^2}{4} k^2 \sin^2\theta_k \left(is \cos\theta_k-is'\cos\theta'_k \right),
}
and then
\eq{\label{App_eq:Phi0_2}
\Phi^{xy}_{xx}(q_z,i\nu_m)=\sum_{ss'}\int\frac{k^2dk d\cos\theta_k}{(2\pi)^2}\frac{n_F(\xi_{s',{\bf k-q}})-n_F(\xi_{s,{\bf k}})}{i\nu_m+\xi_{s',{\bf k-q}}-\xi_{s,{\bf k}}}\bar{G}^{xy}_{xx}(ss';k,\cos\theta_k,q_z).
}
We may decompose $\Phi^{xy}_{xx}(q_z,i\nu_m)$ into several different components $\Phi^{xy}_{xx}(q_z,i\nu_m)=\Phi^{xy,-}_{xx}(q_z,i\nu_m)+\Phi^{xy,+}_{xx}(q_z,i\nu_m)$, where $\pm$ labels the contribution from different $s$ bands. We further decompose $\Phi^{xy,\pm}_{xx}(q_z,i\nu_m)=\Phi^{xy,\pm}_{xx,1}(q_z,i\nu_m)+\Phi^{xy,\pm}_{xx,2}(q_z,i\nu_m)$, where $1$ and $2$ here are for the intra-band and inter-band contribution. Explicitly,
\eqarr{
&&\Phi^{xy,+}_{xx,1}(q_z,i\nu_m)=\int\frac{k^2dk d\cos\theta_k}{(2\pi)^2}\frac{n_F(\xi_{+,{\bf k-q}})-n_F(\xi_{+,{\bf k}})}{i\nu_m+\xi_{+,{\bf k-q}}-\xi_{+,{\bf k}}}\bar{G}^{xy}_{xx}(++;k,\cos\theta_k,q_z)\nonumber\\
&&\Phi^{xy,+}_{xx,2}(q_z,i\nu_m)=\int\frac{k^2dk d\cos\theta_k}{(2\pi)^2}\left(\frac{n_F(\xi_{+,{\bf k-q}})}{i\nu_m+\xi_{+,{\bf k-q}}-\xi_{-,{\bf k}}}\bar{G}^{xy}_{xx}(-+;k,\cos\theta_k,q_z)-\frac{n_F(\xi_{+,{\bf k}})}{i\nu_m+\xi_{-,{\bf k-q}}-\xi_{+,{\bf k}}}\bar{G}^{xy}_{xx}(+-;k,\cos\theta_k,q_z)\right)\nonumber\\
&&\Phi^{xy,-}_{xx,1}(q_z,i\nu_m)=\int\frac{k^2dk d\cos\theta_k}{(2\pi)^2}\frac{n_F(\xi_{-,{\bf k-q}})-n_F(\xi_{-,{\bf k}})}{i\nu_m+\xi_{-,{\bf k-q}}-\xi_{-,{\bf k}}}\bar{G}^{xy}_{xx}(--;k,\cos\theta_k,q_z)\nonumber\\
&&\Phi^{xy,-}_{xx,2}(q_z,i\nu_m)=\int\frac{k^2dk d\cos\theta_k}{(2\pi)^2}\left(\frac{n_F(\xi_{-,{\bf k-q}})}{i\nu_m+\xi_{-,{\bf k-q}}-\xi_{+,{\bf k}}}\bar{G}^{xy}_{xx}(+-;k,\cos\theta_k,q_z)-\frac{n_F(\xi_{-,{\bf k}})}{i\nu_m+\xi_{+,{\bf k-q}}-\xi_{-,{\bf k}}}\bar{G}^{xy}_{xx}(-+;k,\cos\theta_k,q_z)\right)\nonumber\\.
}
Since the above integral involves the singular points, we need to treat it as the Cauchy principal value integral. To numerically evaluate the integral, we need some more simplification. First, the integral over $\theta_k$ can be replaced by the integral over $k'$. All the above integrals takes the form of
\eqarr{
J_1=\int\frac{k^2dk d\cos\theta_k}{(2\pi)^2}\mathcal{G}(k,\cos\theta_k,q_z).
}
With $\int d^3k' \delta({\bf k'-k+q})=\int d^3k'\delta(k'-\sqrt{k^2+q_z^2-2kq_z\cos\theta_k})=1$, we have\cite{zhou2015plasmon}
\eqarr{
&&J_1=\int\frac{k^2dk d\cos\theta_k}{(2\pi)^2}\int d^3k'\delta(k'-\sqrt{k^2+q_z^2-2kq_z\cos\theta_k})\mathcal{G}(k,\cos\theta_k,q_z)\nonumber\\
&&=\frac{1}{(2\pi)^2}\int_0^{\Lambda} dk \int_{|k-q|}^{k+q} dk' \frac{kk'}{q_z}\mathcal{G}(k,\cos\theta_k=\frac{k^2+q_z^2-k'^2}{2kq_z},q_z).
}
Here the integral range of $k'$ is determined by requiring $|\cos\theta_k|\le 1$.
Next one may absorb the velocity $\hbar v_f$ into the definition of the momentum, $\tilde{k}=\hbar v_f k,  \tilde{k}'=\hbar v_f k', q=\hbar v_f q_z$. Thus, one can see that $\mathcal{G}$ does not give any additional factors while the integral $J_1$ should contain an additional pre-factor $\frac{1}{(\hbar v_f)^3}$. Now $\mathcal{G}$ should be a function of $\tilde{k}, \tilde{k}', q$, denoted as $\mathcal{G}(\tilde{k},\tilde{k}',q)$. Finally, one can take the transformation $\tilde{k}=\frac{1}{2}(x+y)$ and $\tilde{k}'=\frac{1}{2}(x-y)$ and the integral can be transformed into
\eqarr{
&&J_1=\frac{1}{(2\pi)^2(\hbar v_f)^3}\int_q^{\Lambda} dx \int_{-q}^{q} dy \frac{x^2-y^2}{8q}\mathcal{G}\left(\tilde{k}=\frac{1}{2}(x+y),\tilde{k}'=\frac{1}{2}(x-y),q\right).
}
Using the equation, we can rewrite the integrals as
\eqarr{
&&\Phi^{xy,+}_{xx,1}(q,\omega+i\eta)=\frac{1}{(2\pi)^2(\hbar v_f)^3}\int_q^{\Lambda} dx \int_{-q}^{q} dy \frac{x^2-y^2}{8q}\frac{n_F(\frac{x-y}{2}-\mu)-n_F(\frac{x+y}{2}-\mu)}{\omega+i\eta-y}\bar{G}^{xy}_{xx}(++;x,y,q)\nonumber\\
&&\Phi^{xy,+}_{xx,2}(q,\omega+i\eta)=\frac{1}{(2\pi)^2(\hbar v_f)^3}\int_q^{\Lambda} dx \int_{-q}^{q} dy \frac{x^2-y^2}{8q}\left(\frac{n_F(\frac{x-y}{2}-\mu)}{\omega+i\eta_m+x}\bar{G}^{xy}_{xx}(-+;x,y,q)\right.\nonumber\\
&&\left.-\frac{n_F(\frac{x+y}{2}-\mu)}{\omega+i\eta-x}\bar{G}^{xy}_{xx}(+-;x,y,q)\right)\nonumber\\
&&\Phi^{xy,-}_{xx,1}(q,\omega+i\eta)=\frac{1}{(2\pi)^2(\hbar v_f)^3}\int_q^{\Lambda} dx \int_{-q}^{q} dy \frac{x^2-y^2}{8q}\frac{n_F(-\frac{x-y}{2}-\mu)-n_F(-\frac{x+y}{2}-\mu)}{\omega+i\eta+y}\bar{G}^{xy}_{xx}(--;x,y,q)\nonumber\\
&&\Phi^{xy,-}_{xx,2}(q,\omega+i\eta)=\frac{1}{(2\pi)^2(\hbar v_f)^3}\int_q^{\Lambda} dx \int_{-q}^{q} dy \frac{x^2-y^2}{8q}\left(\frac{n_F(-\frac{x-y}{2}-\mu)}{\omega+i\eta-x}\bar{G}^{xy}_{xx}(+-;x,y,q)\right.\nonumber\\
&&\left.-\frac{n_F(-\frac{x+y}{2}-\mu)}{\omega+i\eta+x}\bar{G}^{xy}_{xx}(-+;x,y,q)\right)\nonumber\\,
}
where
\eqarr{
&&\bar{G}^{xy}_{xx}(ss';x,y,q)=\frac{(x+y)^2}{16}(1-\cos^2\theta_k)\left(is \cos\theta_k-is'\cos\theta'_k \right),\\
&&\cos\theta_k=\frac{k^2+q_z^2-k'^2}{2kq_z}=\frac{xy+q^2}{(x+y)q},\\
&&\cos\theta'_k=\frac{k\cos\theta_k-q_z}{k'}=\frac{xy-q^2}{(x-y)q},
}
and we have replaced $i\nu_m$ by $\omega+i\eta$. Now one can see that the poles of the integral are determined by $x=\pm \omega$ or $y=\pm \omega$, and the corresponding Cauchy integral is easy to deal with numerically. These expressions form the basis for our numerical calculations.

To recover the unit, one notices that $q, x, y$ in the above expression are in the unit of energy. $\hbar v_f$ is in the unit of $[E\cdot L]$, and thus we need to further choose certain length unit of our system, which is denoted as $a_0$. Then, the unit of energy can be chosen as $\hbar v_f/a_0$. From the expression of $\Phi^{xy}_{xx}$, one can easily find that $\Phi^{xy}_{xx}$ is in the unit of energy density or equivalently $\hbar v_f/a_0^4$. For the numerical calculations, we simply choose $\hbar v_f=1$ and $a_0=1$, and this is also convenient to compare with analytical results in Sec. \ref{App_Sec:Analytical}. The main results of our numerical calculations are summarized in the Fig. 2 in the main text.

Below we provide some more understanding on the analytical aspects of the formalism. We consider the small ${\bf q}$ expansion of Eq. (\ref{App_eq:Phi0_2}) at $i\nu_m=0$, and perform the perturbation expansion. The calculation here is equivalent to the calculations in Sec. \ref{App_Sec:Analytical}, except that we below keep the chemical potential up to $\mu^3$. For a small ${\bf q}=q_z\hat{e}_z$, we have the perturbation expansion
\eqarr{
&&\xi_{s',{\bf k-q}}\approx s'\hbar v_f k-\mu-s'\hbar v_f q\cos\theta_k=\xi_{s',{\bf k}}-s'\hbar v_f q_z\cos\theta_k, \\
&&n_F(\xi_{s',{\bf k-q}})\approx n_F(\xi_{s',{\bf k}})-\frac{\partial n_F}{\partial \xi_{s',{\bf k}}}s'\hbar v_f q_z\cos\theta_k, \\
&&\cos\theta_{k'}\approx \cos\theta_k-\frac{q_z}{k}\sin^2\theta_k,
}
up to the first order in $q_z$. Now let's denote $\Phi^{xy}_{xx}=\sum_{s,s'}\Phi^{xy}_{xx,ss'}$ and evaluate $\Phi^{xy}_{xx,ss'}$ separately. First, let's consider $s=s'$ and we have
\eq{
\bar{G}^{xy}_{xx}(ss;k,\cos\theta_k,q_z)\approx \frac{is}{4}(\hbar v_f)^2\sin^4\theta_k kq_z,
}
which only possesses the linear-$q_z$ term. Thus,
\eqarr{
&&\Phi^{xy}_{xx,ss}=\int\frac{k^2dk d\cos\theta_k}{(2\pi)^2}\frac{\partial n_F}{\partial \xi_{s,{\bf k}}}\bar{G}^{xy}_{xx}(ss;k,\cos\theta_k,q_z)\nonumber\\
&&=\frac{is}{4}(\hbar v_f)^2 q_z \int_{-1}^1d\cos\theta_k \sin^4\theta_k \int_0^{\Lambda} k^3dk \frac{\partial n_F}{\partial \xi_{s,{\bf k}}}.
}
With $\frac{\partial n_F}{\partial \xi_{s,{\bf k}}}=-\frac{\beta e^{\beta \xi_{s,{\bf k}}}}{(1+e^{\beta \xi_{s,{\bf k}}})^2}$ and $x=\beta \xi_{s,{\bf k}}$, we find
\eq{
\Phi^{xy}_{xx,ss}=-i \frac{4 q_z}{15(2\pi)^2} \frac{(k_B T)^3}{(\hbar v_f)^2}\left(\int_{-\beta \mu}^{\infty} dx(x+\beta \mu)^3 \frac{e^x}{(1+e^x)^2} + \int_{\beta\mu}^{\infty}dx (-x+\beta\mu)^3 \frac{e^x}{(1+e^x)^2} \right).
}
With
\eqarr{
&&\int_{\beta \mu}^{\infty} dx x^2\frac{e^x}{(1+e^x)^2}\approx \frac{\pi^2}{6}-\frac{(\beta \mu)^3}{12}\\
&&\int_{\beta \mu}^{\infty} dx \frac{e^x}{(1+e^x)^2}\approx \frac{1}{2}-\frac{\beta\mu}{4}+\frac{(\beta \mu)^3}{48}\\
&&\int_{-\beta \mu}^{\beta\mu} dx (x+\beta\mu)^3 \frac{e^x}{(1+e^x)^2}\approx (\beta \mu)^4,
}
we have
\eq{
\sum_s \Phi^{xy}_{xx,ss}=-i\frac{q_z}{15\pi^2(\hbar v_f)^2}(\mu^3+\mu\pi^2(k_B T)^2).
}

For $s'=-s$,
\eq{
\bar{G}^{xy}_{xx}(s,-s;k,\cos\theta_k,q_z)\approx \frac{is}{4}(\hbar v_f)^2\sin^2\theta_k (2 k^2\cos\theta_k-kq_z\sin^2\theta_k),
}
which contains both $q_z$-independent and linear-$q_z$ terms. In addition,
\eqarr{
&&\frac{n_F(\xi_{-s,{\bf k-q}})-n_F(\xi_{s,{\bf k}})}{\xi_{-s,{\bf k-q}}-\xi_{s,{\bf k}}}\approx \frac{n_F(\xi_{-s,{\bf k}})-n_F(\xi_{s,{\bf k}})+\frac{\partial n_F}{\partial\xi_{-s,{\bf k}}}s\hbar v_f q\cos\theta_k}{-2s\hbar v_f k+s\hbar v_f q\cos\theta_k}\nonumber\\
&&=-\frac{1}{2s\hbar v_f k}\left(n_F(\xi_{-s,{\bf k}})-n_F(\xi_{s,{\bf k}})+q_z\cos\theta_k\left(\frac{\partial n_F}{\partial \xi_{-s,{\bf k}}}s\hbar v_f +\frac{1}{2k}(n_F(\xi_{-s,{\bf k}})-n_F(\xi_{s,{\bf k}}))\right)\right)
}
up to the linear order in $q_z$. Putting all these results together, we have
\eqarr{
&&\Phi^{xy}_{xx,s,-s}=\int\frac{k^2dk d\cos\theta_k}{(2\pi)^2}\frac{(-i\hbar v_f q_z)}{8}\left(-(n_F(\xi_{-s,{\bf k}})-n_F(\xi_{s,{\bf k}}))\sin^4\theta_k\right.\nonumber\\
&&\left.+2k\cos^2\theta_k\sin^2\theta_k\left(\frac{\partial n_F}{\partial \xi_{-s,{\bf k}}}s\hbar v_f +\frac{1}{2k}(n_F(\xi_{-s,{\bf k}})-n_F(\xi_{s,{\bf k}}))\right)\right)\nonumber\\
%&&=\int\frac{k^2dk}{(2\pi)^2}\frac{(-i\hbar v_f q_z)}{15}\left(-2(n_F(\xi_{-s,{\bf k}})-n_F(\xi_{s,{\bf k}}))+k\left(\frac{\partial n_F}{\partial \xi_{-s,{\bf k}}}s\hbar v_f +\frac{1}{2k}(n_F(\xi_{-s,{\bf k}})-n_F(\xi_{s,{\bf k}}))\right)\right)\nonumber\\
&&=\int\frac{k^2dk}{(2\pi)^2}\frac{(-i\hbar v_f q_z)}{15}\left(-\frac{3}{2}(n_F(\xi_{-s,{\bf k}})-n_F(\xi_{s,{\bf k}}))+\frac{\partial n_F}{\partial \xi_{-s,{\bf k}}}s\hbar v_fk\right),
}
where we have used
\eqarr{
&&\int_{-1}^1 d\cos\theta_k \sin^4\theta_k=\frac{16}{15}\\
&&\int_{-1}^1 d\cos\theta_k \sin^2\theta_k\cos^2\theta_k=\frac{4}{15}.
}
In the above expression, since $\sum_s (n_F(\xi_{-s,{\bf k}})-n_F(\xi_{s,{\bf k}}))=0$, the first term will vanish after the summation over $s$. Thus,
\eqarr{
&&\sum_s \Phi^{xy}_{xx,s,-s}=\sum_s\frac{(-is(\hbar v_f)^2 q_z)}{15}\int\frac{dk}{(2\pi)^2}k^3\frac{\partial n_F}{\partial \xi_{-s,{\bf k}}}\nonumber\\
&&=\sum_s\frac{is(\hbar v_f)^2 q_z}{15}\int\frac{dk}{(2\pi)^2}k^3\frac{\beta e^{\beta\xi_{-s,{\bf k}}}}{(e^{\beta\xi_{-s,{\bf k}}}+1)^2}.
}
Now let's define $x=\beta\xi_{-s,{\bf k}}$ and find
\eqarr{
&&\sum_s \Phi^{xy}_{xx,s,-s}=\frac{-iq_z}{15\beta^3 (2\pi \hbar v_f)^2}\left(\int_{\beta\mu}^{\infty} dx (-x+\beta\mu)^3 \frac{e^x}{(1+e^x)^2}+\int_{-\beta\mu}^{\infty} dx (x+\beta\mu)^3 \frac{e^x}{(1+e^x)^2}\right)\nonumber\\
&&\approx \frac{(-i)q_z}{15(2\pi\hbar v_f)^2}(\mu^3+\mu\pi^2(k_B T)^2).
}
Putting all the results together, we have
\eq{\label{App_eq:PhixyxxAna}
\Phi^{xy}_{xx}=\sum_{ss'} \Phi^{xy}_{xx,ss'}=-i\frac{q_z}{12\pi^2(\hbar v_f)^2}(\mu^3+\mu\pi^2(k_B T)^2)=-i\frac{q_z\mu^3}{12\pi^2(\hbar v_f)^2}-i\frac{q_z\mu}{12(\hbar v_f)^2}(k_B T)^2.
}
The second term reproduces the result of Eqs. (\ref{App_eq:parPhql1}) and (\ref{App_eq:F0T}) for the thermal Nieh-Yan anomaly (choosing $\hbar v_f=1$), while the first term is of order $\mu^3$ and thus not included in the calculations of Sec. \ref{App_Sec:Analytical}, which only keep the terms up to linear $\mu$. Therefore, this analytical result is consistent with the results given in Sec. \ref{App_Sec:Analytical}. The expression (\ref{App_eq:PhixyxxAna}) can fit well with our numerical results at small $\mu$, as shown in Fig. 2d in the main text.

\section{Acoustic phonon dynamics in the Kramers-Weyl semimetals}
\label{App_Sec:phonon}
In this section, we will analyze the influence on the effective action (\ref{App_eq:WeffNY2}) on the phonon dynamics. We focus on the acoustic phonons here, and as discussed in Sec. \ref{App_Sec:SymmetryHam}, the $A_0$ and $\Delta$ fields are related to the strain tensor as $A_0(\tilde{q})=C_1\bar{u}(\tilde{q})$ and $\Delta^j_a(\tilde{q})=\frac{1}{\hbar v_f}(g_0\bar{u}(\tilde{q})\delta_{ja}+g_1u_{ja}(\tilde{q}))$, where the strain tensor $u_{ij}$ is treated as a fluctuating field for acoustic phonons and $\bar{u}=\sum_i u_{ii}$. By substituing the forms of $A_0$ and $\Delta$ fields, we find
\eq{
W=\frac{1}{2}\sum_{\tilde{q}}\left(\bar{u}(\tilde{q})\Lambda_0(\tilde{q})\bar{u}(-\tilde{q})+\sum_{i,a}\bar{u}(\tilde{q})
\Lambda^a_{2,i}(\tilde{q})u_{ia}(-\tilde{q})+\sum_{ij,ab}u_{ia}(\tilde{q})\Lambda^{ab}_{3,ij}(\tilde{q})u_{jb}(-\tilde{q})\right),
}
where
\eqarr{
&&\Lambda_0(\tilde{q})=C_1^2\Pi_0(\tilde{q})+\sum_{i,j}\frac{g_0^2}{(\hbar v_f)^2}\Phi^{ij}_{ij}(\tilde{q})-\sum_{i}\frac{C_1g_0}{\hbar v_f}(\Theta^i_i(\tilde{q})+\Theta^i_i(-\tilde{q})),\\
&&\Lambda^a_{2,i}(\tilde{q})=\sum_{j}\frac{g_0g_1}{(\hbar v_f)^2}(\Phi^{ja}_{ji}(\tilde{q})+\Phi^{aj}_{ij}(-\tilde{q}))
-\frac{2C_1g_1}{\hbar v_f}\Theta^a_i(\tilde{q}),\\
&&\Lambda^{ab}_{3,ij}(\tilde{q})=\frac{g_1^2}{(\hbar v_f)^2}\Phi^{ab}_{ij}(\tilde{q}).\label{App_Eq:Lambdaab3ij}
}
We can expand $\Lambda_0(\tilde{q}), \Lambda^a_{2,i}(\tilde{q})$ and $\Lambda^{ab}_{3,ij}(\tilde{q})$ as a function of $\tilde{q}$, $\Lambda(\tilde{q})=\Lambda(0)+(\partial_{\bf q}\Lambda)_{\tilde{q}=0}\cdot{\bf q}+(\partial_{\omega_n}\Lambda)_{\tilde{q}=0}\omega_n+...$. At the zero order term, we have
\eqarr{
&&W^{(0)}=\frac{1}{2}\sum_{\tilde{q}}\left(\bar{u}(\tilde{q})\Lambda_0(0)\bar{u}(-\tilde{q})+\sum_{i,a}\bar{u}(\tilde{q})
\Lambda^a_{2,i}(0)u_{ia}(-\tilde{q})+\sum_{ij,ab}u_{ia}(\tilde{q})\Lambda^{ab}_{3,ij}(0)u_{jb}(-\tilde{q})\right)\nonumber\\
&&=\frac{1}{2}\int_0^{\beta}d\tau\int d^3r \left(\bar{u}({\bf r},\tau)\Lambda_0(0)\bar{u}({\bf r},\tau)+\sum_{i,a}\bar{u}({\bf r},\tau)\Lambda^a_{2,i}(0)u_{ia}({\bf r},\tau)+\sum_{ij,ab}u_{ia}({\bf r},\tau)\Lambda^{ab}_{3,ij}(0)u_{jb}({\bf r},\tau)\right).
}
Here all the terms are quadratic in the strain tensor $u_{ij}$ and let's consider the general form of the effective action for acoustic phonons (elastic wave), which is defined as\cite{barkeshli2012dissipationless,landau2013course}
\eq{\label{App_eq:Sph0}
S_{ph,0}=\frac{1}{2}\int dt d^3r \left(\rho \sum_j \partial_t u_j\partial_t u_j-\sum_{ijkl}\lambda_{ijkl}\partial_i u_j\partial_k u_l\right),
}
where the rank-4 tensor $\lambda_{ijkl}$ is the elastic modulus. It is clear that $W^{(0)}$ just provides the correction of the elastic modulus $\lambda_{ijkl}$.

Next we focus on the first order term with the linear ${\bf q}$ dependence, which takes the form
\eqarr{\label{App_eq:SeffW1}
&&W^{(1)}=\frac{1}{2}\sum_{\tilde{q}}\left(\sum_i(\partial_i\Lambda_0) \bar{u}(\tilde{q})q_i\bar{u}(-\tilde{q})+\sum_{ij,a}(\partial_j\Lambda^a_{2,i})\bar{u}(\tilde{q})
q_j u_{ia}(-\tilde{q})+\sum_{ijl,ab}(\partial_l \Lambda^{ab}_{3,ij})u_{ia}(\tilde{q})q_l u_{jb}(-\tilde{q})\right)\nonumber\\
&&=\frac{1}{2}\int_0^{\beta}d\tau\int d^3r \left(\sum_j (\partial_j\Lambda_0)\bar{u}({\bf r},\tau)\left(i\frac{\partial}{\partial r_j}\bar{u}({\bf r},\tau)\right)+\sum_{ij,a}(\partial_j\Lambda^a_{2,i})\bar{u}({\bf r},\tau)\left(i\frac{\partial}{\partial r_j}u_{ia}({\bf r},\tau)\right)\right.\nonumber\\
&&\left.+\sum_{ijk,ab}(\partial_k\Lambda^{ab}_{3,ij})u_{ia}({\bf r},\tau)\left(i\frac{\partial}{\partial r_k}u_{jb}({\bf r},\tau)\right)\right),
}
where $(\partial_j\Lambda)=\left(\frac{\partial \Lambda}{\partial q_j}\right)_{\tilde{q}=0}$ is just a number. From Eq. (\ref{App_Eq:Lambdaab3ij}), one can see that the last term is just the Nieh-Yan term. Although the above expression seems quite complex, the symmetry gives a strong constraint on the form of the allowed effective action. It is clear that the above action contributes to the following general form of the effective action of acoustic phonons
\eq{
S_{ph,1}=\int dt d^3r \sum_{ijklm}\xi_{ijklm}(\partial_i u_{jk})u_{lm} ,
}
where $\partial_i=\frac{\partial}{\partial r_i}$. Now the question is what is the most general form of the coefficient $\xi_{ijklm}$, a rank-5 tensor, for a uniform isotropic system. Since the strain tensor $u_{ij}$ is symmetric with respect to the indices $i$ and $j$, the coefficient $\xi_{ijklm}$ should satisfy the relations $\xi_{ijklm}=\xi_{ikjlm}=\xi_{ijkml}$. Furthermore, it also satisfies the anti-symmetric relation $\xi_{ijklm}=-\xi_{ilmjk}$. These relations together with the full rotation symmetry will fix the form of $\xi_{ijklm}$.

It is also equivalent to ask how to construct an invariant term with the form of $(\partial_i u_{jk})u_{lm}$. We analyze this problem based on the irreducible representation of the angular momentum. As discussed in Sec. \ref{App_Sec:SymmetryHam}, the symmetric rank-2 tensor $u_{ij}$ can be decomposed to two parts, $\bar{u}$ with the angular momentum {\bf 0} and $u_{J=2,M}$ with the angular momentum {\bf 2}. Since the expression involves two strain tensors, the corresponding angular momentum decomposition can be given by
\eqarr{
&&{\bf 2\otimes 2=4\oplus 3\oplus 2 \oplus 1 \oplus 0},\label{App_eq:decomposition1}\\
&&{\bf 2\otimes 0=2}, \label{App_eq:decomposition2}\\
&&{\bf 0\otimes 0=0}, \label{App_eq:decomposition3}
}
where the bold number represents the total angular momentum. Furthermore, the expression $(\partial_i u_{jk})u_{lm}$ involves one $\partial_i$ in addition to two strain tensors, and $\partial_i$ possesses the angular momentum ${\bf 1}$. To make the term $(\partial_i u_{jk})u_{lm}$ invariant (with the angular momentum ${\bf 0}$), only the irreducible representation with the angular momentum ${\bf 1}$ in ${\bf 2\otimes 2}$ can lead to ${\bf 1\otimes 1= 2\oplus 1 \oplus 0}$. From such angular momentum combination, only a single term is allowed for $\xi_{ijklm}$ in a uniform isotropic system. This term can be explicitly constructed as
\eq{ \label{App_eq:Sph1}
S_{ph,1}=\xi_0\int dt d^3r \sum_{ijklm} \delta_{jl}\epsilon_{ikm}(\partial_i u_{jk})u_{lm} ,
}
so that the parameter $\xi_{ijklm}$ given by
\eq{
\xi_{ijklm}=\frac{\xi_0}{4}(\delta_{jl}\epsilon_{ikm}+\delta_{kl}\epsilon_{ijm}+\delta_{jm}\epsilon_{ikl}
+\delta_{km}\epsilon_{ijl}).
}

We can apply the above analysis to the action $W^{(1)}$, and it is clear that it is impossible to construct an invariant term from the first two terms in Eq. (\ref{App_eq:SeffW1}), and thus the coefficient $\partial_j\Lambda_0$ and $\partial_j\Lambda^a_{2,i}$ must vanish. The effective action then takes the form
\eq{\label{App_eq:SeffW1_1}
W^{(1)}=\frac{1}{2}\int_0^{\beta}d\tau\int d^3r \left(\sum_{ijk,ab}(i\partial_k\Lambda^{ab}_{3,ij})u_{ia}({\bf r},\tau)\left(\frac{\partial}{\partial r_k}u_{jb}({\bf r},\tau)\right)\right),
}
with the coefficient
\eq{
(i\partial_k\Lambda^{ab}_{3,ij})=i\frac{g_1^2}{(\hbar v_f)^2}\left(\partial_k\Phi^{ab}_{ij}(\tilde{q})\right)_{\tilde{q}=0}=i\frac{g_1^2}{(\hbar v_f)^2}\Phi^{ab}_{ij,k}=\frac{g_1^2}{(\hbar v_f)^2}\epsilon^{abk}\delta_{ij}\chi\mu \mathcal{F}_0.
}
Here we have used Eq. \ref{App_eq:parPhql1}. Substituting the coefficient and also changing from imaginary time to real time, we have
\eqarr{\label{App_eq:SeffW1_2}
&&W^{(1)}=-\frac{g_1^2\chi\mu \mathcal{F}_0}{2(\hbar v_f)^2}\sum_{ijk,ab}\int dt d^3r \epsilon^{abk}\delta_{ij} u_{ia}({\bf r},\tau)\left(\frac{\partial}{\partial r_k}u_{jb}({\bf r},\tau)\right)\nonumber\\
&&=\frac{g_1^2\chi\mu \mathcal{F}_0}{2(\hbar v_f)^2}\sum_{ijk,ab}\int dt d^3r \epsilon^{kab}\delta_{ij} \left(\frac{\partial}{\partial r_k}u_{ia}({\bf r},\tau)\right) u_{jb}({\bf r},\tau).
}
Comparing with $S_{ph,1}$ in Eq. (\ref{App_eq:Sph1}), one can see that the Nieh-Yan term can contribute to a new term in the phonon effective action and the corresponding coefficient is given by
\eq{
\xi_0=\frac{g_1^2\chi\mu \mathcal{F}_0}{2(\hbar v_f)^2}=\frac{g_1^2\mu_5 \mathcal{F}_0}{(\hbar v_f)^2}.
}

In the above discussion, we have dropped the term with the form of $u_{ij}\omega_n u_{kl}$ and in the time-domain, this term should have the form of $u_{ij}(\partial_{\tau}u_{kl})$. In the elastic theory, this term is nothing but the viscosity, generally defined as \cite{barkeshli2012dissipationless,avron1995viscosity,landau2013course}
\eq{
S_{ph,2}=\int dt d^3r \sum_{ijkl}\eta_{ijkl}(\partial_t u_{jk})u_{lm}.
}
However, the existence of the viscosity requires the breaking of TR symmetry, while TR symmetry exists in our system. Thus, we will expect the viscosity coefficient $\eta_{ijkl}$ extracted from our model should vanish.

Now we hope to solve the acoustic phonon dispersion (or elastic wave) in our system based on the actions (\ref{App_eq:Sph0}) and (\ref{App_eq:Sph1}). For the uniform isotropic system, there are two independent parameters for the elastic moduli $\lambda_{ijkl}$ and the corresponding equation of motion for the displacement field ${\bf u}$ is given by
\eq{
\frac{d^2}{dt^2}{\bf u}=c_t^2\nabla^2{\bf u}+(c_l^2-c_t^2)\nabla(\nabla\cdot{\bf u}),
}
where $c_l$ and $c_t$ are the velocities of longitudinal and transverse waves. The Nieh-Yan term $S_{ph,1}$ can add an additional term into the above elastic wave equation, as given by
\eqarr{
&&\frac{\delta S_{ph,1}}{\delta u_n}=\frac{\xi_0}{2}\epsilon_{ikm}\left(\delta_{kn}\partial_i\partial_j u_{jm}+\delta_{jn}\partial_i\partial_k u_{jm}-\delta_{mn}\partial_i \partial_j u_{jk}-\delta_{jn}\partial_i\partial_m u_{jk} \right)\nonumber\\
&&=\frac{\xi_0}{2}\left(\epsilon_{inm}\partial_i\partial_j u_{jm}+\epsilon_{ikm}\partial_i\partial_k u_{nm}-\epsilon_{ikn}\partial_i \partial_j u_{jk}-\epsilon_{ikm}\partial_i\partial_m u_{nk} \right)\nonumber\\
&&=\frac{\xi_0}{2}\left(\epsilon_{inm}\partial_i\partial_j u_{jm}-\epsilon_{ikn}\partial_i \partial_j u_{jk} \right)=\frac{\xi_0}{2} \epsilon_{inm}\partial_i\partial_j (\partial_j u_m+\partial_m u_j)\nonumber\\
&&=\frac{\xi_0}{2} \epsilon_{inm}\partial_i (\nabla^2u_m)=-\frac{\xi_0}{2}(\nabla\times(\nabla^2{\bf u}))_n
}
By including the Nieh-Yan term, the equation of motion for the displacement field is now written as
\eq{
\frac{d^2}{dt^2}{\bf u}=c_t^2\nabla^2{\bf u}+(c_l^2-c_t^2)\nabla(\nabla\cdot{\bf u})+\frac{\xi_0}{2}\nabla\times(\nabla^2{\bf u}).
}
We can decompose the displacement field into the longitudinal and transverse components, ${\bf u=u_l+u_t}$. One can show that the longitudinal and transverse components are still decoupled from each other. The longitudinal part does not have any correction from the Nieh-Yan term, while the equation of motion for the transverse part is revised as
\eq{
\frac{d^2}{dt^2}{\bf u}_t=c_t^2\nabla^2{\bf u}_t+\frac{\xi_0}{2}\nabla\times(\nabla^2{\bf u}_t).
}
Let's take ${\bf u}_t={\bf u}_0 e^{i{\bf q\cdot r}-i\omega t}$ and the equation of motion is changed to
\eq{\label{App_eq:WaveEqn1}
\omega^2 {\bf u}_0=c_t^2 q^2 {\bf u}_0+\frac{\xi_0 q^2}{2}i{\bf q}\times{\bf u}_0
}
Now let's choose ${\bf q}=q {\bf \hat{e}_z}$ and ${\bf u}_0=(u_{0x},u_{0y},0)^T$ (the superscript $T$ here represents transpose)  and then the equation of motion (\ref{App_eq:WaveEqn1}) can be written as
\eq{
\omega^2\left(\begin{array}{c}u_{0x}\\u_{0y}\end{array}\right)=
\left(\begin{array}{cc}c_t^2q^2 & -i \frac{\xi_0}{2}q^3 \\ i\frac{\xi_0}{2}q^3 & c_t^2 q^2 \end{array}\right)\left(\begin{array}{c}u_{0x}\\u_{0y}\end{array}\right).
}
This eigen equation can be easily solved. Two branches of dispersion relations for the transverse modes are given by $\omega^t_\pm=\sqrt{c_t^2q^2\pm \frac{|\xi_0|}{2}q^3}=q\sqrt{c_t^2\pm \frac{|\xi_0|}{2}q}$, from which one can see that the Nieh-Yan term mainly corrects the phonon velocity and gives rise to the different velocities for two transverse phonon modes. The corresponding normalized eigen vectors are given by
\eq{
{\bf u}_{0,\pm}=\frac{1}{\sqrt{2}}\left(\begin{array}{c}1\\\pm i\\0\end{array}\right)
}
so that ${\bf u}^{\dag}_{0,s}({\bf k}){\bf u}_{0,s}({\bf k})=1$.
%where $u_0$ represents the amplitude of the displacement field.
The angular momentum of the phonon modes is defined as \cite{hamada2018phonon,zhang2014angular,hamada2020phonon,zhu2018observation,zhang2015chiral}
\eq{\label{App_eq:AMphonon1}
l_{s,i}({\bf q})=\hbar {\bf u}^{\dag}_{0,s}({\bf q})M_i{\bf u}_{0,s}({\bf q})
}
where $s=\pm$, $i=x,y,z$, and $(M_i)_{jk}=(-i)\epsilon_{ijk}$. One can easily show that only $l_{s,z}$ is non-zero and $l_{\pm,z}=\pm\hbar$ for the momentum ${\bf q}=q {\bf \hat{e}_z}$. Due to the full rotation symmetry of our model, the general form of angular momentum for the $s$-mode is given by ${\bf l}_{s}=s \hbar {\bf \hat{q}}$ with ${\bf \hat{q}}=\frac{\bf q}{q}$.

Next let's consider the phonon total angular momentum induced by a temperature gradient. The total phonon angular momentum per volume can be related to the temperature gradient by
\eq{
I^{ph}_i=\alpha_{ij}\frac{\partial T}{\partial r_j}
}
where $i,j=x,y,z$ and ${\bf r}$ is the spatial coordinate. From Ref. \onlinecite{hamada2018phonon}, the response coefficient $\alpha_{ij}$ is derived as
\eq{
\alpha_{ij}=-\frac{\tau}{V}\sum_{{\bf q},s}l_{s,i}v^{ph}_{s,j}\frac{\partial f_0(\omega^t_{s})}{\partial T}
}
in the linear response regime, where $\tau$ is the phonon relaxation time, ${\bf v}^{ph}_{s}$ is the group velocity of $s$-phonon mode and $f_0(\omega^t_{s})$ is the Bose distribution function for phonons. The group velocity of the transverse mode should be given by
\eqarr{
{\bf v}^{ph}_{s}=\frac{\partial \omega^t_{s}}{\partial {\bf q}}=\frac{\partial \omega^t_{s}}{\partial q}\frac{\partial q}{\partial {\bf q}}=
\frac{c_t^2+\frac{3s}{4}|\xi_0|q}{\sqrt{c_t^2+\frac{s}{2}|\xi_0|q}}{\bf \hat{q}}.
}
For our isotropic system, there is only a single non-zero independent component $\alpha_{xx}=\alpha_{yy}=\alpha_{zz}$, so let's evaluate $\alpha_{zz}$, which is given by
\eq{\label{App_eq:alphazz1}
\alpha_{zz}=-\tau\hbar\int\frac{d^3q}{(2\pi)^3}\sum_s \frac{sq_z^2}{q^2} \frac{c_t^2+\frac{3s}{4}|\xi_0|q}{\sqrt{c_t^2+\frac{s}{2}|\xi_0|q}} \frac{\partial f_0}{\partial T}=-\frac{\tau\hbar}{6\pi^2}\int q^2dq\sum_s s \frac{c_t^2+\frac{3s}{4}|\xi_0|q}{\sqrt{c_t^2+\frac{s}{2}|\xi_0|q}} \frac{\partial f_0}{\partial T},
}
where
\eq{
\frac{\partial f_0({\omega})}{\partial T}=k_B \beta^2 \omega\frac{e^{\beta \omega}}{(e^{\beta\omega}-1)^2}.
}
The angular part of the phonon momentum in the above expression can be integrated out since $\frac{\partial f_0({\omega})}{\partial T}$ only depends on the magnitude of ${\bf q}$.
This expression can be evaluated numerically and to get some analytical understanding, we can make the further approximations. According to the Bose distribution $f_0({\omega})$, we know that the phonons are mainly excited in the energy range determined by the temperature $T$, $\omega \ll k_B T$. This means the contribution in the above expression mainly comes from the momentum range $q< \frac{k_BT}{c_t}$. Now let's consider the momentum range, in which the Nieh-Yan term contribution to the velocity $|\xi_0|q$ is much smaller than $c_t^2$. This is naturally satisfied when the temperature $k_B T\ll\frac{c_t^2}{|\xi_0|}$ and in this temperature range, we can treat $|\xi_0|q$ as a perturbation and expand the integrand in Eq. (\ref{App_eq:alphazz1}). First, we have
\eq{
\frac{c_t^2+\frac{3s}{4}|\xi_0|q}{\sqrt{c_t^2+\frac{s}{2}|\xi_0|q}} \approx c_t\left(1+\frac{3s}{4c_t^2}|\xi_0|q\right)\left(1-\frac{s}{4c_t^2}|\xi_0|q\right)\approx c_t\left(1+\frac{s}{2c_t^2}|\xi_0|q\right)
}
and
\eq{
\frac{\partial f_0}{\partial T}=\frac{\partial f_0(\omega_0)}{\partial T}+\frac{\partial^2 f_0}{\partial T\partial\omega}\delta\omega_s,
}
where $\omega_0=c_t q$ and $\delta\omega_s=\frac{s|\xi_0|}{4c_t}q^2$. Let's define $x=\beta \omega_0=\frac{c_t q}{k_BT}$ and then we obtain
\eq{
\frac{\partial f_0(\omega_0)}{\partial T}=k_B \beta^2\omega_0 \frac{e^{\beta \omega_0}}{(e^{\beta\omega_0}-1)^2}=\frac{x}{T}\frac{e^{x}}{(e^{x}-1)^2},
}
and
\eq{
\frac{\partial^2 f_0(\omega_0)}{\partial T\partial \omega}=k_B\beta^2 \frac{e^{\beta \omega_0}}{(e^{\beta\omega_0}-1)^3}\left(e^{\beta\omega_0}(1-\beta\omega_0)-\beta\omega_0-1\right)=\frac{1}{k_B T^2} \frac{e^{x}}{(e^{x}-1)^3}\left(e^{x}(1-x)-x-1\right).
}
Thus, we have
\eqarr{\label{App_eq:alphazz2}
&&\alpha_{zz}\approx -\frac{\tau\hbar}{6\pi^2}\int q^2dq\sum_s s c_t\left(1+\frac{s}{2c_t^2}|\xi_0|q\right) \left(\frac{\partial f_0(\omega_0)}{\partial T}+\frac{\partial^2 f_0}{\partial T\partial\omega}\frac{s|\xi_0|}{4c_t}q^2\right)\nonumber\\
&&= -\frac{\tau\hbar}{6\pi^2}\int q^2dq c_t\left(\frac{\partial f_0(\omega_0)}{\partial T}\frac{|\xi_0|}{c_t^2}q+\frac{\partial^2 f_0}{\partial T\partial\omega}\frac{|\xi_0|}{2c_t}q^2\right)\nonumber\\
&&= -\frac{\tau\hbar}{6\pi^2}\left(\frac{k_B T}{c_t}\right)^3\frac{|\xi_0|}{c_t^2}\int x^2 dx  \left(\frac{\partial f_0(\omega_0)}{\partial T}x k_B T+\frac{\partial^2 f_0}{\partial T\partial\omega}\frac{x^2}{2}\left( k_B T\right)^2\right)\nonumber\\
&&= -\frac{\tau\hbar |\xi_0|k_B}{6\pi^2c_t^5}\left(k_B T\right)^3\int x^4 dx  \left(\frac{e^{x}}{(e^{x}-1)^2} + \frac{1}{2} \frac{e^{x}}{(e^{x}-1)^3}\left(e^{x}(1-x)-x-1\right)\right)\nonumber\\
}
In the above expression, the integral of $x$ is convergent, and thus just gives a number. Therefore, all the temperature dependence is given by the coefficient before the $x$-integral. Since $\xi_0=g_1^2\mu_5 \mathcal{F}_0$ with $\mathcal{F}_0=F_0+F_1(k_BT)^2$, we expect $\alpha_{zz}$ has $T^3$ dependence for the standard Nieh-Yan term and $T^5$ dependence for the thermal Nieh-Yan term. Let's denote $a_0=\int x^4 dx  \left(\frac{e^{x}}{(e^{x}-1)^2} + \frac{1}{2} \frac{e^{x}}{(e^{x}-1)^3}\left(e^{x}(1-x)-x-1\right)\right)=-4\pi^4/15$, and then we have
\eq{
\alpha_{zz}\approx -\frac{\tau\hbar a_0 |\xi_0|k_B}{6\pi^2c_t^5}\left(k_B T\right)^3 = -\frac{\tau\hbar a_0 k_B g_1^2}{6\pi^2c_t^5}\mu_5 \left(F_0 \left(k_B T\right)^3+F_1\left(k_B T\right)^5\right)=\frac{2\tau\hbar \pi^2 k_B g_1^2}{45 c_t^5}\mu_5 \left(F_0 \left(k_B T\right)^3+F_1\left(k_B T\right)^5\right).
}

\section{Electron-electron Interaction}
\label{App_Sec:Interaction}
Since the electron-phonon interaction may be influenced by electron-electron interaction, one may wonder how electron-electron interaction affects the proposed phenomena here. In this section, we will demonstrate that electron-electron interaction is actually decoupled from the phonon modes discussed in this work and thus all the results remain valid even taking into account electron-electron interaction.

To see that, let's start from the effective action $S_{eff}$ in Eq. (\ref{App_eq:Seff2}) with an additional electron-electron Coulomb interaction. We consider the Stratonovich-Hubbard transformation by introducing an additional $\varphi$ field with the effective action
\eq{\label{App_eq:actionSc}
S_C=\int d\tau d^3r \left( \frac{1}{2}(\nabla \varphi({\bf r},\tau))^2 + ie\varphi({\bf r},\tau) \hat{\psi}^{\dag}_{\bf \Gamma_i}({\bf r},\tau) \hat{\psi}_{\bf \Gamma_i}({\bf r},\tau)\right).
}
With the Fourier transform
\eqarr{
&&\hat{\psi}_{\bf \Gamma_i}({\bf r},\tau)=\sum_{\tilde{k}} e^{i{\bf k\cdot r}-i\omega_n\tau} \hat{\psi}_{\bf \Gamma_i}(\tilde{k}),\\
&&\varphi({\bf r},\tau)=\sum_{\tilde{q}} e^{i{\bf q\cdot r}-i\nu_m\tau} \varphi(\tilde{q}),
}
we have
\eq{
S_C=\sum_{\tilde{q}} \frac{q^2}{2} \varphi(\tilde{q})\varphi(-\tilde{q}) +  ie \sum_{\tilde{k},\tilde{q}}\varphi(\tilde{q}) \hat{\psi}^{\dag}_{\bf \Gamma_i}(\tilde{k}) \hat{\psi}_{\bf \Gamma_i}(\tilde{k}-\tilde{q}),
}
where $\sum_{\tilde{k}}=\frac{1}{\beta V}\sum_{{\bf k},i\omega_n}$. Now let's perform the transformation
\eqarr{
&&\tilde{\varphi}(-\tilde{q})=\varphi(-\tilde{q})+i\frac{e}{q^2}\sum_{\tilde{k}}\hat{\psi}^{\dag}_{\bf \Gamma_i}(\tilde{k}) \hat{\psi}_{\bf \Gamma_i}(\tilde{k}-\tilde{q})\\
&&\tilde{\varphi}(\tilde{q})=\varphi(\tilde{q})+i\frac{e}{q^2}\sum_{\tilde{k}}\hat{\psi}^{\dag}_{\bf \Gamma_i}(\tilde{k}) \hat{\psi}_{\bf \Gamma_i}(\tilde{k}+\tilde{q})
}
and
\eqarr{
&&\sum_{\tilde{q}} \frac{q^2}{2} \tilde{\varphi}(\tilde{q})\tilde{\varphi}(-\tilde{q})= \sum_{\tilde{q}} \frac{q^2}{2}\left(\varphi(\tilde{q})+i\frac{e}{q^2}\sum_{\tilde{k}}\hat{\psi}^{\dag}_{\bf \Gamma_i}(\tilde{k}) \hat{\psi}_{\bf \Gamma_i}(\tilde{k}+\tilde{q})\right)\left(\varphi(-\tilde{q})+i\frac{e}{q^2}\sum_{\tilde{k}}\hat{\psi}^{\dag}_{\bf \Gamma_i}(\tilde{k}) \hat{\psi}_{\bf \Gamma_i}(\tilde{k}-\tilde{q})\right)\nonumber\\
&&=\sum_{\tilde{q}} \left(\frac{q^2}{2} \varphi(\tilde{q})\varphi(-\tilde{q})+ie\varphi(\tilde{q})\sum_{\tilde{k}}\hat{\psi}^{\dag}_{\bf \Gamma_i}(\tilde{k}) \hat{\psi}_{\bf \Gamma_i}(\tilde{k}-\tilde{q})-\frac{e^2}{2q^2}\sum_{\tilde{k}\tilde{k}'}\hat{\psi}^{\dag}_{\bf \Gamma_i}(\tilde{k}) \hat{\psi}_{\bf \Gamma_i}(\tilde{k}+\tilde{q})\hat{\psi}^{\dag}_{\bf \Gamma_i}(\tilde{k}') \hat{\psi}_{\bf \Gamma_i}(\tilde{k}'-\tilde{q})
\right).
}
From the above quality, we have
\eq{
S_C=\sum_{\tilde{q}} \frac{q^2}{2} \tilde{\varphi}(\tilde{q})\tilde{\varphi}(-\tilde{q})+\frac{e^2}{2q^2}\sum_{\tilde{q}\tilde{k}\tilde{k}'}\hat{\psi}^{\dag}_{\bf \Gamma_i}(\tilde{k}) \hat{\psi}_{\bf \Gamma_i}(\tilde{k}+\tilde{q})\hat{\psi}^{\dag}_{\bf \Gamma_i}(\tilde{k}') \hat{\psi}_{\bf \Gamma_i}(\tilde{k}'-\tilde{q}),
}
the latter term of which is indeed four-fermion Coulomb interaction with the coefficient $V_q=\frac{e^2}{q^2}$ the Fourier transform of Coulomb interaction in the momentum space (we have set $\epsilon_0=1$).

Now let's compare the second term in $S_C$ (Eq. (\ref{App_eq:actionSc})) with the first term in $S_1$ (Eq. (\ref{App_eq:actionS1})), and one can see that the $A_0$ field is identical to the $-ie\varphi$ field. Similar to the derivation for the effective action $W[A_0,\Delta]$ in Eq. (\ref{App_eq:Weff1}), we consider the effective action for the $\varphi$ and ${\bf \Delta}$ fields (we drop the $A_0$ field here) by integrating out the fermion operators, and the resulting effective action is given by
\eqarr{\label{App_eq:Weff2}
&&W_1[\varphi,{\bf \Delta}]=\frac{1}{2}\sum_{\tilde{q}}\left(q^2\varphi(\tilde{q})\varphi(-\tilde{q})-e^2\Pi_0(\tilde{q})\varphi(\tilde{q})\varphi(-\tilde{q})
+\Phi^{ab}_{ij}(\tilde{q})\Delta^i_a(\tilde{q})\Delta^j_b(-\tilde{q})+ie\Theta^a_i(\tilde{q})\varphi(\tilde{q})\Delta^i_a(-\tilde{q})
+ie\Theta^a_i(-\tilde{q})\Delta^i_a(\tilde{q})\varphi(-\tilde{q})\right)\nonumber\\
&&=\frac{1}{2}\sum_{\tilde{q}}\left(D^{-1}(\tilde{q})\varphi(\tilde{q})\varphi(-\tilde{q})+\Phi^{ab}_{ij}(\tilde{q})\Delta^i_a(\tilde{q})\Delta^j_b(-\tilde{q})
+ie\Theta^a_i(\tilde{q})\varphi(\tilde{q})\Delta^i_a(-\tilde{q})+ie\Theta^a_i(-\tilde{q})\Delta^i_a(\tilde{q})\varphi(-\tilde{q})\right),
}
where the duplicated indices should be summed over. Here $D^{-1}(\tilde{q})=q^2-e^2\Pi_0(\tilde{q})=D_0^{-1}-e^2\Pi_0(\tilde{q})$ defines the full Green's function for the $\varphi$ field.

Next let's consider the path integral
\eq{
Z_1[{\bf \Delta}]=\int\mathcal{D} \varphi e^{-W_1[\varphi,{\bf \Delta}]}
}
and we can perform the transformation
\eqarr{
&&\tilde{\varphi}(\tilde{q})=\varphi(\tilde{q})+ie\Theta^a_i(-\tilde{q})\Delta^i_a(\tilde{q})D(\tilde{q})\\
&&\tilde{\varphi}(-\tilde{q})=\varphi(-\tilde{q})+ie\Theta^a_i(\tilde{q})\Delta^i_a(-\tilde{q})D(-\tilde{q}).
}
Correspondingly,
\eqarr{
&&\frac{1}{2}\sum_{\tilde{q}}D^{-1}(\tilde{q})\tilde{\varphi}(\tilde{q})\tilde{\varphi}(-\tilde{q})=
\frac{1}{2}\sum_{\tilde{q}}D^{-1}(\tilde{q})(\varphi(\tilde{q})+ie\Theta^a_i(-\tilde{q})\Delta^i_a(\tilde{q})D(\tilde{q}))
(\varphi(-\tilde{q})+ie\Theta^b_j(\tilde{q})\Delta^j_b(-\tilde{q})D(-\tilde{q}))\nonumber\\
&&=\frac{1}{2}\sum_{\tilde{q}}\left(D^{-1}(\tilde{q})\varphi(\tilde{q})\varphi(-\tilde{q})+ie\varphi(\tilde{q})\Theta^b_j(\tilde{q})\Delta^j_b(-\tilde{q})
+ie\Theta^a_i(-\tilde{q})\Delta^i_a(\tilde{q})\varphi(-\tilde{q})-e^2\Theta^a_i(-\tilde{q})\Delta^i_a(\tilde{q})D(\tilde{q})
\Theta^b_j(\tilde{q})\Delta^j_b(-\tilde{q})\right)\nonumber\\
}
and
\eqarr{
&&W_1[\varphi,{\bf \Delta}]=\frac{1}{2}\sum_{\tilde{q}}\left(D^{-1}(\tilde{q})\tilde{\varphi}(\tilde{q})\tilde{\varphi}(-\tilde{q})+\left(\Phi^{ab}_{ij}(\tilde{q})
+e^2\Theta^a_i(-\tilde{q})D(\tilde{q})\Theta^b_j(\tilde{q})\right)\Delta^i_a(\tilde{q})\Delta^j_b(-\tilde{q})\right).
}
Now one can see that the $\tilde{\varphi}$ and ${\bf \Delta}$ fields are decoupled from each other. Let's write $Z_1[{\bf \Delta}]\sim e^{-W_2[{\bf \Delta}]}$ and then
\eq{
W_2[{\bf \Delta}]=\frac{1}{2}\sum_{\tilde{q}}\left(\Phi^{ab}_{ij}(\tilde{q})
+e^2\Theta^a_i(-\tilde{q})D(\tilde{q})\Theta^b_j(\tilde{q})\right)\Delta^i_a(\tilde{q})\Delta^j_b(-\tilde{q}).
}
From this expression, we conclude that the full stress-stress correlation function, denoted as $\tilde{\Phi}^{ab}_{ij}(\tilde{q})$, should generally acquire a correction from the electron-electron interaction through
\eq{
\tilde{\Phi}^{ab}_{ij}(\tilde{q})=\Phi^{ab}_{ij}(\tilde{q})+e^2\Theta^a_i(-\tilde{q})D(\tilde{q})\Theta^b_j(\tilde{q}),
}
where the stress-density correlation function $\Theta^a_i(\tilde{q})$ is defined in Eq. (\ref{App_eq:Theta0}).

One can see that the $\Theta^a_i(\tilde{q})$ function couples electrons to acoustic phonons and this function involves one density vertex and one stress tensor vertex from Eq. (\ref{App_eq:Theta0}). Next we will show that for the phonon modes that we are interested in, the $\Theta^a_i(\tilde{q})$ function is always zero if the system has full rotation symmetry. In the above expressions, $a,b,i,j=x,y,z$ which is not convenient for the symmetry analysis. It is more convenient to relabel the strain and stress tensor with the angular momentum, as shown in Eq. (\ref{App_eq:straintensor1})-(\ref{App_eq:stresstensor2}). The rotationally symmetric form of the Hamiltonian (\ref{App_eq:Hamsym2}) is useful for our symmetry analysis below.

To make the problem more concrete, let's consider the phonon momentum along the z-direction, ${\bf q}=q_z\hat{e}_z$, and we first figure out which strain tensors, as well as the stress tensors, are involved in the Nieh-Yan term. To see that, let's write down explicitly the terms in Eq. (\ref{App_eq:Sph1}) along the $q_z\hat{e}_z$ direction, which involves
\eqarr{
&&\delta_{jl}\epsilon_{zkm}(\partial_z u_{jk})u_{lm}=\epsilon_{zxy}(\partial_z u_{jx})u_{jy}+\epsilon_{zyx}(\partial_z u_{jy})u_{jx}\nonumber\\
&&=\left((\partial_z u_{xx})u_{xy}+(\partial_z u_{yx})u_{yy}+(\partial_z u_{zx})u_{zy}\right)-\left((\partial_z u_{xy})u_{xx}+(\partial_z u_{yy})u_{yx}+(\partial_z u_{zy})u_{zx}\right)\nonumber\\
&&=2c\partial_z(u_{xx}-u_{yy})+2(\partial_z u_{zx})u_{zy},
}
in which all the total derivative term has been dropped since they will not contribute to the effective action. Now we notice that only the components $u_{xx}-u_{yy}$, $u_{xy}$, $u_{zx}$ and $u_{zy}$ of the strain tensor are involved for the acoustic phonons along the z direction that we are interested in. From the Eq. (\ref{App_eq:straintensor1})-(\ref{App_eq:straintensor2}), one can see that only $u_{J=2,M}$ with $M=\pm 1, \pm 2$ are involved, which means the the z-direction angular momentum $M$ only takes $\pm 1$ and $\pm 2$, but not $0$. On the other hand, the electron density operator carries z-direction angular momentum $0$. Since the z-direction rotation symmetry is preserved for the momentum ${\bf q}$ along the z direction, we expect elections cannot directly couple to the phonon modes with higher angular momentum.

To make this argument more explicitly, let's define $\mathcal{T}^J_{M}$ as
\eqarr{
&&\mathcal{T}^{2}_2=\frac{1}{2}\left(\mathcal{T}^{x}_x-\mathcal{T}^{y}_y+i(\mathcal{T}^{y}_x+\mathcal{T}^{x}_y\right)\\
&&\mathcal{T}^{2}_{1}=\left(-\frac{1}{2}\right)\left(\mathcal{T}^{z}_x+\mathcal{T}^{x}_z+i(\mathcal{T}^{z}_y+\mathcal{T}^{y}_z)\right)\\
&&\mathcal{T}^{2}_{0}=\sqrt{\frac{1}{6}}\left(2\mathcal{T}^{z}_z-\mathcal{T}^{x}_x-\mathcal{T}^{y}_y\right)\\
&&\mathcal{T}^{2}_{-1}=\frac{1}{2}\left(\mathcal{T}^{z}_x+\mathcal{T}^{x}_z-i(\mathcal{T}^{z}_y+\mathcal{T}^{y}_z)\right)\\
&&\mathcal{T}^{2}_{-2}=\frac{1}{2}\left(\mathcal{T}^{x}_x-\mathcal{T}^{y}_y-i(\mathcal{T}^{y}_x+\mathcal{T}^{x}_y)\right)
}
and the stress-density correlation function can also be re-constructed as
\eq{
\Theta^J_{M}({\bf q},i\nu_m)=\frac{1}{\beta V}\sum_{{\bf k},i\omega_n}Tr_{\sigma}\left(\mathcal{G}_0({\bf k-q},i\omega_n-i\nu_m)
\mathcal{T}^J_{M}({\bf k-q},{\bf k})\mathcal{G}_0({\bf k},i\omega_n)\right)
}
for the angular momentum $J,M$ of the stress operator. Here we also add the density operator $n$ which is an identity. Let's consider a symmetry operator $\mathcal{R}$, which transforms the stress tensor as $\mathcal{R}\mathcal{T}^J_{M}({\bf k-q},{\bf k})\mathcal{R}^{-1}=\sum_{M'}D^J_{MM'}(\mathcal{R})\mathcal{T}^J_{M'}(\mathcal{R}^{-1}({\bf k-q}),\mathcal{R}^{-1}{\bf k})$. The Green function should be invariant under the symmetry operation $\mathcal{R}\mathcal{G}_0({\bf k},i\omega_n)\mathcal{R}^{-1}=\mathcal{G}_0(\mathcal{R}^{-1}{\bf k},i\omega_n)$. Thus, one can insert symmetry operation into the definition of stress-density correlation function and find
\eqarr{
&&\Theta^J_{M}({\bf q},i\nu_m)=\frac{1}{\beta V}\sum_{{\bf k},i\omega_n}Tr_{\sigma}\left(\mathcal{R}\mathcal{G}_0({\bf k-q},i\omega_n-i\nu_m)\mathcal{R}^{-1}\mathcal{R}
\mathcal{T}^J_{M}({\bf k-q},{\bf k})\mathcal{R}^{-1}\mathcal{R}\mathcal{G}_0({\bf k},i\omega_n)\mathcal{R}^{-1}\right)\nonumber\\
&&=\frac{1}{\beta V}\sum_{{\bf k},i\omega_n}Tr_{\sigma}\left(\mathcal{G}_0(\mathcal{R}^{-1}({\bf k-q}),i\omega_n-i\nu_m)\sum_{M'}D^J_{MM'}(\mathcal{R})
\mathcal{T}^J_{M'}(\mathcal{R}^{-1}({\bf k-q}),{\bf k})\mathcal{G}_0(\mathcal{R}^{-1}{\bf k},i\omega_n)\right)\nonumber\\
&&=\frac{1}{\beta V}\sum_{{\bf k}',i\omega_n,M'}D^J_{MM'}(\mathcal{R})Tr_{\sigma}\left(\mathcal{G}_0({\bf k}'-\mathcal{R}^{-1}{\bf q}),i\omega_n-i\nu_m)
\mathcal{T}^J_{M'}({\bf k}'-\mathcal{R}^{-1}{\bf q},{\bf k}')\mathcal{G}_0({\bf k}',i\omega_n)\right)\nonumber\\
&&=\sum_{M'}D^J_{MM'}(\mathcal{R})\Theta^J_{M'}(\mathcal{R}^{-1}{\bf q},i\nu_m).
}
In the above derivation, we have used ${\bf k}'=\mathcal{R}^{-1}{\bf k}$. The above equation gives rise to the symmetry constraint on the form of correlation functions. Along ${\bf q}=q_z\hat{e}_z$, we have $\mathcal{R}^{-1}{\bf q}={\bf q}$ and
\eq{
\Theta^J_{M}(q_z,i\nu_m)=\sum_{M'}D^J_{MM'}(\mathcal{R})\Theta^J_{M'}(q_z,i\nu_m)
}
Now let's consider the orthonormality relation for irreducible representations of a symmetry group, given by $\sum_{\mathcal{R}}D^{\Gamma_j}_{MM'}(\mathcal{R})[D^{\Gamma_j'}_{NN'}(\mathcal{R})]^*=\frac{h}{l_j}\delta_{jj'}\delta_{MN}\delta_{M'N'}$. We should consider the symmetry group formed by the rotation along the z directions and thus only the z-directional angular momentum $M$ is a good quantum number that characterizes the rotation group, while the total angular momentum $J$ is not. Consequently, we expect the orthonormality relation is then given by $\sum_{\mathcal{R}}D^{J}_{MM'}(\mathcal{R})[D^{J'}_{NN'}(\mathcal{R})]^*\propto \delta_{MN}\delta_{M'N'}$. Now let's choose $D^{J'}_{NN'}$ to be $J'=0$ (identity representation), so $N=N'=0$ and $D^{0}_{00}=1$ for any $\mathcal{R}$. Then the orthonormality relation takes the form $\sum_{\mathcal{R}}D^{J}_{MM'}(\mathcal{R})\propto \delta_{M0}\delta_{M'0}$, which leads to
\eqarr{
&&\sum_{\mathcal{R}}\Theta^J_{M}(q_z,i\nu_m)=\sum_{M'}\sum_{\mathcal{R}}D^J_{MM'}(\mathcal{R})\Theta^J_{M'}(q_z,i\nu_m)\propto \delta_{M0}\nonumber\\
&&\rightarrow \Theta^J_{M}(q_z,i\nu_m)\propto \delta_{M0}.
}
So the correlation function $\Theta^J_{M}$ can only be nonzero for the z-directional angular momentum $M=0$. On the other hand, as we have shown above, the Nieh-Yan anomaly term only involves the strain tensor with $M=\pm 1, \pm 2$, and this concludes that the electron-electron Coulomb interaction will not contribute to the acoustic phonon modes, whose dynamics is influenced by the Nieh-Yan anomaly.

\end{widetext}
\end{appendix}


\begin{thebibliography}{92}%
\makeatletter
\providecommand \@ifxundefined [1]{%
 \@ifx{#1\undefined}
}%
\providecommand \@ifnum [1]{%
 \ifnum #1\expandafter \@firstoftwo
 \else \expandafter \@secondoftwo
 \fi
}%
\providecommand \@ifx [1]{%
 \ifx #1\expandafter \@firstoftwo
 \else \expandafter \@secondoftwo
 \fi
}%
\providecommand \natexlab [1]{#1}%
\providecommand \enquote  [1]{``#1''}%
\providecommand \bibnamefont  [1]{#1}%
\providecommand \bibfnamefont [1]{#1}%
\providecommand \citenamefont [1]{#1}%
\providecommand \href@noop [0]{\@secondoftwo}%
\providecommand \href [0]{\begingroup \@sanitize@url \@href}%
\providecommand \@href[1]{\@@startlink{#1}\@@href}%
\providecommand \@@href[1]{\endgroup#1\@@endlink}%
\providecommand \@sanitize@url [0]{\catcode `\\12\catcode `\$12\catcode
  `\&12\catcode `\#12\catcode `\^12\catcode `\_12\catcode `\%12\relax}%
\providecommand \@@startlink[1]{}%
\providecommand \@@endlink[0]{}%
\providecommand \url  [0]{\begingroup\@sanitize@url \@url }%
\providecommand \@url [1]{\endgroup\@href {#1}{\urlprefix }}%
\providecommand \urlprefix  [0]{URL }%
\providecommand \Eprint [0]{\href }%
\providecommand \doibase [0]{http://dx.doi.org/}%
\providecommand \selectlanguage [0]{\@gobble}%
\providecommand \bibinfo  [0]{\@secondoftwo}%
\providecommand \bibfield  [0]{\@secondoftwo}%
\providecommand \translation [1]{[#1]}%
\providecommand \BibitemOpen [0]{}%
\providecommand \bibitemStop [0]{}%
\providecommand \bibitemNoStop [0]{.\EOS\space}%
\providecommand \EOS [0]{\spacefactor3000\relax}%
\providecommand \BibitemShut  [1]{\csname bibitem#1\endcsname}%
\let\auto@bib@innerbib\@empty
%</preamble>
\bibitem [{\citenamefont {Peskin}(2018)}]{peskin2018introduction}%
  \BibitemOpen
  \bibfield  {author} {\bibinfo {author} {\bibfnamefont {M.}~\bibnamefont
  {Peskin}},\ }\href@noop {} {\emph {\bibinfo {title} {An introduction to
  quantum field theory}}}\ (\bibinfo  {publisher} {CRC press},\ \bibinfo {year}
  {2018})\BibitemShut {NoStop}%
\bibitem [{\citenamefont {Fujikawa}\ \emph {et~al.}(2004)\citenamefont
  {Fujikawa}, \citenamefont {Fujikawa}, \citenamefont {Suzuki} \emph
  {et~al.}}]{fujikawa2004path}%
  \BibitemOpen
  \bibfield  {author} {\bibinfo {author} {\bibfnamefont {K.}~\bibnamefont
  {Fujikawa}}, \bibinfo {author} {\bibfnamefont {K.}~\bibnamefont {Fujikawa}},
  \bibinfo {author} {\bibfnamefont {H.}~\bibnamefont {Suzuki}},  \emph
  {et~al.},\ }\href@noop {} {\emph {\bibinfo {title} {Path integrals and
  quantum anomalies}}},\ \bibinfo {number} {122}\ (\bibinfo  {publisher}
  {Oxford University Press on Demand},\ \bibinfo {year} {2004})\BibitemShut
  {NoStop}%
\bibitem [{\citenamefont {Bertlmann}(2000)}]{bertlmann2000anomalies}%
  \BibitemOpen
  \bibfield  {author} {\bibinfo {author} {\bibfnamefont {R.~A.}\ \bibnamefont
  {Bertlmann}},\ }\href@noop {} {\emph {\bibinfo {title} {Anomalies in quantum
  field theory}}},\ Vol.~\bibinfo {volume} {91}\ (\bibinfo  {publisher} {Oxford
  university press},\ \bibinfo {year} {2000})\BibitemShut {NoStop}%
\bibitem [{\citenamefont {Fukushima}\ \emph {et~al.}(2008)\citenamefont
  {Fukushima}, \citenamefont {Kharzeev},\ and\ \citenamefont
  {Warringa}}]{fukushima2008chiral}%
  \BibitemOpen
  \bibfield  {author} {\bibinfo {author} {\bibfnamefont {K.}~\bibnamefont
  {Fukushima}}, \bibinfo {author} {\bibfnamefont {D.~E.}\ \bibnamefont
  {Kharzeev}}, \ and\ \bibinfo {author} {\bibfnamefont {H.~J.}\ \bibnamefont
  {Warringa}},\ }\href@noop {} {\bibfield  {journal} {\bibinfo  {journal}
  {Physical Review D}\ }\textbf {\bibinfo {volume} {78}},\ \bibinfo {pages}
  {074033} (\bibinfo {year} {2008})}\BibitemShut {NoStop}%
\bibitem [{\citenamefont {Delbourgo}\ and\ \citenamefont
  {Salam}(1972)}]{delbourgo1972gravitational}%
  \BibitemOpen
  \bibfield  {author} {\bibinfo {author} {\bibfnamefont {R.}~\bibnamefont
  {Delbourgo}}\ and\ \bibinfo {author} {\bibfnamefont {A.}~\bibnamefont
  {Salam}},\ }\href@noop {} {\bibfield  {journal} {\bibinfo  {journal} {Physics
  Letters B}\ }\textbf {\bibinfo {volume} {40}},\ \bibinfo {pages} {381}
  (\bibinfo {year} {1972})}\BibitemShut {NoStop}%
\bibitem [{\citenamefont {Eguchi}\ and\ \citenamefont
  {Freund}(1976)}]{eguchi1976quantum}%
  \BibitemOpen
  \bibfield  {author} {\bibinfo {author} {\bibfnamefont {T.}~\bibnamefont
  {Eguchi}}\ and\ \bibinfo {author} {\bibfnamefont {P.~G.}\ \bibnamefont
  {Freund}},\ }\href@noop {} {\bibfield  {journal} {\bibinfo  {journal}
  {Physical Review Letters}\ }\textbf {\bibinfo {volume} {37}},\ \bibinfo
  {pages} {1251} (\bibinfo {year} {1976})}\BibitemShut {NoStop}%
\bibitem [{\citenamefont {Khaidukov}\ and\ \citenamefont
  {Zubkov}(2018)}]{khaidukov2018chiral}%
  \BibitemOpen
  \bibfield  {author} {\bibinfo {author} {\bibfnamefont {Z.~V.}\ \bibnamefont
  {Khaidukov}}\ and\ \bibinfo {author} {\bibfnamefont {M.}~\bibnamefont
  {Zubkov}},\ }\href@noop {} {\bibfield  {journal} {\bibinfo  {journal} {JETP
  letters}\ }\textbf {\bibinfo {volume} {108}},\ \bibinfo {pages} {670}
  (\bibinfo {year} {2018})}\BibitemShut {NoStop}%
\bibitem [{\citenamefont {Sumiyoshi}\ and\ \citenamefont
  {Fujimoto}(2016)}]{sumiyoshi2016torsional}%
  \BibitemOpen
  \bibfield  {author} {\bibinfo {author} {\bibfnamefont {H.}~\bibnamefont
  {Sumiyoshi}}\ and\ \bibinfo {author} {\bibfnamefont {S.}~\bibnamefont
  {Fujimoto}},\ }\href@noop {} {\bibfield  {journal} {\bibinfo  {journal}
  {Physical review letters}\ }\textbf {\bibinfo {volume} {116}},\ \bibinfo
  {pages} {166601} (\bibinfo {year} {2016})}\BibitemShut {NoStop}%
\bibitem [{\citenamefont {Imaki}\ and\ \citenamefont
  {Yamamoto}(2019)}]{imaki2019lattice}%
  \BibitemOpen
  \bibfield  {author} {\bibinfo {author} {\bibfnamefont {S.}~\bibnamefont
  {Imaki}}\ and\ \bibinfo {author} {\bibfnamefont {A.}~\bibnamefont
  {Yamamoto}},\ }\href@noop {} {\bibfield  {journal} {\bibinfo  {journal}
  {Physical Review D}\ }\textbf {\bibinfo {volume} {100}},\ \bibinfo {pages}
  {054509} (\bibinfo {year} {2019})}\BibitemShut {NoStop}%
\bibitem [{\citenamefont {Nieh}\ and\ \citenamefont
  {Yan}(1982{\natexlab{a}})}]{nieh1982quantized}%
  \BibitemOpen
  \bibfield  {author} {\bibinfo {author} {\bibfnamefont {H.}~\bibnamefont
  {Nieh}}\ and\ \bibinfo {author} {\bibfnamefont {M.}~\bibnamefont {Yan}},\
  }\href@noop {} {\bibfield  {journal} {\bibinfo  {journal} {Annals of
  Physics}\ }\textbf {\bibinfo {volume} {138}},\ \bibinfo {pages} {237}
  (\bibinfo {year} {1982}{\natexlab{a}})}\BibitemShut {NoStop}%
\bibitem [{\citenamefont {Nieh}\ and\ \citenamefont
  {Yan}(1982{\natexlab{b}})}]{nieh1982identity}%
  \BibitemOpen
  \bibfield  {author} {\bibinfo {author} {\bibfnamefont {H.}~\bibnamefont
  {Nieh}}\ and\ \bibinfo {author} {\bibfnamefont {M.}~\bibnamefont {Yan}},\
  }\href@noop {} {\bibfield  {journal} {\bibinfo  {journal} {Journal of
  Mathematical Physics}\ }\textbf {\bibinfo {volume} {23}},\ \bibinfo {pages}
  {373} (\bibinfo {year} {1982}{\natexlab{b}})}\BibitemShut {NoStop}%
\bibitem [{\citenamefont {Chandia}\ and\ \citenamefont
  {Zanelli}(1997)}]{chandia1997topological}%
  \BibitemOpen
  \bibfield  {author} {\bibinfo {author} {\bibfnamefont {O.}~\bibnamefont
  {Chandia}}\ and\ \bibinfo {author} {\bibfnamefont {J.}~\bibnamefont
  {Zanelli}},\ }\href@noop {} {\bibfield  {journal} {\bibinfo  {journal}
  {Physical Review D}\ }\textbf {\bibinfo {volume} {55}},\ \bibinfo {pages}
  {7580} (\bibinfo {year} {1997})}\BibitemShut {NoStop}%
\bibitem [{\citenamefont {Nieh}(2007)}]{nieh2007torsional}%
  \BibitemOpen
  \bibfield  {author} {\bibinfo {author} {\bibfnamefont {H.}~\bibnamefont
  {Nieh}},\ }\href@noop {} {\bibfield  {journal} {\bibinfo  {journal}
  {International Journal of Modern Physics A}\ }\textbf {\bibinfo {volume}
  {22}},\ \bibinfo {pages} {5237} (\bibinfo {year} {2007})}\BibitemShut
  {NoStop}%
\bibitem [{\citenamefont {Armitage}\ \emph {et~al.}(2018)\citenamefont
  {Armitage}, \citenamefont {Mele},\ and\ \citenamefont
  {Vishwanath}}]{armitage2018weyl}%
  \BibitemOpen
  \bibfield  {author} {\bibinfo {author} {\bibfnamefont {N.}~\bibnamefont
  {Armitage}}, \bibinfo {author} {\bibfnamefont {E.}~\bibnamefont {Mele}}, \
  and\ \bibinfo {author} {\bibfnamefont {A.}~\bibnamefont {Vishwanath}},\
  }\href@noop {} {\bibfield  {journal} {\bibinfo  {journal} {Reviews of Modern
  Physics}\ }\textbf {\bibinfo {volume} {90}},\ \bibinfo {pages} {015001}
  (\bibinfo {year} {2018})}\BibitemShut {NoStop}%
\bibitem [{\citenamefont {Yan}\ and\ \citenamefont
  {Felser}(2017)}]{yan2017topological}%
  \BibitemOpen
  \bibfield  {author} {\bibinfo {author} {\bibfnamefont {B.}~\bibnamefont
  {Yan}}\ and\ \bibinfo {author} {\bibfnamefont {C.}~\bibnamefont {Felser}},\
  }\href@noop {} {\bibfield  {journal} {\bibinfo  {journal} {Annual Review of
  Condensed Matter Physics}\ }\textbf {\bibinfo {volume} {8}},\ \bibinfo
  {pages} {337} (\bibinfo {year} {2017})}\BibitemShut {NoStop}%
\bibitem [{\citenamefont {Hosur}\ and\ \citenamefont
  {Qi}(2013)}]{hosur2013recent}%
  \BibitemOpen
  \bibfield  {author} {\bibinfo {author} {\bibfnamefont {P.}~\bibnamefont
  {Hosur}}\ and\ \bibinfo {author} {\bibfnamefont {X.}~\bibnamefont {Qi}},\
  }\href@noop {} {\bibfield  {journal} {\bibinfo  {journal} {Comptes Rendus
  Physique}\ }\textbf {\bibinfo {volume} {14}},\ \bibinfo {pages} {857}
  (\bibinfo {year} {2013})}\BibitemShut {NoStop}%
\bibitem [{\citenamefont {Vafek}\ and\ \citenamefont
  {Vishwanath}(2014)}]{vafek2014dirac}%
  \BibitemOpen
  \bibfield  {author} {\bibinfo {author} {\bibfnamefont {O.}~\bibnamefont
  {Vafek}}\ and\ \bibinfo {author} {\bibfnamefont {A.}~\bibnamefont
  {Vishwanath}},\ }\href@noop {} {\bibfield  {journal} {\bibinfo  {journal}
  {Annu. Rev. Condens. Matter Phys.}\ }\textbf {\bibinfo {volume} {5}},\
  \bibinfo {pages} {83} (\bibinfo {year} {2014})}\BibitemShut {NoStop}%
\bibitem [{\citenamefont {Hasan}\ \emph {et~al.}(2017)\citenamefont {Hasan},
  \citenamefont {Xu}, \citenamefont {Belopolski},\ and\ \citenamefont
  {Huang}}]{hasan2017discovery}%
  \BibitemOpen
  \bibfield  {author} {\bibinfo {author} {\bibfnamefont {M.~Z.}\ \bibnamefont
  {Hasan}}, \bibinfo {author} {\bibfnamefont {S.-Y.}\ \bibnamefont {Xu}},
  \bibinfo {author} {\bibfnamefont {I.}~\bibnamefont {Belopolski}}, \ and\
  \bibinfo {author} {\bibfnamefont {S.-M.}\ \bibnamefont {Huang}},\ }\href@noop
  {} {\bibfield  {journal} {\bibinfo  {journal} {Annual Review of Condensed
  Matter Physics}\ }\textbf {\bibinfo {volume} {8}},\ \bibinfo {pages} {289}
  (\bibinfo {year} {2017})}\BibitemShut {NoStop}%
\bibitem [{\citenamefont {Lv}\ \emph {et~al.}(2015)\citenamefont {Lv},
  \citenamefont {Weng}, \citenamefont {Fu}, \citenamefont {Wang}, \citenamefont
  {Miao}, \citenamefont {Ma}, \citenamefont {Richard}, \citenamefont {Huang},
  \citenamefont {Zhao}, \citenamefont {Chen} \emph
  {et~al.}}]{lv2015experimental}%
  \BibitemOpen
  \bibfield  {author} {\bibinfo {author} {\bibfnamefont {B.}~\bibnamefont
  {Lv}}, \bibinfo {author} {\bibfnamefont {H.}~\bibnamefont {Weng}}, \bibinfo
  {author} {\bibfnamefont {B.}~\bibnamefont {Fu}}, \bibinfo {author}
  {\bibfnamefont {X.~P.}\ \bibnamefont {Wang}}, \bibinfo {author}
  {\bibfnamefont {H.}~\bibnamefont {Miao}}, \bibinfo {author} {\bibfnamefont
  {J.}~\bibnamefont {Ma}}, \bibinfo {author} {\bibfnamefont {P.}~\bibnamefont
  {Richard}}, \bibinfo {author} {\bibfnamefont {X.}~\bibnamefont {Huang}},
  \bibinfo {author} {\bibfnamefont {L.}~\bibnamefont {Zhao}}, \bibinfo {author}
  {\bibfnamefont {G.}~\bibnamefont {Chen}},  \emph {et~al.},\ }\href@noop {}
  {\bibfield  {journal} {\bibinfo  {journal} {Physical Review X}\ }\textbf
  {\bibinfo {volume} {5}},\ \bibinfo {pages} {031013} (\bibinfo {year}
  {2015})}\BibitemShut {NoStop}%
\bibitem [{\citenamefont {Yang}\ \emph {et~al.}(2015)\citenamefont {Yang},
  \citenamefont {Liu}, \citenamefont {Sun}, \citenamefont {Peng}, \citenamefont
  {Yang}, \citenamefont {Zhang}, \citenamefont {Zhou}, \citenamefont {Zhang},
  \citenamefont {Guo}, \citenamefont {Rahn} \emph {et~al.}}]{yang2015weyl}%
  \BibitemOpen
  \bibfield  {author} {\bibinfo {author} {\bibfnamefont {L.}~\bibnamefont
  {Yang}}, \bibinfo {author} {\bibfnamefont {Z.}~\bibnamefont {Liu}}, \bibinfo
  {author} {\bibfnamefont {Y.}~\bibnamefont {Sun}}, \bibinfo {author}
  {\bibfnamefont {H.}~\bibnamefont {Peng}}, \bibinfo {author} {\bibfnamefont
  {H.}~\bibnamefont {Yang}}, \bibinfo {author} {\bibfnamefont {T.}~\bibnamefont
  {Zhang}}, \bibinfo {author} {\bibfnamefont {B.}~\bibnamefont {Zhou}},
  \bibinfo {author} {\bibfnamefont {Y.}~\bibnamefont {Zhang}}, \bibinfo
  {author} {\bibfnamefont {Y.}~\bibnamefont {Guo}}, \bibinfo {author}
  {\bibfnamefont {M.}~\bibnamefont {Rahn}},  \emph {et~al.},\ }\href@noop {}
  {\bibfield  {journal} {\bibinfo  {journal} {Nature physics}\ }\textbf
  {\bibinfo {volume} {11}},\ \bibinfo {pages} {728} (\bibinfo {year}
  {2015})}\BibitemShut {NoStop}%
\bibitem [{\citenamefont {Liu}\ \emph {et~al.}(2019)\citenamefont {Liu},
  \citenamefont {Liang}, \citenamefont {Liu}, \citenamefont {Xu}, \citenamefont
  {Li}, \citenamefont {Chen}, \citenamefont {Pei}, \citenamefont {Shi},
  \citenamefont {Mo}, \citenamefont {Dudin} \emph {et~al.}}]{liu2019magnetic}%
  \BibitemOpen
  \bibfield  {author} {\bibinfo {author} {\bibfnamefont {D.}~\bibnamefont
  {Liu}}, \bibinfo {author} {\bibfnamefont {A.}~\bibnamefont {Liang}}, \bibinfo
  {author} {\bibfnamefont {E.}~\bibnamefont {Liu}}, \bibinfo {author}
  {\bibfnamefont {Q.}~\bibnamefont {Xu}}, \bibinfo {author} {\bibfnamefont
  {Y.}~\bibnamefont {Li}}, \bibinfo {author} {\bibfnamefont {C.}~\bibnamefont
  {Chen}}, \bibinfo {author} {\bibfnamefont {D.}~\bibnamefont {Pei}}, \bibinfo
  {author} {\bibfnamefont {W.}~\bibnamefont {Shi}}, \bibinfo {author}
  {\bibfnamefont {S.}~\bibnamefont {Mo}}, \bibinfo {author} {\bibfnamefont
  {P.}~\bibnamefont {Dudin}},  \emph {et~al.},\ }\href@noop {} {\bibfield
  {journal} {\bibinfo  {journal} {Science}\ }\textbf {\bibinfo {volume}
  {365}},\ \bibinfo {pages} {1282} (\bibinfo {year} {2019})}\BibitemShut
  {NoStop}%
\bibitem [{\citenamefont {Xu}\ \emph {et~al.}(2015{\natexlab{a}})\citenamefont
  {Xu}, \citenamefont {Belopolski}, \citenamefont {Alidoust}, \citenamefont
  {Neupane}, \citenamefont {Bian}, \citenamefont {Zhang}, \citenamefont
  {Sankar}, \citenamefont {Chang}, \citenamefont {Yuan}, \citenamefont {Lee}
  \emph {et~al.}}]{xu2015discovery}%
  \BibitemOpen
  \bibfield  {author} {\bibinfo {author} {\bibfnamefont {S.-Y.}\ \bibnamefont
  {Xu}}, \bibinfo {author} {\bibfnamefont {I.}~\bibnamefont {Belopolski}},
  \bibinfo {author} {\bibfnamefont {N.}~\bibnamefont {Alidoust}}, \bibinfo
  {author} {\bibfnamefont {M.}~\bibnamefont {Neupane}}, \bibinfo {author}
  {\bibfnamefont {G.}~\bibnamefont {Bian}}, \bibinfo {author} {\bibfnamefont
  {C.}~\bibnamefont {Zhang}}, \bibinfo {author} {\bibfnamefont
  {R.}~\bibnamefont {Sankar}}, \bibinfo {author} {\bibfnamefont
  {G.}~\bibnamefont {Chang}}, \bibinfo {author} {\bibfnamefont
  {Z.}~\bibnamefont {Yuan}}, \bibinfo {author} {\bibfnamefont {C.-C.}\
  \bibnamefont {Lee}},  \emph {et~al.},\ }\href@noop {} {\bibfield  {journal}
  {\bibinfo  {journal} {Science}\ }\textbf {\bibinfo {volume} {349}},\ \bibinfo
  {pages} {613} (\bibinfo {year} {2015}{\natexlab{a}})}\BibitemShut {NoStop}%
\bibitem [{\citenamefont {Xu}\ \emph {et~al.}(2015{\natexlab{b}})\citenamefont
  {Xu}, \citenamefont {Belopolski}, \citenamefont {Sanchez}, \citenamefont
  {Zhang}, \citenamefont {Chang}, \citenamefont {Guo}, \citenamefont {Bian},
  \citenamefont {Yuan}, \citenamefont {Lu}, \citenamefont {Chang} \emph
  {et~al.}}]{xu2015experimental}%
  \BibitemOpen
  \bibfield  {author} {\bibinfo {author} {\bibfnamefont {S.-Y.}\ \bibnamefont
  {Xu}}, \bibinfo {author} {\bibfnamefont {I.}~\bibnamefont {Belopolski}},
  \bibinfo {author} {\bibfnamefont {D.~S.}\ \bibnamefont {Sanchez}}, \bibinfo
  {author} {\bibfnamefont {C.}~\bibnamefont {Zhang}}, \bibinfo {author}
  {\bibfnamefont {G.}~\bibnamefont {Chang}}, \bibinfo {author} {\bibfnamefont
  {C.}~\bibnamefont {Guo}}, \bibinfo {author} {\bibfnamefont {G.}~\bibnamefont
  {Bian}}, \bibinfo {author} {\bibfnamefont {Z.}~\bibnamefont {Yuan}}, \bibinfo
  {author} {\bibfnamefont {H.}~\bibnamefont {Lu}}, \bibinfo {author}
  {\bibfnamefont {T.-R.}\ \bibnamefont {Chang}},  \emph {et~al.},\ }\href@noop
  {} {\bibfield  {journal} {\bibinfo  {journal} {Science advances}\ }\textbf
  {\bibinfo {volume} {1}},\ \bibinfo {pages} {e1501092} (\bibinfo {year}
  {2015}{\natexlab{b}})}\BibitemShut {NoStop}%
\bibitem [{\citenamefont {Huang}\ \emph {et~al.}(2015)\citenamefont {Huang},
  \citenamefont {Zhao}, \citenamefont {Long}, \citenamefont {Wang},
  \citenamefont {Chen}, \citenamefont {Yang}, \citenamefont {Liang},
  \citenamefont {Xue}, \citenamefont {Weng}, \citenamefont {Fang} \emph
  {et~al.}}]{huang2015observation}%
  \BibitemOpen
  \bibfield  {author} {\bibinfo {author} {\bibfnamefont {X.}~\bibnamefont
  {Huang}}, \bibinfo {author} {\bibfnamefont {L.}~\bibnamefont {Zhao}},
  \bibinfo {author} {\bibfnamefont {Y.}~\bibnamefont {Long}}, \bibinfo {author}
  {\bibfnamefont {P.}~\bibnamefont {Wang}}, \bibinfo {author} {\bibfnamefont
  {D.}~\bibnamefont {Chen}}, \bibinfo {author} {\bibfnamefont {Z.}~\bibnamefont
  {Yang}}, \bibinfo {author} {\bibfnamefont {H.}~\bibnamefont {Liang}},
  \bibinfo {author} {\bibfnamefont {M.}~\bibnamefont {Xue}}, \bibinfo {author}
  {\bibfnamefont {H.}~\bibnamefont {Weng}}, \bibinfo {author} {\bibfnamefont
  {Z.}~\bibnamefont {Fang}},  \emph {et~al.},\ }\href@noop {} {\bibfield
  {journal} {\bibinfo  {journal} {Physical Review X}\ }\textbf {\bibinfo
  {volume} {5}},\ \bibinfo {pages} {031023} (\bibinfo {year}
  {2015})}\BibitemShut {NoStop}%
\bibitem [{\citenamefont {Burkov}(2015)}]{burkov2015chiral}%
  \BibitemOpen
  \bibfield  {author} {\bibinfo {author} {\bibfnamefont {A.}~\bibnamefont
  {Burkov}},\ }\href@noop {} {\bibfield  {journal} {\bibinfo  {journal}
  {Journal of Physics: Condensed Matter}\ }\textbf {\bibinfo {volume} {27}},\
  \bibinfo {pages} {113201} (\bibinfo {year} {2015})}\BibitemShut {NoStop}%
\bibitem [{\citenamefont {Gooth}\ \emph {et~al.}(2017)\citenamefont {Gooth},
  \citenamefont {Niemann}, \citenamefont {Meng}, \citenamefont {Grushin},
  \citenamefont {Landsteiner}, \citenamefont {Gotsmann}, \citenamefont
  {Menges}, \citenamefont {Schmidt}, \citenamefont {Shekhar}, \citenamefont
  {S{\"u}{\ss}} \emph {et~al.}}]{gooth2017experimental}%
  \BibitemOpen
  \bibfield  {author} {\bibinfo {author} {\bibfnamefont {J.}~\bibnamefont
  {Gooth}}, \bibinfo {author} {\bibfnamefont {A.~C.}\ \bibnamefont {Niemann}},
  \bibinfo {author} {\bibfnamefont {T.}~\bibnamefont {Meng}}, \bibinfo {author}
  {\bibfnamefont {A.~G.}\ \bibnamefont {Grushin}}, \bibinfo {author}
  {\bibfnamefont {K.}~\bibnamefont {Landsteiner}}, \bibinfo {author}
  {\bibfnamefont {B.}~\bibnamefont {Gotsmann}}, \bibinfo {author}
  {\bibfnamefont {F.}~\bibnamefont {Menges}}, \bibinfo {author} {\bibfnamefont
  {M.}~\bibnamefont {Schmidt}}, \bibinfo {author} {\bibfnamefont
  {C.}~\bibnamefont {Shekhar}}, \bibinfo {author} {\bibfnamefont
  {V.}~\bibnamefont {S{\"u}{\ss}}},  \emph {et~al.},\ }\href@noop {} {\bibfield
   {journal} {\bibinfo  {journal} {Nature}\ }\textbf {\bibinfo {volume}
  {547}},\ \bibinfo {pages} {324} (\bibinfo {year} {2017})}\BibitemShut
  {NoStop}%
\bibitem [{\citenamefont {Liu}\ \emph {et~al.}(2013)\citenamefont {Liu},
  \citenamefont {Ye},\ and\ \citenamefont {Qi}}]{liu2013chiral}%
  \BibitemOpen
  \bibfield  {author} {\bibinfo {author} {\bibfnamefont {C.-X.}\ \bibnamefont
  {Liu}}, \bibinfo {author} {\bibfnamefont {P.}~\bibnamefont {Ye}}, \ and\
  \bibinfo {author} {\bibfnamefont {X.-L.}\ \bibnamefont {Qi}},\ }\href@noop {}
  {\bibfield  {journal} {\bibinfo  {journal} {Physical Review B}\ }\textbf
  {\bibinfo {volume} {87}},\ \bibinfo {pages} {235306} (\bibinfo {year}
  {2013})}\BibitemShut {NoStop}%
\bibitem [{\citenamefont {Pikulin}\ \emph {et~al.}(2016)\citenamefont
  {Pikulin}, \citenamefont {Chen},\ and\ \citenamefont
  {Franz}}]{pikulin2016chiral}%
  \BibitemOpen
  \bibfield  {author} {\bibinfo {author} {\bibfnamefont {D.}~\bibnamefont
  {Pikulin}}, \bibinfo {author} {\bibfnamefont {A.}~\bibnamefont {Chen}}, \
  and\ \bibinfo {author} {\bibfnamefont {M.}~\bibnamefont {Franz}},\
  }\href@noop {} {\bibfield  {journal} {\bibinfo  {journal} {Physical Review
  X}\ }\textbf {\bibinfo {volume} {6}},\ \bibinfo {pages} {041021} (\bibinfo
  {year} {2016})}\BibitemShut {NoStop}%
\bibitem [{\citenamefont {Grushin}\ \emph {et~al.}(2016)\citenamefont
  {Grushin}, \citenamefont {Venderbos}, \citenamefont {Vishwanath},\ and\
  \citenamefont {Ilan}}]{grushin2016inhomogeneous}%
  \BibitemOpen
  \bibfield  {author} {\bibinfo {author} {\bibfnamefont {A.~G.}\ \bibnamefont
  {Grushin}}, \bibinfo {author} {\bibfnamefont {J.~W.}\ \bibnamefont
  {Venderbos}}, \bibinfo {author} {\bibfnamefont {A.}~\bibnamefont
  {Vishwanath}}, \ and\ \bibinfo {author} {\bibfnamefont {R.}~\bibnamefont
  {Ilan}},\ }\href@noop {} {\bibfield  {journal} {\bibinfo  {journal} {Physical
  Review X}\ }\textbf {\bibinfo {volume} {6}},\ \bibinfo {pages} {041046}
  (\bibinfo {year} {2016})}\BibitemShut {NoStop}%
\bibitem [{\citenamefont {Jia}\ \emph {et~al.}(2019)\citenamefont {Jia},
  \citenamefont {Zhang}, \citenamefont {Gao}, \citenamefont {Guo},
  \citenamefont {Yang}, \citenamefont {Hu}, \citenamefont {Bi}, \citenamefont
  {Xiang}, \citenamefont {Liu},\ and\ \citenamefont
  {Zhang}}]{jia2019observation}%
  \BibitemOpen
  \bibfield  {author} {\bibinfo {author} {\bibfnamefont {H.}~\bibnamefont
  {Jia}}, \bibinfo {author} {\bibfnamefont {R.}~\bibnamefont {Zhang}}, \bibinfo
  {author} {\bibfnamefont {W.}~\bibnamefont {Gao}}, \bibinfo {author}
  {\bibfnamefont {Q.}~\bibnamefont {Guo}}, \bibinfo {author} {\bibfnamefont
  {B.}~\bibnamefont {Yang}}, \bibinfo {author} {\bibfnamefont {J.}~\bibnamefont
  {Hu}}, \bibinfo {author} {\bibfnamefont {Y.}~\bibnamefont {Bi}}, \bibinfo
  {author} {\bibfnamefont {Y.}~\bibnamefont {Xiang}}, \bibinfo {author}
  {\bibfnamefont {C.}~\bibnamefont {Liu}}, \ and\ \bibinfo {author}
  {\bibfnamefont {S.}~\bibnamefont {Zhang}},\ }\href@noop {} {\bibfield
  {journal} {\bibinfo  {journal} {Science}\ }\textbf {\bibinfo {volume}
  {363}},\ \bibinfo {pages} {148} (\bibinfo {year} {2019})}\BibitemShut
  {NoStop}%
\bibitem [{\citenamefont {Peri}\ \emph {et~al.}(2019)\citenamefont {Peri},
  \citenamefont {Serra-Garcia}, \citenamefont {Ilan},\ and\ \citenamefont
  {Huber}}]{peri2019axial}%
  \BibitemOpen
  \bibfield  {author} {\bibinfo {author} {\bibfnamefont {V.}~\bibnamefont
  {Peri}}, \bibinfo {author} {\bibfnamefont {M.}~\bibnamefont {Serra-Garcia}},
  \bibinfo {author} {\bibfnamefont {R.}~\bibnamefont {Ilan}}, \ and\ \bibinfo
  {author} {\bibfnamefont {S.~D.}\ \bibnamefont {Huber}},\ }\href@noop {}
  {\bibfield  {journal} {\bibinfo  {journal} {Nature Physics}\ }\textbf
  {\bibinfo {volume} {15}},\ \bibinfo {pages} {357} (\bibinfo {year}
  {2019})}\BibitemShut {NoStop}%
\bibitem [{\citenamefont {Ilan}\ \emph {et~al.}(2020)\citenamefont {Ilan},
  \citenamefont {Grushin},\ and\ \citenamefont {Pikulin}}]{ilan2020}%
  \BibitemOpen
  \bibfield  {author} {\bibinfo {author} {\bibfnamefont {R.}~\bibnamefont
  {Ilan}}, \bibinfo {author} {\bibfnamefont {A.~G.}\ \bibnamefont {Grushin}}, \
  and\ \bibinfo {author} {\bibfnamefont {D.~I.}\ \bibnamefont {Pikulin}},\
  }\href {https://doi.org/10.1038/s42254-019-0121-8} {\bibfield  {journal}
  {\bibinfo  {journal} {Nature Reviews Physics}\ }\textbf {\bibinfo {volume}
  {2}},\ \bibinfo {pages} {29} (\bibinfo {year} {2020})}\BibitemShut {NoStop}%
\bibitem [{\citenamefont {Weinberg}(1972)}]{weinberg1972gravitation}%
  \BibitemOpen
  \bibfield  {author} {\bibinfo {author} {\bibfnamefont {S.}~\bibnamefont
  {Weinberg}},\ }\href@noop {} {\  (\bibinfo {year} {1972})}\BibitemShut
  {NoStop}%
\bibitem [{\citenamefont {Parker}\ and\ \citenamefont
  {Toms}(2009)}]{parker2009quantum}%
  \BibitemOpen
  \bibfield  {author} {\bibinfo {author} {\bibfnamefont {L.}~\bibnamefont
  {Parker}}\ and\ \bibinfo {author} {\bibfnamefont {D.}~\bibnamefont {Toms}},\
  }\href@noop {} {\emph {\bibinfo {title} {Quantum field theory in curved
  spacetime: quantized fields and gravity}}}\ (\bibinfo  {publisher} {Cambridge
  university press},\ \bibinfo {year} {2009})\BibitemShut {NoStop}%
\bibitem [{\citenamefont {Liang}\ and\ \citenamefont
  {Ojanen}(2019)}]{liang2019curved}%
  \BibitemOpen
  \bibfield  {author} {\bibinfo {author} {\bibfnamefont {L.}~\bibnamefont
  {Liang}}\ and\ \bibinfo {author} {\bibfnamefont {T.}~\bibnamefont {Ojanen}},\
  }\href@noop {} {\bibfield  {journal} {\bibinfo  {journal} {Physical Review
  Research}\ }\textbf {\bibinfo {volume} {1}},\ \bibinfo {pages} {032006}
  (\bibinfo {year} {2019})}\BibitemShut {NoStop}%
\bibitem [{\citenamefont {Weststr{\"o}m}\ and\ \citenamefont
  {Ojanen}(2017)}]{weststrom2017designer}%
  \BibitemOpen
  \bibfield  {author} {\bibinfo {author} {\bibfnamefont {A.}~\bibnamefont
  {Weststr{\"o}m}}\ and\ \bibinfo {author} {\bibfnamefont {T.}~\bibnamefont
  {Ojanen}},\ }\href@noop {} {\bibfield  {journal} {\bibinfo  {journal}
  {Physical Review X}\ }\textbf {\bibinfo {volume} {7}},\ \bibinfo {pages}
  {041026} (\bibinfo {year} {2017})}\BibitemShut {NoStop}%
\bibitem [{\citenamefont {Jia}\ \emph {et~al.}(2020)\citenamefont {Jia},
  \citenamefont {Zhang}, \citenamefont {Gao}, \citenamefont {Zhang},\ and\
  \citenamefont {Chan}}]{jia2020chiral}%
  \BibitemOpen
  \bibfield  {author} {\bibinfo {author} {\bibfnamefont {H.}~\bibnamefont
  {Jia}}, \bibinfo {author} {\bibfnamefont {R.-Y.}\ \bibnamefont {Zhang}},
  \bibinfo {author} {\bibfnamefont {W.}~\bibnamefont {Gao}}, \bibinfo {author}
  {\bibfnamefont {S.}~\bibnamefont {Zhang}}, \ and\ \bibinfo {author}
  {\bibfnamefont {C.}~\bibnamefont {Chan}},\ }\href@noop {} {\bibfield
  {journal} {\bibinfo  {journal} {arXiv preprint arXiv:2009.05954}\ } (\bibinfo
  {year} {2020})}\BibitemShut {NoStop}%
\bibitem [{\citenamefont {Zubkov}(2015)}]{zubkov2015emergent}%
  \BibitemOpen
  \bibfield  {author} {\bibinfo {author} {\bibfnamefont {M.}~\bibnamefont
  {Zubkov}},\ }\href@noop {} {\bibfield  {journal} {\bibinfo  {journal} {Annals
  of Physics}\ }\textbf {\bibinfo {volume} {360}},\ \bibinfo {pages} {655}
  (\bibinfo {year} {2015})}\BibitemShut {NoStop}%
\bibitem [{\citenamefont {Cortijo}\ and\ \citenamefont
  {Zubkov}(2016)}]{cortijo2016emergent}%
  \BibitemOpen
  \bibfield  {author} {\bibinfo {author} {\bibfnamefont {A.}~\bibnamefont
  {Cortijo}}\ and\ \bibinfo {author} {\bibfnamefont {M.}~\bibnamefont
  {Zubkov}},\ }\href@noop {} {\bibfield  {journal} {\bibinfo  {journal} {Annals
  of Physics}\ }\textbf {\bibinfo {volume} {366}},\ \bibinfo {pages} {45}
  (\bibinfo {year} {2016})}\BibitemShut {NoStop}%
\bibitem [{\citenamefont {Hughes}\ \emph {et~al.}(2013)\citenamefont {Hughes},
  \citenamefont {Leigh},\ and\ \citenamefont {Parrikar}}]{hughes2013torsional}%
  \BibitemOpen
  \bibfield  {author} {\bibinfo {author} {\bibfnamefont {T.~L.}\ \bibnamefont
  {Hughes}}, \bibinfo {author} {\bibfnamefont {R.~G.}\ \bibnamefont {Leigh}}, \
  and\ \bibinfo {author} {\bibfnamefont {O.}~\bibnamefont {Parrikar}},\
  }\href@noop {} {\bibfield  {journal} {\bibinfo  {journal} {Physical Review
  D}\ }\textbf {\bibinfo {volume} {88}},\ \bibinfo {pages} {025040} (\bibinfo
  {year} {2013})}\BibitemShut {NoStop}%
\bibitem [{\citenamefont {Parrikar}\ \emph {et~al.}(2014)\citenamefont
  {Parrikar}, \citenamefont {Hughes},\ and\ \citenamefont
  {Leigh}}]{parrikar2014torsion}%
  \BibitemOpen
  \bibfield  {author} {\bibinfo {author} {\bibfnamefont {O.}~\bibnamefont
  {Parrikar}}, \bibinfo {author} {\bibfnamefont {T.~L.}\ \bibnamefont
  {Hughes}}, \ and\ \bibinfo {author} {\bibfnamefont {R.~G.}\ \bibnamefont
  {Leigh}},\ }\href@noop {} {\bibfield  {journal} {\bibinfo  {journal}
  {Physical Review D}\ }\textbf {\bibinfo {volume} {90}},\ \bibinfo {pages}
  {105004} (\bibinfo {year} {2014})}\BibitemShut {NoStop}%
\bibitem [{\citenamefont {Huang}\ \emph
  {et~al.}(2020{\natexlab{a}})\citenamefont {Huang}, \citenamefont {Han},\ and\
  \citenamefont {Stone}}]{huang2020nieh}%
  \BibitemOpen
  \bibfield  {author} {\bibinfo {author} {\bibfnamefont {Z.-M.}\ \bibnamefont
  {Huang}}, \bibinfo {author} {\bibfnamefont {B.}~\bibnamefont {Han}}, \ and\
  \bibinfo {author} {\bibfnamefont {M.}~\bibnamefont {Stone}},\ }\href@noop {}
  {\bibfield  {journal} {\bibinfo  {journal} {Physical Review B}\ }\textbf
  {\bibinfo {volume} {101}},\ \bibinfo {pages} {125201} (\bibinfo {year}
  {2020}{\natexlab{a}})}\BibitemShut {NoStop}%
\bibitem [{\citenamefont {Nissinen}\ and\ \citenamefont
  {Volovik}(2019)}]{nissinen2019thermal}%
  \BibitemOpen
  \bibfield  {author} {\bibinfo {author} {\bibfnamefont {J.}~\bibnamefont
  {Nissinen}}\ and\ \bibinfo {author} {\bibfnamefont {G.~E.}\ \bibnamefont
  {Volovik}},\ }\href@noop {} {\bibfield  {journal} {\bibinfo  {journal} {JETP
  Letters}\ }\textbf {\bibinfo {volume} {110}},\ \bibinfo {pages} {789}
  (\bibinfo {year} {2019})}\BibitemShut {NoStop}%
\bibitem [{\citenamefont {Nissinen}(2020)}]{nissinen2020emergent}%
  \BibitemOpen
  \bibfield  {author} {\bibinfo {author} {\bibfnamefont {J.}~\bibnamefont
  {Nissinen}},\ }\href@noop {} {\bibfield  {journal} {\bibinfo  {journal}
  {Physical review letters}\ }\textbf {\bibinfo {volume} {124}},\ \bibinfo
  {pages} {117002} (\bibinfo {year} {2020})}\BibitemShut {NoStop}%
\bibitem [{\citenamefont {Nissinen}\ and\ \citenamefont
  {Volovik}(2020)}]{nissinen2020thermal}%
  \BibitemOpen
  \bibfield  {author} {\bibinfo {author} {\bibfnamefont {J.}~\bibnamefont
  {Nissinen}}\ and\ \bibinfo {author} {\bibfnamefont {G.}~\bibnamefont
  {Volovik}},\ }\href@noop {} {\bibfield  {journal} {\bibinfo  {journal}
  {Physical Review Research}\ }\textbf {\bibinfo {volume} {2}},\ \bibinfo
  {pages} {033269} (\bibinfo {year} {2020})}\BibitemShut {NoStop}%
\bibitem [{\citenamefont {Laurila}\ and\ \citenamefont
  {Nissinen}(2020)}]{laurila2020torsional}%
  \BibitemOpen
  \bibfield  {author} {\bibinfo {author} {\bibfnamefont {S.}~\bibnamefont
  {Laurila}}\ and\ \bibinfo {author} {\bibfnamefont {J.}~\bibnamefont
  {Nissinen}},\ }\href@noop {} {\bibfield  {journal} {\bibinfo  {journal}
  {Physical Review B}\ }\textbf {\bibinfo {volume} {102}},\ \bibinfo {pages}
  {235163} (\bibinfo {year} {2020})}\BibitemShut {NoStop}%
\bibitem [{\citenamefont {Huang}\ \emph
  {et~al.}(2020{\natexlab{b}})\citenamefont {Huang}, \citenamefont {Han},\ and\
  \citenamefont {Stone}}]{huang2020hamiltonian}%
  \BibitemOpen
  \bibfield  {author} {\bibinfo {author} {\bibfnamefont {Z.-M.}\ \bibnamefont
  {Huang}}, \bibinfo {author} {\bibfnamefont {B.}~\bibnamefont {Han}}, \ and\
  \bibinfo {author} {\bibfnamefont {M.}~\bibnamefont {Stone}},\ }\href@noop {}
  {\bibfield  {journal} {\bibinfo  {journal} {Physical Review B}\ }\textbf
  {\bibinfo {volume} {101}},\ \bibinfo {pages} {165201} (\bibinfo {year}
  {2020}{\natexlab{b}})}\BibitemShut {NoStop}%
\bibitem [{\citenamefont {Huang}\ and\ \citenamefont
  {Han}(2020)}]{huang2020torsional}%
  \BibitemOpen
  \bibfield  {author} {\bibinfo {author} {\bibfnamefont {Z.-M.}\ \bibnamefont
  {Huang}}\ and\ \bibinfo {author} {\bibfnamefont {B.}~\bibnamefont {Han}},\
  }\href@noop {} {\bibfield  {journal} {\bibinfo  {journal} {arXiv preprint
  arXiv:2003.04853}\ } (\bibinfo {year} {2020})}\BibitemShut {NoStop}%
\bibitem [{\citenamefont {Liang}\ and\ \citenamefont
  {Ojanen}(2020)}]{liang2020topological}%
  \BibitemOpen
  \bibfield  {author} {\bibinfo {author} {\bibfnamefont {L.}~\bibnamefont
  {Liang}}\ and\ \bibinfo {author} {\bibfnamefont {T.}~\bibnamefont {Ojanen}},\
  }\href@noop {} {\bibfield  {journal} {\bibinfo  {journal} {Physical Review
  Research}\ }\textbf {\bibinfo {volume} {2}},\ \bibinfo {pages} {022016}
  (\bibinfo {year} {2020})}\BibitemShut {NoStop}%
\bibitem [{\citenamefont {Ferreiros}\ \emph {et~al.}(2019)\citenamefont
  {Ferreiros}, \citenamefont {Kedem}, \citenamefont {Bergholtz},\ and\
  \citenamefont {Bardarson}}]{ferreiros2019mixed}%
  \BibitemOpen
  \bibfield  {author} {\bibinfo {author} {\bibfnamefont {Y.}~\bibnamefont
  {Ferreiros}}, \bibinfo {author} {\bibfnamefont {Y.}~\bibnamefont {Kedem}},
  \bibinfo {author} {\bibfnamefont {E.~J.}\ \bibnamefont {Bergholtz}}, \ and\
  \bibinfo {author} {\bibfnamefont {J.~H.}\ \bibnamefont {Bardarson}},\
  }\href@noop {} {\bibfield  {journal} {\bibinfo  {journal} {Physical review
  letters}\ }\textbf {\bibinfo {volume} {122}},\ \bibinfo {pages} {056601}
  (\bibinfo {year} {2019})}\BibitemShut {NoStop}%
\bibitem [{\citenamefont {Chang}\ \emph {et~al.}(2018)\citenamefont {Chang},
  \citenamefont {Wieder}, \citenamefont {Schindler}, \citenamefont {Sanchez},
  \citenamefont {Belopolski}, \citenamefont {Huang}, \citenamefont {Singh},
  \citenamefont {Wu}, \citenamefont {Chang}, \citenamefont {Neupert} \emph
  {et~al.}}]{chang2018topological}%
  \BibitemOpen
  \bibfield  {author} {\bibinfo {author} {\bibfnamefont {G.}~\bibnamefont
  {Chang}}, \bibinfo {author} {\bibfnamefont {B.~J.}\ \bibnamefont {Wieder}},
  \bibinfo {author} {\bibfnamefont {F.}~\bibnamefont {Schindler}}, \bibinfo
  {author} {\bibfnamefont {D.~S.}\ \bibnamefont {Sanchez}}, \bibinfo {author}
  {\bibfnamefont {I.}~\bibnamefont {Belopolski}}, \bibinfo {author}
  {\bibfnamefont {S.-M.}\ \bibnamefont {Huang}}, \bibinfo {author}
  {\bibfnamefont {B.}~\bibnamefont {Singh}}, \bibinfo {author} {\bibfnamefont
  {D.}~\bibnamefont {Wu}}, \bibinfo {author} {\bibfnamefont {T.-R.}\
  \bibnamefont {Chang}}, \bibinfo {author} {\bibfnamefont {T.}~\bibnamefont
  {Neupert}},  \emph {et~al.},\ }\href@noop {} {\bibfield  {journal} {\bibinfo
  {journal} {Nature materials}\ }\textbf {\bibinfo {volume} {17}},\ \bibinfo
  {pages} {978} (\bibinfo {year} {2018})}\BibitemShut {NoStop}%
\bibitem [{\citenamefont {Sanchez}\ \emph {et~al.}(2019)\citenamefont
  {Sanchez}, \citenamefont {Belopolski}, \citenamefont {Cochran}, \citenamefont
  {Xu}, \citenamefont {Yin}, \citenamefont {Chang}, \citenamefont {Xie},
  \citenamefont {Manna}, \citenamefont {S{\"u}{\ss}}, \citenamefont {Huang}
  \emph {et~al.}}]{sanchez2019topological}%
  \BibitemOpen
  \bibfield  {author} {\bibinfo {author} {\bibfnamefont {D.~S.}\ \bibnamefont
  {Sanchez}}, \bibinfo {author} {\bibfnamefont {I.}~\bibnamefont {Belopolski}},
  \bibinfo {author} {\bibfnamefont {T.~A.}\ \bibnamefont {Cochran}}, \bibinfo
  {author} {\bibfnamefont {X.}~\bibnamefont {Xu}}, \bibinfo {author}
  {\bibfnamefont {J.-X.}\ \bibnamefont {Yin}}, \bibinfo {author} {\bibfnamefont
  {G.}~\bibnamefont {Chang}}, \bibinfo {author} {\bibfnamefont
  {W.}~\bibnamefont {Xie}}, \bibinfo {author} {\bibfnamefont {K.}~\bibnamefont
  {Manna}}, \bibinfo {author} {\bibfnamefont {V.}~\bibnamefont {S{\"u}{\ss}}},
  \bibinfo {author} {\bibfnamefont {C.-Y.}\ \bibnamefont {Huang}},  \emph
  {et~al.},\ }\href@noop {} {\bibfield  {journal} {\bibinfo  {journal}
  {Nature}\ }\textbf {\bibinfo {volume} {567}},\ \bibinfo {pages} {500}
  (\bibinfo {year} {2019})}\BibitemShut {NoStop}%
\bibitem [{\citenamefont {Schr{\"o}ter}\ \emph {et~al.}(2019)\citenamefont
  {Schr{\"o}ter}, \citenamefont {Pei}, \citenamefont {Vergniory}, \citenamefont
  {Sun}, \citenamefont {Manna}, \citenamefont {De~Juan}, \citenamefont
  {Krieger}, \citenamefont {S{\"u}ss}, \citenamefont {Schmidt}, \citenamefont
  {Dudin} \emph {et~al.}}]{schroter2019chiral}%
  \BibitemOpen
  \bibfield  {author} {\bibinfo {author} {\bibfnamefont {N.~B.}\ \bibnamefont
  {Schr{\"o}ter}}, \bibinfo {author} {\bibfnamefont {D.}~\bibnamefont {Pei}},
  \bibinfo {author} {\bibfnamefont {M.~G.}\ \bibnamefont {Vergniory}}, \bibinfo
  {author} {\bibfnamefont {Y.}~\bibnamefont {Sun}}, \bibinfo {author}
  {\bibfnamefont {K.}~\bibnamefont {Manna}}, \bibinfo {author} {\bibfnamefont
  {F.}~\bibnamefont {De~Juan}}, \bibinfo {author} {\bibfnamefont {J.~A.}\
  \bibnamefont {Krieger}}, \bibinfo {author} {\bibfnamefont {V.}~\bibnamefont
  {S{\"u}ss}}, \bibinfo {author} {\bibfnamefont {M.}~\bibnamefont {Schmidt}},
  \bibinfo {author} {\bibfnamefont {P.}~\bibnamefont {Dudin}},  \emph
  {et~al.},\ }\href@noop {} {\bibfield  {journal} {\bibinfo  {journal} {Nature
  Physics}\ }\textbf {\bibinfo {volume} {15}},\ \bibinfo {pages} {759}
  (\bibinfo {year} {2019})}\BibitemShut {NoStop}%
\bibitem [{\citenamefont {Schr{\"o}ter}\ \emph {et~al.}(2020)\citenamefont
  {Schr{\"o}ter}, \citenamefont {Stolz}, \citenamefont {Manna}, \citenamefont
  {De~Juan}, \citenamefont {Vergniory}, \citenamefont {Krieger}, \citenamefont
  {Pei}, \citenamefont {Schmitt}, \citenamefont {Dudin}, \citenamefont {Kim}
  \emph {et~al.}}]{schroter2020observation}%
  \BibitemOpen
  \bibfield  {author} {\bibinfo {author} {\bibfnamefont {N.~B.}\ \bibnamefont
  {Schr{\"o}ter}}, \bibinfo {author} {\bibfnamefont {S.}~\bibnamefont {Stolz}},
  \bibinfo {author} {\bibfnamefont {K.}~\bibnamefont {Manna}}, \bibinfo
  {author} {\bibfnamefont {F.}~\bibnamefont {De~Juan}}, \bibinfo {author}
  {\bibfnamefont {M.~G.}\ \bibnamefont {Vergniory}}, \bibinfo {author}
  {\bibfnamefont {J.~A.}\ \bibnamefont {Krieger}}, \bibinfo {author}
  {\bibfnamefont {D.}~\bibnamefont {Pei}}, \bibinfo {author} {\bibfnamefont
  {T.}~\bibnamefont {Schmitt}}, \bibinfo {author} {\bibfnamefont
  {P.}~\bibnamefont {Dudin}}, \bibinfo {author} {\bibfnamefont {T.~K.}\
  \bibnamefont {Kim}},  \emph {et~al.},\ }\href@noop {} {\bibfield  {journal}
  {\bibinfo  {journal} {Science}\ }\textbf {\bibinfo {volume} {369}},\ \bibinfo
  {pages} {179} (\bibinfo {year} {2020})}\BibitemShut {NoStop}%
\bibitem [{\citenamefont {Li}\ \emph {et~al.}(2019)\citenamefont {Li},
  \citenamefont {Xu}, \citenamefont {Rao}, \citenamefont {Zhou}, \citenamefont
  {Wang}, \citenamefont {Zhou}, \citenamefont {Tian}, \citenamefont {Gao},
  \citenamefont {Li}, \citenamefont {Huang} \emph {et~al.}}]{li2019chiral}%
  \BibitemOpen
  \bibfield  {author} {\bibinfo {author} {\bibfnamefont {H.}~\bibnamefont
  {Li}}, \bibinfo {author} {\bibfnamefont {S.}~\bibnamefont {Xu}}, \bibinfo
  {author} {\bibfnamefont {Z.-C.}\ \bibnamefont {Rao}}, \bibinfo {author}
  {\bibfnamefont {L.-Q.}\ \bibnamefont {Zhou}}, \bibinfo {author}
  {\bibfnamefont {Z.-J.}\ \bibnamefont {Wang}}, \bibinfo {author}
  {\bibfnamefont {S.-M.}\ \bibnamefont {Zhou}}, \bibinfo {author}
  {\bibfnamefont {S.-J.}\ \bibnamefont {Tian}}, \bibinfo {author}
  {\bibfnamefont {S.-Y.}\ \bibnamefont {Gao}}, \bibinfo {author} {\bibfnamefont
  {J.-J.}\ \bibnamefont {Li}}, \bibinfo {author} {\bibfnamefont {Y.-B.}\
  \bibnamefont {Huang}},  \emph {et~al.},\ }\href@noop {} {\bibfield  {journal}
  {\bibinfo  {journal} {Nature communications}\ }\textbf {\bibinfo {volume}
  {10}},\ \bibinfo {pages} {1} (\bibinfo {year} {2019})}\BibitemShut {NoStop}%
\bibitem [{\citenamefont {Hu}\ \emph {et~al.}(2021)\citenamefont {Hu},
  \citenamefont {Yu}, \citenamefont {Garate},\ and\ \citenamefont
  {Liu}}]{hu2021phonon}%
  \BibitemOpen
  \bibfield  {author} {\bibinfo {author} {\bibfnamefont {L.-H.}\ \bibnamefont
  {Hu}}, \bibinfo {author} {\bibfnamefont {J.}~\bibnamefont {Yu}}, \bibinfo
  {author} {\bibfnamefont {I.}~\bibnamefont {Garate}}, \ and\ \bibinfo {author}
  {\bibfnamefont {C.-X.}\ \bibnamefont {Liu}},\ }\href@noop {} {\bibfield
  {journal} {\bibinfo  {journal} {arXiv preprint arXiv:2104.02270}\ } (\bibinfo
  {year} {2021})}\BibitemShut {NoStop}%
\bibitem [{\citenamefont {Hamada}\ \emph {et~al.}(2018)\citenamefont {Hamada},
  \citenamefont {Minamitani}, \citenamefont {Hirayama},\ and\ \citenamefont
  {Murakami}}]{hamada2018phonon}%
  \BibitemOpen
  \bibfield  {author} {\bibinfo {author} {\bibfnamefont {M.}~\bibnamefont
  {Hamada}}, \bibinfo {author} {\bibfnamefont {E.}~\bibnamefont {Minamitani}},
  \bibinfo {author} {\bibfnamefont {M.}~\bibnamefont {Hirayama}}, \ and\
  \bibinfo {author} {\bibfnamefont {S.}~\bibnamefont {Murakami}},\ }\href@noop
  {} {\bibfield  {journal} {\bibinfo  {journal} {Physical review letters}\
  }\textbf {\bibinfo {volume} {121}},\ \bibinfo {pages} {175301} (\bibinfo
  {year} {2018})}\BibitemShut {NoStop}%
\bibitem [{\citenamefont {He}\ \emph {et~al.}(2019)\citenamefont {He},
  \citenamefont {Xu},\ and\ \citenamefont {Law}}]{he2019kramers}%
  \BibitemOpen
  \bibfield  {author} {\bibinfo {author} {\bibfnamefont {W.-Y.}\ \bibnamefont
  {He}}, \bibinfo {author} {\bibfnamefont {X.~Y.}\ \bibnamefont {Xu}}, \ and\
  \bibinfo {author} {\bibfnamefont {K.}~\bibnamefont {Law}},\ }\href@noop {}
  {\bibfield  {journal} {\bibinfo  {journal} {arXiv preprint arXiv:1905.12575}\
  } (\bibinfo {year} {2019})}\BibitemShut {NoStop}%
\bibitem [{\citenamefont {Hayashi}\ and\ \citenamefont
  {Shirafuji}(1979)}]{hayashi1979new}%
  \BibitemOpen
  \bibfield  {author} {\bibinfo {author} {\bibfnamefont {K.}~\bibnamefont
  {Hayashi}}\ and\ \bibinfo {author} {\bibfnamefont {T.}~\bibnamefont
  {Shirafuji}},\ }\href@noop {} {\bibfield  {journal} {\bibinfo  {journal}
  {Physical Review D}\ }\textbf {\bibinfo {volume} {19}},\ \bibinfo {pages}
  {3524} (\bibinfo {year} {1979})}\BibitemShut {NoStop}%
\bibitem [{\citenamefont {Zhang}\ and\ \citenamefont
  {Niu}(2014)}]{zhang2014angular}%
  \BibitemOpen
  \bibfield  {author} {\bibinfo {author} {\bibfnamefont {L.}~\bibnamefont
  {Zhang}}\ and\ \bibinfo {author} {\bibfnamefont {Q.}~\bibnamefont {Niu}},\
  }\href@noop {} {\bibfield  {journal} {\bibinfo  {journal} {Physical Review
  Letters}\ }\textbf {\bibinfo {volume} {112}},\ \bibinfo {pages} {085503}
  (\bibinfo {year} {2014})}\BibitemShut {NoStop}%
\bibitem [{\citenamefont {Hamada}\ and\ \citenamefont
  {Murakami}(2020)}]{hamada2020phonon}%
  \BibitemOpen
  \bibfield  {author} {\bibinfo {author} {\bibfnamefont {M.}~\bibnamefont
  {Hamada}}\ and\ \bibinfo {author} {\bibfnamefont {S.}~\bibnamefont
  {Murakami}},\ }\href@noop {} {\bibfield  {journal} {\bibinfo  {journal}
  {Physical Review B}\ }\textbf {\bibinfo {volume} {101}},\ \bibinfo {pages}
  {144306} (\bibinfo {year} {2020})}\BibitemShut {NoStop}%
\bibitem [{\citenamefont {Zhu}\ \emph {et~al.}(2018)\citenamefont {Zhu},
  \citenamefont {Yi}, \citenamefont {Li}, \citenamefont {Xiao}, \citenamefont
  {Zhang}, \citenamefont {Yang}, \citenamefont {Kaindl}, \citenamefont {Li},
  \citenamefont {Wang},\ and\ \citenamefont {Zhang}}]{zhu2018observation}%
  \BibitemOpen
  \bibfield  {author} {\bibinfo {author} {\bibfnamefont {H.}~\bibnamefont
  {Zhu}}, \bibinfo {author} {\bibfnamefont {J.}~\bibnamefont {Yi}}, \bibinfo
  {author} {\bibfnamefont {M.-Y.}\ \bibnamefont {Li}}, \bibinfo {author}
  {\bibfnamefont {J.}~\bibnamefont {Xiao}}, \bibinfo {author} {\bibfnamefont
  {L.}~\bibnamefont {Zhang}}, \bibinfo {author} {\bibfnamefont {C.-W.}\
  \bibnamefont {Yang}}, \bibinfo {author} {\bibfnamefont {R.~A.}\ \bibnamefont
  {Kaindl}}, \bibinfo {author} {\bibfnamefont {L.-J.}\ \bibnamefont {Li}},
  \bibinfo {author} {\bibfnamefont {Y.}~\bibnamefont {Wang}}, \ and\ \bibinfo
  {author} {\bibfnamefont {X.}~\bibnamefont {Zhang}},\ }\href@noop {}
  {\bibfield  {journal} {\bibinfo  {journal} {Science}\ }\textbf {\bibinfo
  {volume} {359}},\ \bibinfo {pages} {579} (\bibinfo {year}
  {2018})}\BibitemShut {NoStop}%
\bibitem [{\citenamefont {Zhang}\ and\ \citenamefont
  {Niu}(2015)}]{zhang2015chiral}%
  \BibitemOpen
  \bibfield  {author} {\bibinfo {author} {\bibfnamefont {L.}~\bibnamefont
  {Zhang}}\ and\ \bibinfo {author} {\bibfnamefont {Q.}~\bibnamefont {Niu}},\
  }\href@noop {} {\bibfield  {journal} {\bibinfo  {journal} {Physical review
  letters}\ }\textbf {\bibinfo {volume} {115}},\ \bibinfo {pages} {115502}
  (\bibinfo {year} {2015})}\BibitemShut {NoStop}%
\bibitem [{\citenamefont {Edelstein}(1990)}]{edelstein1990spin}%
  \BibitemOpen
  \bibfield  {author} {\bibinfo {author} {\bibfnamefont {V.~M.}\ \bibnamefont
  {Edelstein}},\ }\href@noop {} {\bibfield  {journal} {\bibinfo  {journal}
  {Solid State Communications}\ }\textbf {\bibinfo {volume} {73}},\ \bibinfo
  {pages} {233} (\bibinfo {year} {1990})}\BibitemShut {NoStop}%
\bibitem [{\citenamefont {Rinkel}\ \emph {et~al.}(2017)\citenamefont {Rinkel},
  \citenamefont {Lopes},\ and\ \citenamefont {Garate}}]{rinkel2017signatures}%
  \BibitemOpen
  \bibfield  {author} {\bibinfo {author} {\bibfnamefont {P.}~\bibnamefont
  {Rinkel}}, \bibinfo {author} {\bibfnamefont {P.~L.}\ \bibnamefont {Lopes}}, \
  and\ \bibinfo {author} {\bibfnamefont {I.}~\bibnamefont {Garate}},\
  }\href@noop {} {\bibfield  {journal} {\bibinfo  {journal} {Physical review
  letters}\ }\textbf {\bibinfo {volume} {119}},\ \bibinfo {pages} {107401}
  (\bibinfo {year} {2017})}\BibitemShut {NoStop}%
\bibitem [{\citenamefont {Rinkel}\ \emph {et~al.}(2019)\citenamefont {Rinkel},
  \citenamefont {Lopes},\ and\ \citenamefont {Garate}}]{rinkel2019influence}%
  \BibitemOpen
  \bibfield  {author} {\bibinfo {author} {\bibfnamefont {P.}~\bibnamefont
  {Rinkel}}, \bibinfo {author} {\bibfnamefont {P.~L.}\ \bibnamefont {Lopes}}, \
  and\ \bibinfo {author} {\bibfnamefont {I.}~\bibnamefont {Garate}},\
  }\href@noop {} {\bibfield  {journal} {\bibinfo  {journal} {Physical Review
  B}\ }\textbf {\bibinfo {volume} {99}},\ \bibinfo {pages} {144301} (\bibinfo
  {year} {2019})}\BibitemShut {NoStop}%
\bibitem [{\citenamefont {Heidari}\ \emph {et~al.}(2019)\citenamefont
  {Heidari}, \citenamefont {Cortijo},\ and\ \citenamefont
  {Asgari}}]{heidari2019hall}%
  \BibitemOpen
  \bibfield  {author} {\bibinfo {author} {\bibfnamefont {S.}~\bibnamefont
  {Heidari}}, \bibinfo {author} {\bibfnamefont {A.}~\bibnamefont {Cortijo}}, \
  and\ \bibinfo {author} {\bibfnamefont {R.}~\bibnamefont {Asgari}},\
  }\href@noop {} {\bibfield  {journal} {\bibinfo  {journal} {Physical Review
  B}\ }\textbf {\bibinfo {volume} {100}},\ \bibinfo {pages} {165427} (\bibinfo
  {year} {2019})}\BibitemShut {NoStop}%
\bibitem [{\citenamefont {Chernodub}\ and\ \citenamefont
  {Vozmediano}(2019)}]{chernodub2019chiral}%
  \BibitemOpen
  \bibfield  {author} {\bibinfo {author} {\bibfnamefont {M.}~\bibnamefont
  {Chernodub}}\ and\ \bibinfo {author} {\bibfnamefont {M.~A.}\ \bibnamefont
  {Vozmediano}},\ }\href@noop {} {\bibfield  {journal} {\bibinfo  {journal}
  {Physical Review Research}\ }\textbf {\bibinfo {volume} {1}},\ \bibinfo
  {pages} {032040} (\bibinfo {year} {2019})}\BibitemShut {NoStop}%
\bibitem [{\citenamefont {Yuan}\ \emph {et~al.}(2020)\citenamefont {Yuan},
  \citenamefont {Zhang}, \citenamefont {Zhang}, \citenamefont {Yan},
  \citenamefont {Lyu}, \citenamefont {Zhang}, \citenamefont {Li}, \citenamefont
  {Song}, \citenamefont {Zhao}, \citenamefont {Leng} \emph
  {et~al.}}]{yuan2020discovery}%
  \BibitemOpen
  \bibfield  {author} {\bibinfo {author} {\bibfnamefont {X.}~\bibnamefont
  {Yuan}}, \bibinfo {author} {\bibfnamefont {C.}~\bibnamefont {Zhang}},
  \bibinfo {author} {\bibfnamefont {Y.}~\bibnamefont {Zhang}}, \bibinfo
  {author} {\bibfnamefont {Z.}~\bibnamefont {Yan}}, \bibinfo {author}
  {\bibfnamefont {T.}~\bibnamefont {Lyu}}, \bibinfo {author} {\bibfnamefont
  {M.}~\bibnamefont {Zhang}}, \bibinfo {author} {\bibfnamefont
  {Z.}~\bibnamefont {Li}}, \bibinfo {author} {\bibfnamefont {C.}~\bibnamefont
  {Song}}, \bibinfo {author} {\bibfnamefont {M.}~\bibnamefont {Zhao}}, \bibinfo
  {author} {\bibfnamefont {P.}~\bibnamefont {Leng}},  \emph {et~al.},\
  }\href@noop {} {\bibfield  {journal} {\bibinfo  {journal} {Nature
  communications}\ }\textbf {\bibinfo {volume} {11}},\ \bibinfo {pages} {1}
  (\bibinfo {year} {2020})}\BibitemShut {NoStop}%
\bibitem [{\citenamefont {Sengupta}\ \emph {et~al.}(2020)\citenamefont
  {Sengupta}, \citenamefont {Lhachemi},\ and\ \citenamefont
  {Garate}}]{sengupta2020phonon}%
  \BibitemOpen
  \bibfield  {author} {\bibinfo {author} {\bibfnamefont {S.}~\bibnamefont
  {Sengupta}}, \bibinfo {author} {\bibfnamefont {M.~N.~Y.}\ \bibnamefont
  {Lhachemi}}, \ and\ \bibinfo {author} {\bibfnamefont {I.}~\bibnamefont
  {Garate}},\ }\href@noop {} {\bibfield  {journal} {\bibinfo  {journal}
  {Physical Review Letters}\ }\textbf {\bibinfo {volume} {125}},\ \bibinfo
  {pages} {146402} (\bibinfo {year} {2020})}\BibitemShut {NoStop}%
\bibitem [{\citenamefont {Sukhachov}\ and\ \citenamefont
  {Glazman}(2021)}]{sukhachov2021anomalous}%
  \BibitemOpen
  \bibfield  {author} {\bibinfo {author} {\bibfnamefont {P.}~\bibnamefont
  {Sukhachov}}\ and\ \bibinfo {author} {\bibfnamefont {L.}~\bibnamefont
  {Glazman}},\ }\href@noop {} {\bibfield  {journal} {\bibinfo  {journal} {arXiv
  preprint arXiv:2102.04510}\ } (\bibinfo {year} {2021})}\BibitemShut {NoStop}%
\bibitem [{\citenamefont {Nguyen}\ \emph {et~al.}(2020)\citenamefont {Nguyen},
  \citenamefont {Han}, \citenamefont {Andrejevic}, \citenamefont {Pablo-Pedro},
  \citenamefont {Apte}, \citenamefont {Tsurimaki}, \citenamefont {Ding},
  \citenamefont {Zhang}, \citenamefont {Alatas}, \citenamefont {Alp} \emph
  {et~al.}}]{nguyen2020topological}%
  \BibitemOpen
  \bibfield  {author} {\bibinfo {author} {\bibfnamefont {T.}~\bibnamefont
  {Nguyen}}, \bibinfo {author} {\bibfnamefont {F.}~\bibnamefont {Han}},
  \bibinfo {author} {\bibfnamefont {N.}~\bibnamefont {Andrejevic}}, \bibinfo
  {author} {\bibfnamefont {R.}~\bibnamefont {Pablo-Pedro}}, \bibinfo {author}
  {\bibfnamefont {A.}~\bibnamefont {Apte}}, \bibinfo {author} {\bibfnamefont
  {Y.}~\bibnamefont {Tsurimaki}}, \bibinfo {author} {\bibfnamefont
  {Z.}~\bibnamefont {Ding}}, \bibinfo {author} {\bibfnamefont {K.}~\bibnamefont
  {Zhang}}, \bibinfo {author} {\bibfnamefont {A.}~\bibnamefont {Alatas}},
  \bibinfo {author} {\bibfnamefont {E.~E.}\ \bibnamefont {Alp}},  \emph
  {et~al.},\ }\href@noop {} {\bibfield  {journal} {\bibinfo  {journal}
  {Physical review letters}\ }\textbf {\bibinfo {volume} {124}},\ \bibinfo
  {pages} {236401} (\bibinfo {year} {2020})}\BibitemShut {NoStop}%
\bibitem [{\citenamefont {Yue}\ \emph {et~al.}(2019)\citenamefont {Yue},
  \citenamefont {Chorsi}, \citenamefont {Goyal}, \citenamefont {Schumann},
  \citenamefont {Yang}, \citenamefont {Xu}, \citenamefont {Deng}, \citenamefont
  {Stemmer}, \citenamefont {Schuller},\ and\ \citenamefont
  {Liao}}]{yue2019soft}%
  \BibitemOpen
  \bibfield  {author} {\bibinfo {author} {\bibfnamefont {S.}~\bibnamefont
  {Yue}}, \bibinfo {author} {\bibfnamefont {H.~T.}\ \bibnamefont {Chorsi}},
  \bibinfo {author} {\bibfnamefont {M.}~\bibnamefont {Goyal}}, \bibinfo
  {author} {\bibfnamefont {T.}~\bibnamefont {Schumann}}, \bibinfo {author}
  {\bibfnamefont {R.}~\bibnamefont {Yang}}, \bibinfo {author} {\bibfnamefont
  {T.}~\bibnamefont {Xu}}, \bibinfo {author} {\bibfnamefont {B.}~\bibnamefont
  {Deng}}, \bibinfo {author} {\bibfnamefont {S.}~\bibnamefont {Stemmer}},
  \bibinfo {author} {\bibfnamefont {J.~A.}\ \bibnamefont {Schuller}}, \ and\
  \bibinfo {author} {\bibfnamefont {B.}~\bibnamefont {Liao}},\ }\href@noop {}
  {\bibfield  {journal} {\bibinfo  {journal} {Physical Review Research}\
  }\textbf {\bibinfo {volume} {1}},\ \bibinfo {pages} {033101} (\bibinfo {year}
  {2019})}\BibitemShut {NoStop}%
\bibitem [{\citenamefont {Mentink}\ \emph {et~al.}(2019)\citenamefont
  {Mentink}, \citenamefont {Katsnelson},\ and\ \citenamefont
  {Lemeshko}}]{mentink2019quantum}%
  \BibitemOpen
  \bibfield  {author} {\bibinfo {author} {\bibfnamefont {J.~H.}\ \bibnamefont
  {Mentink}}, \bibinfo {author} {\bibfnamefont {M.}~\bibnamefont {Katsnelson}},
  \ and\ \bibinfo {author} {\bibfnamefont {M.}~\bibnamefont {Lemeshko}},\
  }\href@noop {} {\bibfield  {journal} {\bibinfo  {journal} {Physical Review
  B}\ }\textbf {\bibinfo {volume} {99}},\ \bibinfo {pages} {064428} (\bibinfo
  {year} {2019})}\BibitemShut {NoStop}%
\bibitem [{\citenamefont {Garanin}\ and\ \citenamefont
  {Chudnovsky}(2015)}]{garanin2015angular}%
  \BibitemOpen
  \bibfield  {author} {\bibinfo {author} {\bibfnamefont {D.}~\bibnamefont
  {Garanin}}\ and\ \bibinfo {author} {\bibfnamefont {E.}~\bibnamefont
  {Chudnovsky}},\ }\href@noop {} {\bibfield  {journal} {\bibinfo  {journal}
  {Physical Review B}\ }\textbf {\bibinfo {volume} {92}},\ \bibinfo {pages}
  {024421} (\bibinfo {year} {2015})}\BibitemShut {NoStop}%
\bibitem [{\citenamefont {Bauke}\ \emph {et~al.}(2014)\citenamefont {Bauke},
  \citenamefont {Ahrens}, \citenamefont {Keitel},\ and\ \citenamefont
  {Grobe}}]{bauke2014electron}%
  \BibitemOpen
  \bibfield  {author} {\bibinfo {author} {\bibfnamefont {H.}~\bibnamefont
  {Bauke}}, \bibinfo {author} {\bibfnamefont {S.}~\bibnamefont {Ahrens}},
  \bibinfo {author} {\bibfnamefont {C.~H.}\ \bibnamefont {Keitel}}, \ and\
  \bibinfo {author} {\bibfnamefont {R.}~\bibnamefont {Grobe}},\ }\href@noop {}
  {\bibfield  {journal} {\bibinfo  {journal} {New Journal of Physics}\ }\textbf
  {\bibinfo {volume} {16}},\ \bibinfo {pages} {103028} (\bibinfo {year}
  {2014})}\BibitemShut {NoStop}%
\bibitem [{\citenamefont {Nakane}\ and\ \citenamefont
  {Kohno}(2018)}]{nakane2018angular}%
  \BibitemOpen
  \bibfield  {author} {\bibinfo {author} {\bibfnamefont {J.~J.}\ \bibnamefont
  {Nakane}}\ and\ \bibinfo {author} {\bibfnamefont {H.}~\bibnamefont {Kohno}},\
  }\href@noop {} {\bibfield  {journal} {\bibinfo  {journal} {Physical Review
  B}\ }\textbf {\bibinfo {volume} {97}},\ \bibinfo {pages} {174403} (\bibinfo
  {year} {2018})}\BibitemShut {NoStop}%
\bibitem [{\citenamefont {Holanda}\ \emph {et~al.}(2018)\citenamefont
  {Holanda}, \citenamefont {Maior}, \citenamefont {Azevedo},\ and\
  \citenamefont {Rezende}}]{holanda2018detecting}%
  \BibitemOpen
  \bibfield  {author} {\bibinfo {author} {\bibfnamefont {J.}~\bibnamefont
  {Holanda}}, \bibinfo {author} {\bibfnamefont {D.}~\bibnamefont {Maior}},
  \bibinfo {author} {\bibfnamefont {A.}~\bibnamefont {Azevedo}}, \ and\
  \bibinfo {author} {\bibfnamefont {S.}~\bibnamefont {Rezende}},\ }\href@noop
  {} {\bibfield  {journal} {\bibinfo  {journal} {Nature Physics}\ }\textbf
  {\bibinfo {volume} {14}},\ \bibinfo {pages} {500} (\bibinfo {year}
  {2018})}\BibitemShut {NoStop}%
\bibitem [{\citenamefont {Thingstad}\ \emph {et~al.}(2019)\citenamefont
  {Thingstad}, \citenamefont {Kamra}, \citenamefont {Brataas},\ and\
  \citenamefont {Sudb{\o}}}]{thingstad2019chiral}%
  \BibitemOpen
  \bibfield  {author} {\bibinfo {author} {\bibfnamefont {E.}~\bibnamefont
  {Thingstad}}, \bibinfo {author} {\bibfnamefont {A.}~\bibnamefont {Kamra}},
  \bibinfo {author} {\bibfnamefont {A.}~\bibnamefont {Brataas}}, \ and\
  \bibinfo {author} {\bibfnamefont {A.}~\bibnamefont {Sudb{\o}}},\ }\href@noop
  {} {\bibfield  {journal} {\bibinfo  {journal} {Physical review letters}\
  }\textbf {\bibinfo {volume} {122}},\ \bibinfo {pages} {107201} (\bibinfo
  {year} {2019})}\BibitemShut {NoStop}%
\bibitem [{\citenamefont {Park}\ and\ \citenamefont
  {Yang}(2020)}]{park2020phonon}%
  \BibitemOpen
  \bibfield  {author} {\bibinfo {author} {\bibfnamefont {S.}~\bibnamefont
  {Park}}\ and\ \bibinfo {author} {\bibfnamefont {B.-J.}\ \bibnamefont
  {Yang}},\ }\href@noop {} {\bibfield  {journal} {\bibinfo  {journal} {Nano
  Letters}\ }\textbf {\bibinfo {volume} {20}},\ \bibinfo {pages} {7694}
  (\bibinfo {year} {2020})}\BibitemShut {NoStop}%
\bibitem [{\citenamefont {Juraschek}\ and\ \citenamefont
  {Spaldin}(2019)}]{juraschek2019orbital}%
  \BibitemOpen
  \bibfield  {author} {\bibinfo {author} {\bibfnamefont {D.~M.}\ \bibnamefont
  {Juraschek}}\ and\ \bibinfo {author} {\bibfnamefont {N.~A.}\ \bibnamefont
  {Spaldin}},\ }\href@noop {} {\bibfield  {journal} {\bibinfo  {journal}
  {Physical Review Materials}\ }\textbf {\bibinfo {volume} {3}},\ \bibinfo
  {pages} {064405} (\bibinfo {year} {2019})}\BibitemShut {NoStop}%
\bibitem [{\citenamefont {Xiao}\ \emph {et~al.}(2016)\citenamefont {Xiao},
  \citenamefont {Li},\ and\ \citenamefont {Ma}}]{xiao2016thermoelectric}%
  \BibitemOpen
  \bibfield  {author} {\bibinfo {author} {\bibfnamefont {C.}~\bibnamefont
  {Xiao}}, \bibinfo {author} {\bibfnamefont {D.}~\bibnamefont {Li}}, \ and\
  \bibinfo {author} {\bibfnamefont {Z.}~\bibnamefont {Ma}},\ }\href@noop {}
  {\bibfield  {journal} {\bibinfo  {journal} {Frontiers of Physics}\ }\textbf
  {\bibinfo {volume} {11}},\ \bibinfo {pages} {117201} (\bibinfo {year}
  {2016})}\BibitemShut {NoStop}%
\bibitem [{\citenamefont {Dyrda{\l}}\ \emph {et~al.}(2013)\citenamefont
  {Dyrda{\l}}, \citenamefont {Inglot}, \citenamefont {Dugaev},\ and\
  \citenamefont {Barna{\'s}}}]{dyrdal2013thermally}%
  \BibitemOpen
  \bibfield  {author} {\bibinfo {author} {\bibfnamefont {A.}~\bibnamefont
  {Dyrda{\l}}}, \bibinfo {author} {\bibfnamefont {M.}~\bibnamefont {Inglot}},
  \bibinfo {author} {\bibfnamefont {V.}~\bibnamefont {Dugaev}}, \ and\ \bibinfo
  {author} {\bibfnamefont {J.}~\bibnamefont {Barna{\'s}}},\ }\href@noop {}
  {\bibfield  {journal} {\bibinfo  {journal} {Physical Review B}\ }\textbf
  {\bibinfo {volume} {87}},\ \bibinfo {pages} {245309} (\bibinfo {year}
  {2013})}\BibitemShut {NoStop}%
\bibitem [{\citenamefont {Dyrda{\l}}\ \emph {et~al.}(2018)\citenamefont
  {Dyrda{\l}}, \citenamefont {Barna{\'s}}, \citenamefont {Dugaev},\ and\
  \citenamefont {Berakdar}}]{dyrdal2018thermally}%
  \BibitemOpen
  \bibfield  {author} {\bibinfo {author} {\bibfnamefont {A.}~\bibnamefont
  {Dyrda{\l}}}, \bibinfo {author} {\bibfnamefont {J.}~\bibnamefont
  {Barna{\'s}}}, \bibinfo {author} {\bibfnamefont {V.}~\bibnamefont {Dugaev}},
  \ and\ \bibinfo {author} {\bibfnamefont {J.}~\bibnamefont {Berakdar}},\
  }\href@noop {} {\bibfield  {journal} {\bibinfo  {journal} {Physical Review
  B}\ }\textbf {\bibinfo {volume} {98}},\ \bibinfo {pages} {075307} (\bibinfo
  {year} {2018})}\BibitemShut {NoStop}%
\bibitem [{\citenamefont {Winkler}\ \emph {et~al.}(2003)\citenamefont
  {Winkler}, \citenamefont {Papadakis}, \citenamefont {De~Poortere},\ and\
  \citenamefont {Shayegan}}]{winkler2003spin}%
  \BibitemOpen
  \bibfield  {author} {\bibinfo {author} {\bibfnamefont {R.}~\bibnamefont
  {Winkler}}, \bibinfo {author} {\bibfnamefont {S.}~\bibnamefont {Papadakis}},
  \bibinfo {author} {\bibfnamefont {E.}~\bibnamefont {De~Poortere}}, \ and\
  \bibinfo {author} {\bibfnamefont {M.}~\bibnamefont {Shayegan}},\ }\href@noop
  {} {\emph {\bibinfo {title} {Spin-Orbit Coupling in Two-Dimensional Electron
  and Hole Systems}}},\ Vol.~\bibinfo {volume} {41}\ (\bibinfo  {publisher}
  {Springer},\ \bibinfo {year} {2003})\BibitemShut {NoStop}%
\bibitem [{\citenamefont {Dresselhaus}\ \emph {et~al.}(2007)\citenamefont
  {Dresselhaus}, \citenamefont {Dresselhaus},\ and\ \citenamefont
  {Jorio}}]{dresselhaus2007group}%
  \BibitemOpen
  \bibfield  {author} {\bibinfo {author} {\bibfnamefont {M.~S.}\ \bibnamefont
  {Dresselhaus}}, \bibinfo {author} {\bibfnamefont {G.}~\bibnamefont
  {Dresselhaus}}, \ and\ \bibinfo {author} {\bibfnamefont {A.}~\bibnamefont
  {Jorio}},\ }\href@noop {} {\emph {\bibinfo {title} {Group theory: application
  to the physics of condensed matter}}}\ (\bibinfo  {publisher} {Springer
  Science \& Business Media},\ \bibinfo {year} {2007})\BibitemShut {NoStop}%
\bibitem [{\citenamefont {Eguchi}\ \emph {et~al.}(1980)\citenamefont {Eguchi},
  \citenamefont {Gilkey},\ and\ \citenamefont
  {Hanson}}]{eguchi1980gravitation}%
  \BibitemOpen
  \bibfield  {author} {\bibinfo {author} {\bibfnamefont {T.}~\bibnamefont
  {Eguchi}}, \bibinfo {author} {\bibfnamefont {P.~B.}\ \bibnamefont {Gilkey}},
  \ and\ \bibinfo {author} {\bibfnamefont {A.~J.}\ \bibnamefont {Hanson}},\
  }\href@noop {} {\bibfield  {journal} {\bibinfo  {journal} {Physics reports}\
  }\textbf {\bibinfo {volume} {66}},\ \bibinfo {pages} {213} (\bibinfo {year}
  {1980})}\BibitemShut {NoStop}%
\bibitem [{\citenamefont {Hehl}\ \emph {et~al.}(1976)\citenamefont {Hehl},
  \citenamefont {Von~der Heyde}, \citenamefont {Kerlick},\ and\ \citenamefont
  {Nester}}]{hehl1976general}%
  \BibitemOpen
  \bibfield  {author} {\bibinfo {author} {\bibfnamefont {F.~W.}\ \bibnamefont
  {Hehl}}, \bibinfo {author} {\bibfnamefont {P.}~\bibnamefont {Von~der Heyde}},
  \bibinfo {author} {\bibfnamefont {G.~D.}\ \bibnamefont {Kerlick}}, \ and\
  \bibinfo {author} {\bibfnamefont {J.~M.}\ \bibnamefont {Nester}},\
  }\href@noop {} {\bibfield  {journal} {\bibinfo  {journal} {Reviews of Modern
  Physics}\ }\textbf {\bibinfo {volume} {48}},\ \bibinfo {pages} {393}
  (\bibinfo {year} {1976})}\BibitemShut {NoStop}%
\bibitem [{\citenamefont {Zhou}\ \emph {et~al.}(2015)\citenamefont {Zhou},
  \citenamefont {Chang},\ and\ \citenamefont {Xiao}}]{zhou2015plasmon}%
  \BibitemOpen
  \bibfield  {author} {\bibinfo {author} {\bibfnamefont {J.}~\bibnamefont
  {Zhou}}, \bibinfo {author} {\bibfnamefont {H.-R.}\ \bibnamefont {Chang}}, \
  and\ \bibinfo {author} {\bibfnamefont {D.}~\bibnamefont {Xiao}},\ }\href@noop
  {} {\bibfield  {journal} {\bibinfo  {journal} {Physical Review B}\ }\textbf
  {\bibinfo {volume} {91}},\ \bibinfo {pages} {035114} (\bibinfo {year}
  {2015})}\BibitemShut {NoStop}%
\bibitem [{\citenamefont {Barkeshli}\ \emph {et~al.}(2012)\citenamefont
  {Barkeshli}, \citenamefont {Chung},\ and\ \citenamefont
  {Qi}}]{barkeshli2012dissipationless}%
  \BibitemOpen
  \bibfield  {author} {\bibinfo {author} {\bibfnamefont {M.}~\bibnamefont
  {Barkeshli}}, \bibinfo {author} {\bibfnamefont {S.~B.}\ \bibnamefont
  {Chung}}, \ and\ \bibinfo {author} {\bibfnamefont {X.-L.}\ \bibnamefont
  {Qi}},\ }\href@noop {} {\bibfield  {journal} {\bibinfo  {journal} {Physical
  Review B}\ }\textbf {\bibinfo {volume} {85}},\ \bibinfo {pages} {245107}
  (\bibinfo {year} {2012})}\BibitemShut {NoStop}%
\bibitem [{\citenamefont {Landau}\ \emph {et~al.}(1986)\citenamefont {Landau},
  \citenamefont {Pitaevskii}, \citenamefont {Kosevich},\ and\ \citenamefont
  {Lifshitz}}]{landau2013course}%
  \BibitemOpen
  \bibfield  {author} {\bibinfo {author} {\bibfnamefont {L.~D.}\ \bibnamefont
  {Landau}}, \bibinfo {author} {\bibfnamefont {L.~P.}\ \bibnamefont
  {Pitaevskii}}, \bibinfo {author} {\bibfnamefont {A.~M.}\ \bibnamefont
  {Kosevich}}, \ and\ \bibinfo {author} {\bibfnamefont {E.~M.}\ \bibnamefont
  {Lifshitz}},\ }\href@noop {} {\emph {\bibinfo {title} {Theory of Elasticity:
  Volume 7 (Course of Theoretical Physics)}}}\ (\bibinfo  {publisher}
  {Elsevier},\ \bibinfo {year} {1986})\BibitemShut {NoStop}%
\bibitem [{\citenamefont {Avron}\ \emph {et~al.}(1995)\citenamefont {Avron},
  \citenamefont {Seiler},\ and\ \citenamefont {Zograf}}]{avron1995viscosity}%
  \BibitemOpen
  \bibfield  {author} {\bibinfo {author} {\bibfnamefont {J.}~\bibnamefont
  {Avron}}, \bibinfo {author} {\bibfnamefont {R.}~\bibnamefont {Seiler}}, \
  and\ \bibinfo {author} {\bibfnamefont {P.~G.}\ \bibnamefont {Zograf}},\
  }\href@noop {} {\bibfield  {journal} {\bibinfo  {journal} {Physical review
  letters}\ }\textbf {\bibinfo {volume} {75}},\ \bibinfo {pages} {697}
  (\bibinfo {year} {1995})}\BibitemShut {NoStop}%
\end{thebibliography}
\end{document}